\documentclass[fleqn,usenatbib]{mnras}

\usepackage[T1]{fontenc}

\DeclareRobustCommand{\VAN}[3]{#2}
\let\VANthebibliography\thebibliography
\def\thebibliography{\DeclareRobustCommand{\VAN}[3]{##3}\VANthebibliography}

\usepackage{graphicx}	% Including figure files
\usepackage{amsmath}	% Advanced maths commands
\usepackage{amssymb}	% Extra maths symbols
\usepackage{xcolor}

\usepackage{booktabs} % cmidrule 
\newcommand{\beq}{\begin{equation}}
\newcommand{\eeq}{\end{equation}}
\newcommand{\beqn}{\begin{equation*}}
\newcommand{\eeqn}{\end{equation*}}
\newcommand{\rhoenv}{\rho_{\rm env}}

\newcommand{\Tenv}{T_{\rm env}}

\newcommand{\Tiso}{T_{\rm iso}}

\newcommand{\rcool}{r_{\rm cool}}

\newcommand{\Tinit}{T_{\rm init}}
\newcommand{\Pinit}{P_{\rm init}}
\newcommand{\fiso}{f_{\rm iso}}
\newcommand{\rdamp}{r_{\rm damp}}
\newcommand{\rhoinit}{\rho_{\rm init}}
\newcommand{\mconv}{\mathcal{M}_{\rm turb}}
\newcommand{\rcirc}{r_{\rm circ}}
\newcommand{\avercirc}{\overline{r}_{\rm circ}}
\newcommand{\risco}{r_{\rm ISCO}}
\newcommand{\jisco}{j_{\rm ISCO}}
\newcommand{\jrand}{j_{\rm rand}}
\newcommand{\vjrand}{\bmath{j}_{\rm rand}}
\newcommand{\ujrand}{\hat{{\jmath}}_{\rm rand}}
\newcommand{\avej}{\overline{\jmath}_{\rm rand}}
\newcommand{\avejvec}{\bmath{\overline{\jmath}}_{\rm rand}}
\newcommand{\vc}{v_{\rm c}}
\newcommand{\ath}{\texttt{Athena++ }}
\newcommand{\mesa}{\texttt{MESA }}

\newcommand{\mesans}{\texttt{MESA}}
\newcommand{\msun}{M_\odot}
\newcommand{\rsun}{R_\odot}

\newcommand{\ujacc}{\hat{\jmath}_{\rm acc}}
\newcommand{\vjacc}{\bmath{j}_{\rm acc}}
\newcommand{\jacc}{j_{\rm acc}}

\definecolor{andrea}{HTML}{FF376A}

\title[Convective supergiant collapse]{Numerical Simulations of the Random Angular Momentum in Convection: Implications for Supergiant Collapse to Form Black Holes}

\author[A. Antoni and E. Quataert]{
Andrea Antoni$^{1}$\thanks{E-mail: aantoni@berkeley.edu}\href{https://orcid.org/0000-0003-3062-4773}{\includegraphics[width=9pt]{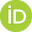}}
and Eliot Quataert$^{1,2}$\href{https://orcid.org/0000-0001-9185-5044}{\includegraphics[width=9pt]{ORCID32}}
\\
$^{1}$Astronomy Department and Theoretical Astrophysics Center, University of California, Berkeley, CA 94720, USA\\
$^{2}$Department of Astrophysical Sciences, Princeton University, Princeton, NJ 08544, USA\\
}

\date{Accepted XXX. Received YYY; in original form ZZZ}

\pubyear{2021}

\begin{document}
\label{firstpage}
\pagerange{\pageref{firstpage}--\pageref{lastpage}}
\maketitle

\begin{abstract}
%single paragraph not more than 250 words (200 words for Letters).
%No references should appear in the abstract.
During the core collapse of massive stars that do not undergo a canonical energetic explosion, some of the hydrogen envelope of a red supergiant (RSG) progenitor may infall onto the newborn black hole (BH).  Within the \texttt{Athena++} framework, we perform three-dimensional, hydrodynamical simulations of idealized models of supergiant convection and collapse in order to assess whether the infall of the convective envelope can give rise to rotationally-supported material, even if the star has zero angular momentum overall. Our dimensionless, polytropic models are applicable to the optically-thick hydrogen envelope of non-rotating RSGs and cover a factor of 20 in stellar radius.  At all radii, the specific angular momentum due to random convective flows implies associated circularization radii of 10 - 1500 times the innermost stable circular orbit of the BH.  During collapse, the angular momentum vector of the convective flows is approximately conserved and is slowly varying on the timescale relevant to forming disks at small radii.  Our results indicate that otherwise failed explosions of RSGs lead to the formation of rotationally-supported flows that are capable of driving outflows to large radii and powering observable transients. When the BH is able to accrete most of the hydrogen envelope, the final BH spin parameter is $\sim$ 0.5, even though the star is non-rotating. For fractional accretion of the envelope, the spin parameter is generally lower and never exceeds 0.8. We discuss the implications of our results for transients produced by RSG collapse to a black hole.  
\end{abstract}

\begin{keywords}
black hole physics -- convection -- stars: massive -- supernovae: general.
\end{keywords}

%%%%%%%%%%%%%%%%%%%%%%%%%%%%%%%%%%%%%%%%%%%%%%%%%%

%%%%%%%%%%%%%%%%% BODY OF PAPER %%%%%%%%%%%%%%%%%%

\section{Introduction}
\label{sec:intro}
It is plausible that in some fraction of massive stellar deaths,  iron core collapse does not lead to a successful neutrino-driven supernova \citep[SN;][]{2008ApJ...679..639Z,2011ApJ...730...70O,2012ApJ...757...69U,2016ApJ...821...38S}. Instead, core collapse results in a failed supernova (FSN) in which the proto-neutron star (PNS) quickly collapses to a black hole (BH) and a large fraction of the star accretes onto the newborn BH.  The outcome of stellar core collapse depends on the final pre-SN structure of the star in a complex way \citep{2011ApJ...730...70O,2012ApJ...757...69U,2016ApJ...818..124E,2020MNRAS.492.2578S} which probably cannot be disentangled  without a statistically significant sample of (4$\pi$) three-dimensional (3D) simulations of both the pre-SN \citep{2015ApJ...808L..21C,2020ApJ...890...94Y,2020ApJ...901...33F} and post-bounce \citep{2020MNRAS.491.2715B,2020MNRAS.494.4665P} phases of core collapse.  The result of the complex relationship between the explodability of a massive star and its pre-SN evolution is that explodability is likely neither a monotonic nor a singled-valued function of zero-age main sequence (ZAMS) mass \citep{2018ApJ...860...93S}. 

The angular momentum content of a collapsing star has important consequences for the outcome of core collapse, including the properties of the remnant, the amount of material returned to the stellar environment, and the transient signal produced.  Collapse of rotating Wolf Rayet (WR) stars in successful or failed SN (i.e. Type II or Type I collapsars, respectively) may power some long gamma-ray bursts \citep{1993ApJ...405..273W,1999ApJ...524..262M}. Accretion of high angular momentum material in more extended rotating stars, e.g. red and blue supergiants (RSGs and BSGs, respectively), in weak or failed SN may be responsible for ultra-long gamma-ray transients \citep{2012ApJ...752...32W,2012MNRAS.419L...1Q,2018ApJ...859...48P}.  

Even in the absence of net angular momentum, FSN can still give rise to detectable transients.    During the PNS phase, neutrino losses reduce the gravitational mass of the PNS by $\sim 0.1-0.3$ $\msun$ over a few seconds.  The envelope of the star, which is suddenly over-pressured due to the nearly instantaneous reduction of the gravitational potential, reacts on a dynamical time.  A sound pulse forms and steepens into a weak shock that may unbind a portion of the envelope of the star \citep{1980Ap&SS..69..115N,2013ApJ...769..109L,2018MNRAS.476.2366F}.  For RSG progenitors, the ejection of several $\msun$ of the hydrogen envelope results in a $\approx$ few day shock breakout followed by a $\approx 300 - 400$ day transient of $\approx 10^{39} - 10^{40}$ erg s$^{-1}$ powered by hydrogen recombination in the expanding ejecta \citep{2013ApJ...769..109L,2013ApJ...768L..14P,2017ApJ...845..103L,2018MNRAS.476.2366F}.   The brightening and subsequent disappearance of a RSG progenitor in the FSN candidate event N6946-BH1 is broadly consistent with this model \citep{2015MNRAS.450.3289G,2017MNRAS.468.4968A,2020arXiv200715658B,2021arXiv210403318N}.  The neutrino-mass-loss mechanism can also launch weak shocks in yellow supergiants (YSGs), BSGs, and WRs, but the ejected masses are much lower and some progenitors eject no unbound mass at all \citep{2018MNRAS.476.2366F}.   For semi-analytic work related to mass ejection from weak shocks, see \citet{2018MNRAS.477.1225C,2018ApJ...863..158C,2019ApJ...874...58C,2019ApJ...878..150R,2021MNRAS.501.4266L}.

A known limitation of \citet{2013ApJ...769..109L} and \cite{2018MNRAS.476.2366F}'s calculations of how much neutrino losses occur prior to BH formation was the use of analytic functions to model the neutrino emission. \citet{2021ApJ...911....6I} improves on this earlier work by using general-relativistic neutrino radiation-hydrodynamic simulations to model the evolution of the inner core of each progenitor to BH-formation using three different equations-of-state (EOSs) for the PNS.  Self-consistently modeling the core collapse for a given progenitor is important because the neutrino-loss function and time to BH formation set the amount of neutrino energy radiated and thus the energy of the outgoing weak shock. For their RSG progenitor \citeauthor{2021ApJ...911....6I} find that this more careful treatment of BH-formation and a softer equation of state reduces the ejected mass by a factor of a few and means that $1-3$~$\msun$ of the hydrogen envelope remains bound and will accrete onto the newborn BH.  For comparison, the characteristic mass loss from their YSG and lower-mass BSG progenitors is $\approx 10^{-2}\msun$ while the remaining BSG and WR progenitors eject $\approx 10^{-4} \msun$ in unbound mass. This work suggests that the infall of a large fraction of the hydrogen envelope is common in FSN of RSGs and YSGs. The accretion of this material is the focus of our study.

Rotation is not the only source of angular momentum in FSN progenitors. The random velocity fields in convective zones have an associated angular momentum because in each radial shell there is a mean `horizontal' velocity perpendicular to the radial axis even if the net angular momentum of the star is zero.   For each convective zone, the key question is whether the mean specific angular momentum arising due to convection, $\jrand$, is larger than $\jisco$, the Keplerian specific angular momentum associated with the innermost stable circular orbit (ISCO) of the BH.  \citet{2014MNRAS.439.4011G} first explored the angular momentum present in the convective regions of pre-SN stars in the context of jet-driven SN \citep{2010MNRAS.401.2793S,2011MNRAS.416.1697P}. 
They derived an analytical estimate for $\jrand$ contained in a shell of material with randomly oriented velocities of magnitude $\vc$, where $\vc$ is the convective velocity.  For a sample of pre-SN stars computed with \mesa \citep{2011ApJS..192....3P,2013ApJS..208....4P,2015ApJS..220...15P,2018ApJS..234...34P,2019ApJS..243...10P}, their analytical estimate predicts significant angular momentum content in the helium and hydrogen layers of their supergiant models (specifically $\jrand > 10^5 \jisco$ in the hydrogen envelopes of their RSG and YSG models).  They argue that the fallback of the helium layer is sufficient to drive a canonical ($> 10^{51}$ erg) SN, thus removing the possibility of a FSN.

\citet*{Quataert2019}, on the other hand, derived an analytical scaling for $\jrand$ as a function of $\vc$ that is at least two orders-of-magnitude smaller than the result of \citet{2014MNRAS.439.4011G}.  To validate their expression for $\jrand$, \citeauthor{Quataert2019} performed Boussinesq simulations of convection in a Cartesian slab of material, which confirmed their analytic scaling within a factor of 2.  \citet{2016ApJ...827...40G} carried out 3D simulations of the convective helium zones in massive stars; their simulations are also consistent with \citeauthor{Quataert2019}'s estimate (to within a factor of a few).   When applied to \mesa models of a RSG and a YSG, \citeauthor{Quataert2019} found that $\jrand$ is significant only in the hydrogen envelope. That is, material interior to the base of the hydrogen envelope could accrete spherically onto the BH, but the hydrogen envelope had sufficient angular momentum due to random convective flows that the material would circularize beyond the ISCO of the BH.  

The aim of our work is to determine whether the collapse of a non-rotating supergiant envelope in a FSN results in rotationally-supported material outside the ISCO of the newborn BH.    We perform two numerical experiments.  In the first, we simulate convection in a spherical polytropic envelope in order to relate the convective velocity, $\vc$, to the magnitude of the specific angular momentum, $\jrand$, arising from the random motion of the convective fluid.  In the second experiment, we follow the collapse of the envelope in order to study the evolution of $\jrand$ during infall and to measure the time-dependent angular momentum vector of the material accreted onto the BH.  Our study thus extends the results of \citet{Quataert2019} to the spherical geometry of a star and determines how well $\jrand$ of the convective shells is conserved during infall.  

This paper is organized as follows.  Section \ref{sec:background} derives the analytical scaling for $\jrand$ of \citet{Quataert2019}.  We describe our supergiant model and simulation methods in Section \ref{sec:methods}.  Sections \ref{sec:convectresults} and \ref{sec:collapseresults}, respectively, present the results of our convection and collapse simulations.  We place these results in the context of supergiants and FSN in Section \ref{sec:discussion}.  Our summary and conclusions are given in Section \ref{sec:conclusion}.  Appendix \ref{sec:appxpredict} describes how accretion rates in the collapse simulations can be predicted from snapshots of the flow prior to collapse using a simple test problem (we apply this technique to convective flows in Section \ref{sec:predict}). 

%%%%%%%%%%% BACKGROUND %%%%%%%%%%%%%%%%%%%%%%%%%%%%%%%%%%%
\section{Background}
\label{sec:background}
\citet{Quataert2019}'s analytical estimate for $j_{\rm rand}$ will be useful in what follows, so we review the derivation here. Consider a spherical shell of the convective envelope at radius $r$ and with thickness $H$, where $H$ is the local pressure scale height.  The velocities of the convective material have components along $r$, which do not give rise to angular momentum, and `horizontal' components (perpendicular to $r$) that do.  These velocities are randomly oriented so a sum over an infinite number of eddies should result in zero net angular momentum. But the volume of the shell is finite so there is a finite number of eddies at radius $r$, given approximately by
$$N_{\rm edd} \sim  \frac{4\pi r^2 H}{H^3} \sim 4\pi\frac{r^2}{H^2}.$$  
The convective velocity, $\vc$, and the magnitude of the mean velocity vector are related by 
$\|\langle\vec{v}\rangle\| \sim \vc / \sqrt{N_{\rm edd}}$. So to a factor of order unity, the mean horizontal velocity is related to the convective velocity by,
\beq
v_h \approx \frac{\vc}{\sqrt{4\pi}}\frac{H}{r}.
\eeq
This implies a specific angular momentum content of 
\beq
\jrand \sim r v_h \sim \frac{\vc H}{\sqrt{4\pi}}.
\label{eq:jrand}
\eeq
Or, normalizing to typical values for the convective regions of supergiants,
\beq
\jrand \sim  6 \times 10^{18} \, \bigg(\frac{r}{1000\rsun}\bigg)\bigg(\frac{\vc}{10 \,{\rm km}/{\rm s}}\bigg)\bigg(\frac{H/r}{0.3}\bigg) \, {\rm cm}^2 \, {\rm s}^{-1}
\label{eq:jrand_quataert}
\eeq
For comparison,
\beq
\jisco = 1.15 - 3.46 \frac{G M_\bullet}{c} \approx 0.5 - 1.5 \times 10^{17} \bigg(\frac{M_\bullet}{10\msun} \bigg)\, {\rm cm}^2 \, {\rm s}^{-1}
\label{eq:jisco}
\eeq
 where $M_\bullet$ is the mass of the BH.  The lower limit corresponds to maximum BH spin (and prograde orbits) and the upper limit corresponds to zero BH spin. 

The upper panel of fig.~3 of \citet{Quataert2019} plots eq.~\eqref{eq:jrand} for \mesa models of a $15 \msun$ RSG and a $22\msun$ YSG.  Material interior to the hydrogen envelope has $\jrand \ll \jisco$, implying spherical accretion of the helium shell.  The hydrogen envelope of both the RSG and YSG models have $ 1 \lesssim \jrand / \jisco \lesssim 10$ and therefore may have sufficient angular momentum to feed a rotationally-supported structure outside the BH horizon. 
 
We emphasize that, in the preceding picture, the total angular momentum of the star is zero. Random velocities give rise to non-zero mean angular momentum in a given shell even though the sum over the star is zero.  The key question we study in this paper is the random angular momentum flows in spherical convective zones and how, and whether, the envelope material restructures itself during infall. 

%%%%%%%%%%%%%%%%%%%%%%%%%%%%%%%%%%%%%%%%%%%%%%%%%%%%%%%%
%%%%%%%%%%%% Section 3 %%%%%%%%%%%%%%%%%%%%%%%%%%%%%%%%%
%%%%%%%%%%%%%%%%%%%%%%%%%%%%%%%%%%%%%%%%%%%%%%%%%%%%%%%%
\section{Numerical Methods}
\label{sec:methods}
We model the collapse of non-rotating, convective supergiant envelopes in 3D using 
\ath\footnote{Version 19.0, \href{https://princetonuniversity.github.io/athena}{https://princetonuniversity.github.io/athena}} \citep{2020ApJS..249....4S}, an Eulerian hydrodynamic code based on \texttt{Athena} \citep{2008ApJS..178..137S}.   As the star is non-rotating, there is no symmetry axis to motivate the use of spherical-polar coordinates.  We instead perform our simulations in a Cartesian grid. All simulations use third-order Runge-Kutta time integration (\texttt{integrator = rk3}), piecewise parabolic spatial reconstruction (\texttt{xorder = 3}), and the Harten-Lax-van Leer contact Riemann solver (\verb|--flux hllc|).  Our supergiant model initially has radial symmetry and we use nested static mesh refinement to allow resolution to fall off approximately logarithmically with radial distance from the center of the star. 

We use a gamma-law EOS and neglect the self-gravity of the gas.  We do not include radiation transport, but we do include a simplified cooling function at the surface of the star to remove heat from the rising convective fluid.   We include a heating source term at small radii to drive convection in the envelope and we run our simulations until thermal equilibrium is reached.  Our simulations solve a model problem that is similar to RSGs in many key ways.  We have a convective polytrope with a density and pressure profile that is representative of the hydrogen envelope of RSGs with roughly the right convective velocities. 

One of the main goals of this work is to study the collapse phase of FSN. In particular, we aim to understand whether there is reshuffling of the angular momentum during the infall or whether the angular momentum content of the convective zone prior to collapse is a good proxy for the accreted angular momentum during collapse.  We use our simulations of convection as the initial conditions for the collapse calculations and our main focus is on the change (or not) in angular momentum content of the material from the convection zone through the collapse. 

\subsection{Equations Solved} %%%%%%%%%%%%%%%%%%%%%%%%%%%%%%%%%%%%%%%%%%%%%
We solve the equations of inviscid hydrodynamics 

\beq
\frac{\upartial \rho}{\upartial t} + \nabla\cdot(\rho \bmath{v}) = 0
\eeq
\beq
\frac{\upartial (\rho{\bmath v})}{\upartial t} + \nabla\cdot(\rho \bmath{v}\bmath{v} + P \mathbfss{I}) = - \rho \nabla \Phi + \rho \bmath{a}_{\rm damp}
\label{eq:momentum}
\eeq
\beq
\frac{\upartial E}{\upartial t} + \nabla\cdot\big[(E + P)\bmath{v}\big] = -\rho \bmath{v}\cdot \nabla \Phi - Q_{\rm cool} + Q_{\rm heat} + \rho \bmath{v}\cdot \bmath{a}_{\rm damp}
\label{eq:energy}
\eeq
where $\rho$ is the mass density, $\rho \bmath{v}$ is the momentum density, $E = \epsilon + \rho \bmath{v}\cdot\bmath{v}/2$ is the total energy density, $\epsilon$ is the thermal energy density, $P$ is the gas pressure, and \mathbfss{I} is the identity tensor. The source terms for gravity, cooling, heating, and damping ($\Phi$, $Q_{\rm cool}$, $Q_{\rm heat}$, and $\bmath{a}_{\rm damp}$, respectively) are defined in Sec.~\ref{sec:sourceterms}. We adopt an ideal gas EOS
\beq
\epsilon = \frac{P}{\gamma -1}
\eeq
with adiabatic index $\gamma$. 

\subsection{Supergiant Model} %%%%%%%%%%%%%%%%%%%%%%%%%%%%%%%%%%%%%%%%%%%%%%
In this section we introduce our semi-analytic model for the supergiant envelope and atmosphere.  Section \ref{sec:initialization} will describe how this model is mapped to \ath to initialize the grid.  

%\subsubsection{Initial Hydrostatic Equilibrium Profile}
As discussed in Section \ref{sec:background}, we expect everything interior to the base of the hydrogen envelope to have $\jrand \ll \jisco$ and to be accreted onto the core at the start of collapse.  We represent all of this mass, $M$, with a Plummer potential 
\beq
\Phi(r) = -\frac{GM}{(r^n + a^n)^\frac{1}{n}}
\label{eq:potential}
\eeq
with softening length $a$ and index $n$. We take $n = 8$, so that $\Phi(r)$ converges to a point mass potential for $r \gtrsim 1.4 a$.  

Supergiant envelopes have roughly power-law density profiles with nearly constant entropy, so for $r \gtrsim a$ we will model the hydrogen envelope with a density profile of the form $\rho \propto r^{-b}$.  For RSGs and YSGs, $1.5 \lesssim b \lesssim 2.5$ \citep[see][figs.~7 and 12]{2018ApJ...863..158C}.  For an ideal gas, a power-law density profile gives the envelope a temperature profile of $T \propto r^{-1}$.  To mimic the stellar photosphere where cooling is rapid, the temperature is smoothly dropped to an isothermal value for $r \gtrsim \rcool$.  The resultant temperature profile has the functional form 

\beq
\Tinit(r) = \frac{\Tenv(r)}{2}\big[1 - F(r)\big] + \frac{\Tiso}{2}\big[1+F(r)\big]
\label{eq:temperartureprof}
\eeq
where
\beq
\Tenv(r) = \frac{\mu m_p}{k_b (b + 1)} \frac{GM}{(r^n + a^n)^\frac{1}{n}}, 
\label{eq:Tenv} 
\eeq
\beq
\Tiso = \fiso\Tenv(\rcool + 0.5r_0),
\label{eq:tiso}
\eeq
and 
\beq
F(r) = \tanh\bigg[\frac{r - \rcool - 0.5r_0}{0.9r_0}\bigg].
\label{eq:tanhfunc}
\eeq
In the last expression, $r_0$ is the characteristic length scale in our model.  In our code units, $r_0 = 1$ (Sec.~\ref{sec:units}) and our simulations adopt $\rcool = 5 r_0$, $b = 2.1$, $\fiso=0.3$, and $a \le 0.16 r_0$ (Sec.~\ref{sec:parameters}).  

For $r \lesssim \rcool$, $\Tinit \sim \Tenv$, which is the power-law temperature profile in the envelope that results from assuming an ideal gas EOS and choosing the density profile for $r \lesssim \rcool$ to take the form 
\beq
\rhoenv(r) = \rho_0 \big[(r_0^n + a^n)/(r^n + a^n)\big]^\frac{b}{n},
\label{eq:rhoenv}
\eeq
where $\rho_0$ is the density at $r_0$ and  $\rho \propto r^{-b}$ for $r \gtrsim a$.  By contrast, for $r \gtrsim r_{\rm cool}$, $\Tinit \sim \Tiso =$ constant. 
 
The constants in eqs.~\eqref{eq:tiso} and \eqref{eq:tanhfunc} set the location and width of the transition region between the envelope temperature profile, $\Tenv(r)$, and the isothermal temperature, $\Tiso$. The shift of $0.5r_0$ ensures that the transition region does not begin until $\sim \rcool$ (otherwise the temperature profile would depart from a power law interior to $\rcool$).  With $\rcool = 5r_0$, the transition occurs between $5r_0$ and $7r_0$.   When scaling our supergiant model to the physical parameters of a star, we will associate this transition region with the photosphere in the stellar model (see Sec.~\ref{sec:units}). In our simulations, we also assume that convective material that reaches $r > \rcool$ can cool.  See Section \ref{sec:cooling} for a description of our cooling function implementation.

Given $\Tinit(r)$, the ideal gas law, and $\Phi(r)$, the equation of hydrostatic equilibrium takes the form
\beq
\frac{d\Pinit}{dr} = -\frac{\mu m_p}{k_b}\frac{\Pinit}{\Tinit}\frac{d\Phi}{dr}.
\label{eq:hse_integrate}
\eeq
We integrate eq.~\eqref{eq:hse_integrate} numerically to obtain the initial pressure profile, $\Pinit(r)$. The initial density profile, $\rhoinit(r)$ is obtained from the ideal gas law.  

\begin{figure*}
	\includegraphics[width=0.95\textwidth]{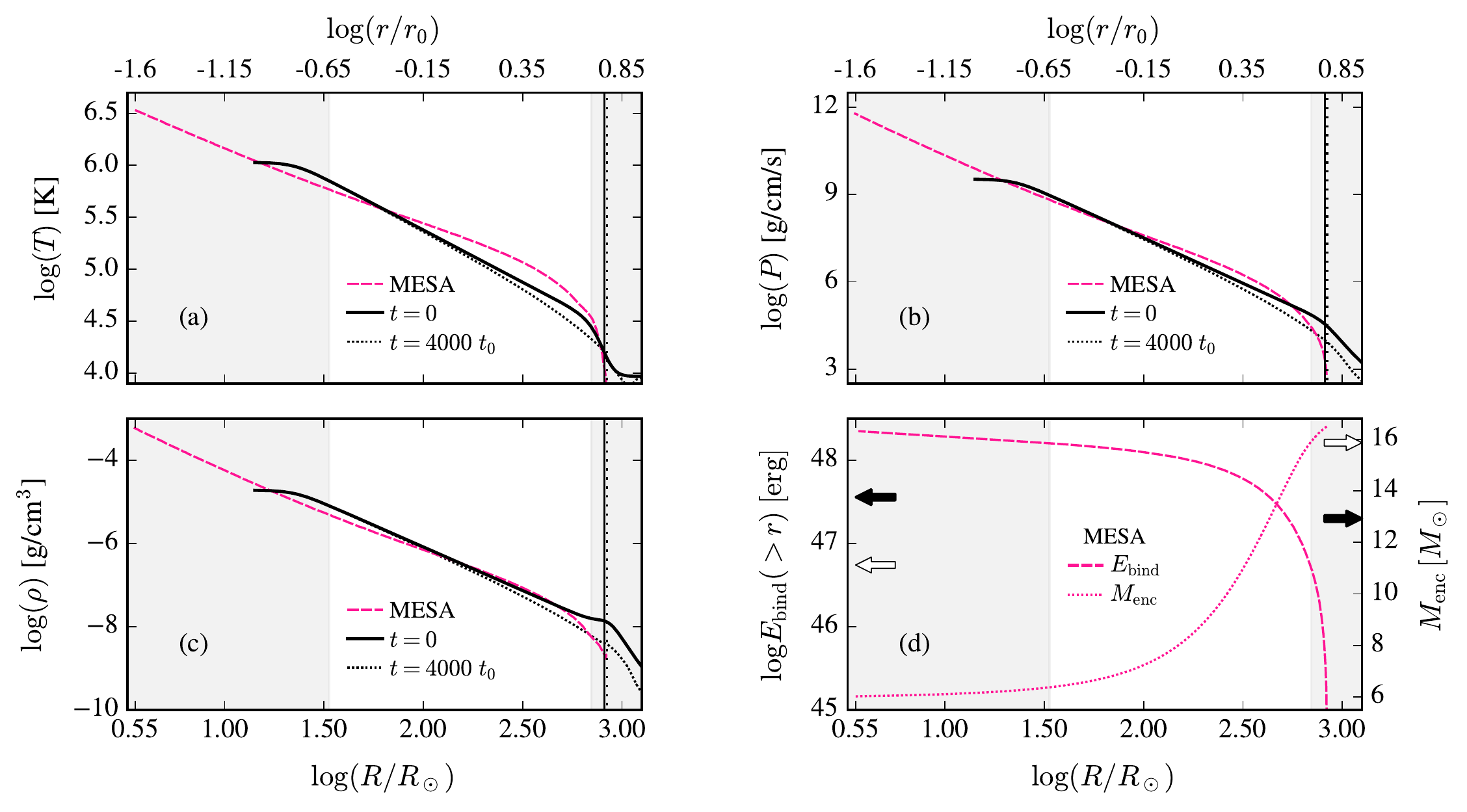}
    \caption{{\bf Panels (a)-(c):} Comparison of our \ath model (with Plummer softening length $a = 0.16$) to an 18$\msun$ (at ZAMS) supergiant model, computed with \mesans, near the end of oxygen burning.  The pink, dashed curve in each panel shows the \mesa model from the base of the convective hydrogen envelope at $R\approx 3.5\rsun$ to the photosphere at $r_{\rm ph} = 840\rsun$ (this is the outer $10.49\msun$ of the now $16.5\msun$ star). The black, solid lines are radial profiles of our \ath model showing temperature (a), pressure (b), and mass density (c) at initialization ($t = 0$), scaled to the \mesa model as described in Sec.~\ref{sec:units}. The black, dotted lines for $t =4000 t_0$ show the same quantities after the simulation achieves thermal equilibrium (see Sec.~\ref{sec:fid}). The bottom axes show radii in physical units and the upper $x$-axes give the original code units for reference. The vertical dotted line and vertical solid line, respectively, show $r_{\rm ph}$ and the $\tau = c / c_s$ radius ($R = 818\rsun$) in the \mesa model. To the left of these lines, the envelope is optically thick and roughly adiabatic.  Grey shading indicates where $r < r_{\rm heat, out} = 1.5 a$ or $r > r_{\rm cool}$.  The physically important part of our simulation domain, $r_{\rm heat, out} < r < \rcool$, falls between these two regions. The \ath model shown thus covers a range of $33 \rsun$ to $700 \rsun$ in radius for this \mesa model. Our polytropic envelope with $b = 2.1$ reproduces the power-law hydrogen envelope of the \mesa model well. {\bf Panel (d)}: The dashed line shows the binding energy of the envelope exterior to $R$ (left-hand $y$-axis) and the dotted line shows the mass enclosed at each $R$ (right-hand $y$-axis) for the \mesa model. The open and filled arrows on the mass axis mark the mass coordinates outside of which the envelope would be unbound due to a weak shock with the kinetic energy marked by a corresponding arrow on the binding energy axis. These shock energies bound the range of values found for RSGs by \citet{2021ApJ...911....6I}. If the weak shock has energy of $3.6 \times 10^{47}$ erg (filled arrows), then roughly $3.6\msun$ could be ejected, leaving $\sim 6.9\msun$ of the hydrogen envelope to collapse onto the BH.}
\label{fig:initial_profiles}
\end{figure*}

The functions $\Tinit(r)$, $\Pinit(r)$, and $\rhoinit(r)$ are plotted versus radius in panels (a)-(c) of Fig.~\ref{fig:initial_profiles} (black, solid curves).  The upper $x$-axes give $r$ in code units of $r_0$. The curves adopt $\rcool = 5r_0$ and $b = 2.1$, as we do in all of our simulations, and $a=0.16r_0$ as in our fiducial model. See Sec.~\ref{sec:units} for a complete description of the figure.

\subsection{Source Terms}
\label{sec:sourceterms}
In this section, we define the gravity, heating, cooling, and damping source terms appearing in the momentum and energy equations (eqs.~\ref{eq:momentum} and \ref{eq:energy}, respectively). All of these source terms are active during our convection simulations. Only gravity is active during our collapse simulations. 

\subsubsection{Gravity}
We neglect the self-gravity of the envelope gas, so $\Phi$ is simply eq.~\eqref{eq:potential}, implemented in Cartesian coordinates. 

\subsubsection{Heating}
\label{sec:heating}
We drive convection by including a heating source term in eq.~\eqref{eq:energy} that operates in the region $r_{\rm heat, in} \lesssim r \lesssim r_{\rm heat, out}$.  The constant, volume-integrated heating rate is
\beq
L_{\rm heat} = f_{\rm heat}\times 4\pi r_0^2 \rho_0 c_{s,0}^3
\label{eq:Lheat}
\eeq
where $c_{s,0}$ is the adiabatic sound speed at $r_0$ and $f_{\rm heat}$ is a constant.  The heating source term is then 
\beq
Q_{\rm heat} = \frac{L_{\rm heat}}{4V}\bigg[1 +\tanh\bigg(\frac{r - r_{\rm heat, in}}{0.03r_0}\bigg)\bigg]\bigg[1 -\tanh\bigg(\frac{r - r_{\rm heat, out}}{0.03r_0}\bigg)\bigg]
\label{eq:qcool}
\eeq
where $V$ is the volume of the heating region and the $\tanh$ factors smooth the boundaries of the heating annulus. Our convection simulations adopt $r_{\rm heat, in} = 0.9a$ and $r_{\rm heat, out} = 1.5a$. We choose $f_{\rm heat} = 6 \times 10^{-4}$ to roughly reproduce the convective Mach numbers of RSG envelopes (see Section \ref{sec:fid_mach}).

\subsubsection{Cooling}
\label{sec:cooling}
We mimic the stellar photosphere by including a cooling term in eq.~\eqref{eq:energy} given by
\beq
Q_{\rm cool} = 	\frac{\rho k_b\big[T -\Tinit\big]}{2\mu m_p(\gamma - 1)\tau_{\rm cool}}\bigg[1 +\tanh\bigg(\frac{r - \rcool}{0.3r_0}\bigg)\bigg]
\label{eq:qcool}
\eeq
if $T > \Tinit$ and $Q_{\rm cool} = 0$ if $T \le \Tinit$, where $\tau_{\rm cool} = f_{\rm cool}\sqrt{{\rcool^3}/{GM}}$ and the spatial dependence contained in the $\frac{1}{2}[1 + \tanh{x}]$ factor ensures that the cooling function only operates at $r \gtrsim \rcool$.  As hot convective material rises beyond $\rcool$, its temperature is brought back to $\Tinit(r)$, as is the case for convective material that reaches the photosphere in real stars.  The timescale over which this cooling occurs, $\tau_{\rm cool}$, is a fraction, $f_{\rm cool}$, of the dynamical time at $\rcool$. We set $f_{\rm cool} = 0.01$ in all simulations.

We include cooling and place $\rcool$ well inside the simulation domain because the rising convective parcels carry finite angular momentum.  Cooled convective material sinks back into the envelope region rather than exiting the domain and increasing the net angular momentum in the domain, as would occur if we instead extended the convective envelope to the grid boundaries.  

\subsubsection{Damping}
\label{sec:damping}
While cooling allows most of the convective material to return to $r \lesssim \rcool$, convection in this region still drives waves and some matter to larger radii, causing mass and angular momentum to leave the domain. This results in non-zero total angular momentum in the computation domain over the very long simulation times required to achieve thermal equilibrium.  For numerical reasons, we are limited in how steep we can make the density gradient at $\rcool$ (parameterized by $\fiso$). In real stars, the density contrast near the photosphere is much larger, limiting mass and momentum loss relative to our simple model. Instead, we damp motions at large radii in order to limit the mass and angular momentum that leaves the domain.  We apply damping beyond a radius of $\rdamp$ with $\rdamp$ sufficiently larger than $\rcool$ to avoid damping in the region where nearly all of the cooling happens. 

The damping source term in eqs.~\eqref{eq:momentum} and \eqref{eq:energy} is implemented as
\beq
\bmath{a}_{\rm damp} = -\frac{1}{2}\bigg(\frac{\bmath{v}\cdot\hat{r}}{\tau_{\rm damp}} \bigg) \bigg[1+ \tanh\bigg(\frac{r - \rdamp}{0.3r_0}\bigg)\bigg]\hat{r}
\eeq
where $\tau_{\rm damp} = f_{\rm damp}\sqrt{\rdamp^3/GM}$ is the damping timescale and the $\frac{1}{2}[1+\tanh x]$ factor limits damping to $r\gtrsim \rdamp$.

In all simulations, we set $\rdamp = 8.0r_0$ and $f_{\rm damp} = 0.05$, so that $\tau_{\rm cool} = 0.16 \tau_{\rm damp}$ and $\tau_{\rm damp}$ is similar enough to the dynamical time at $\rdamp$ to avoid strong reflection of waves back into the region of interest in our simulations.

In our tests, inclusion of the damping term did not effect the simulation results for $r < \rcool$. In particular, it did not modify the specific angular momentum profiles in this region which are the key measurement for this study.  The damping simply prevents our star-in-a-box from expanding too much over the simulation time and limits the loss of mass from the simulation domain.  

By including the region of significant cooling well within the domain boundaries and by damping motions at large $r$, we keep the total specific angular momentum in the box, which starts at zero, below $3 \times 10^{-4}$ (in our code units of $j_0$, see below). This is $1-2$ orders-of-magnitude smaller than the mean specific angular momentum in any one shell (see Fig.~ \ref{fig:pts_jrand_mag}), so the change in total angular momentum in the box is negligible for our purposes.

\subsection{Boundary Conditions}
\subsubsection{Outer Boundaries}
In all simulations, we use a box size $(40r_0)^3$ centered at $x = y = z = 0$.  At the outer boundaries, $x = y = z = \pm 20 r_0$, we use zero-gradient, diode boundary conditions that allow material to leave but not enter the domain.  That is, the density, pressure, and velocity components of the ghost zones are copied from the last grid cell of the computational domain, but with the following exception.  The velocity component perpendicular to the boundary is set to zero if the direction represents inflow.  

\subsubsection{Absorbing Sink Inner Boundary}
In our simulations of collapse, we activate an absorbing sink centered at $\mathbf{r}=\mathbf{0}$.  The radius of the sink, $r_s$, is a runtime parameter.  When the sink is active, the density and pressure inside the sink are set to $10^{-2}$ times the average values of the density and pressure just outside the sink and the fluid velocities inside the sink are set to zero.  This allows material to flow into the sink unimpeded, representing perfect accretion without feedback.  The sink is not active during our convection simulations, which do not have any inner boundary condition, since we are using a Cartesian grid. Instead, the Plummer potential ensures that our converging flows are well-behaved as $r\to 0$.

\subsection{Initialization}
\label{sec:initialization}
Grid variables are initialized in \ath by specifying $\rho$, $E$ and the Cartesian velocity components $\bmath{v} = (v_x, v_y, v_z)$.  The initial density is $\rhoinit(\mathbf{r})$. The model is in hydrostatic equilibrium, so $\bmath{v}(\mathbf{r},t=0) = \mathbf{0}$.  Therefore, the initial total energy density is  $E(\mathbf{r},t=0) = {\Pinit(\mathbf{r})}/{(\gamma -1)}$.  

\subsection{Code Units and Comparison to \mesa Model}
\label{sec:units}
In our simulations, we set $GM =  (r_0^n + a^n)^\frac{1}{n}$ with $r_0 = 1$.  This yields a characteristic velocity of
\beq
v_0 = \bigg(\frac{GM}{r_0}\bigg)^{1/2} = \bigg[\frac{(r_0^n + a^n)^\frac{1}{n}}{r_0}\bigg]^{1/2} \approx 1
\eeq 
at $r_0$ and a time unit of  
\beq
t_0 = \bigg(\frac{r_0^3}{GM}\bigg)^{1/2} = \Bigg[\frac{r_0^3}{ (r_0^n + a^n)^\frac{1}{n}}\Bigg]^{1/2} \approx 1
\eeq
where in each expression the last equality assumes $a^n \ll1$.  In these units, the adiabatic sound speed at $r_0$ is 
\beq
c_{s,0} = \sqrt{\gamma/(b+1)}
\eeq
and the Keplerian specific angular momentum at $r_0$ is
\beq
j_0 = v_0 r_0 = \sqrt{GMr_0} = 1.
\eeq
The unit of gas mass is $m_0 = \rho_0 r_0^3$, so mass accretion rates are reported in units of 
$\dot{m}_0 \equiv m_0/t_0 = {\rho_0 r_0^3}/{t_0} = \rho_0 r_0^{3/2}(GM)^{1/2}$.

Our dimensionless setup can be scaled to a wide range of RSG and YSG envelopes whose convective Mach number profiles are similar to those in Fig.~\ref{fig:brunt_mturb_mesa} by identifying the photosphere radius as the radius of our isothermal transition at $r \approx 6r_0$.  Rescaling is done most simply by computing the quantities $r_{\rm ph}$, $M_\bullet$, and $\rho_\star$ from the stellar model. These are, respectively, the photosphere radius, the mass interior to the hydrogen envelope, and the density at $r_{\rm ph}/6$.   Code units are then scaled to the star by setting $M = M_\bullet$, $r_0 = r_{\rm ph} / 6$, and $\rho_0 = \rho_\star$ giving, 

\beq
r_0 = 1.16\times10^{13} \bigg(\frac{r_{\rm ph}}{1000\rsun}\bigg) {\rm cm}
\label{eq:r0_star}
\eeq

\beq
t_0 = 0.034 \bigg(\frac{r_{\rm ph}}{1000\rsun}\bigg)^{{3}/{2}}\bigg(\frac{10 \msun}{M_\bullet}\bigg)^{{1}/{2}} \text{yr}
\label{eq:t0_star}
\eeq

\begin{align}
j_0  =  1.24 \times 10^{20} \bigg(\frac{r_{\rm ph}}{1000\rsun}\bigg)^{1/2}\bigg(\frac{M_\bullet}{10\msun}\bigg)^{1/2} {\rm cm}^2 \, {\rm s}^{-1}
\label{eq:j0_star}
\end{align}

\beq
\dot{m}_0 = 2.27 \bigg(\frac{\rho_\star}{10^{-7} \text{g}/\text{cm}^{3}}\bigg)\bigg(\frac{r_{\rm ph}}{1000\rsun}\bigg)^{{3}/{2}} \bigg(\frac{M_\bullet}{10 \msun}\bigg)^{{1}/{2}}\msun \, {\rm yr}^{-1}.
\label{eq:mdot0_star}
\eeq
Panels (a)-(c) of Fig.~\ref{fig:initial_profiles} show the initial profiles for our \ath setup (black, solid curves) scaled to an $18\msun$ (at ZAMS) star computed with \mesa (pink, dashed curves). The \mesa model, which adopts Solar metallicity and a mixing-length $\alpha$ of 1.5, is shown at the end of oxygen burning, at which time the star is a $16.5\msun$ RSG. The \ath curves have been scaled from dimensionless code units using the values $r_{\rm ph} = 840 \rsun$, $M_\bullet = 6 \msun$, and $\rho_\star = 4.0 \times 10^{-7}$ g cm$^{-3}$ obtained from the \mesa model.  The upper $x$-axis shows $r$ in the original code units, for reference.  The vertical dotted line in each panel marks $r_{\rm ph}$.  The $\tau = c/c_s$ radius (vertical solid line) is $818 \rsun$. To the left of this line, the \mesa model is optically thick and fluid motions in the envelope are roughly adiabatic as we assume in our simulated envelope.  

Panel (d)  of Fig.~\ref{fig:initial_profiles} shows, for reference, the binding energy of the envelope exterior to $R$ (dashed line, left-hand $y$-axis) and the mass enclosed at each $R$ (dotted line, right-hand $y$-axis) for the \mesa model.  The open and filled arrows on the left, respectively, mark $0.55 \times 10^{47}$ and $3.6\times 10^{47}$ erg, which are roughly the lowest and highest of the shock kinetic energies obtained for RSG and YSG progenitors by \citet{2021ApJ...911....6I} in their studies of mass-ejection in FSN (this range of energies covers three different NS equations-of-state, one quite stiff and one quite soft). The mass that is lost is roughly the material exterior to where  the shock kinetic energy is equal to the binding energy. The arrows on the right show the mass coordinates that correspond to the two shock energies. If the weak shock has energy of $3.6 \times 10^{47}$ erg (filled arrow at left), then material exterior to the mass coordinate of $12.9\msun$ (filled arrow at right) would be ejected and roughly $\sim 6.9\msun$ of the hydrogen envelope would remain bound and collapse onto the BH.  For the lower shock energy in Fig.~\ref{fig:initial_profiles}, even more of the hydrogen envelope remains bound. Most of the hydrogen envelope that we simulate is thus likely to remain bound and accrete onto the BH in a FSN.

\subsection{Summary of Runtime Parameters and Simulations Performed}
\label{sec:parameters}
In all simulations, we adopt a Plummer index of $n = 8$ and a density power-law index of $b = 2.1$. Cooling occurs beyond a radius of $\rcool = 5r_0$ and damping occurs outside of $\rdamp = 8r_0$.  The cooling timescale fraction is $f_{\rm cool} = 0.01$ and the damping timescale fraction is $f_{\rm damp} = 0.05$.  We adopt $f_{\rm heat} = 0.0006$ for the heating-rate parameter.  The adiabatic index in all simulations is $\gamma = 1.4762$, which gives a flat entropy profile in the envelope at initialization for our choice of $b$.

\begin{table}
	\centering
	\caption{Table of convection simulations.}
	\begin{tabular}{p{8mm}p{6mm}p{7mm}p{8mm}cccc} 
	  \hline
	  Convect &  $\,\,a$   & Base$^{a}$  & Refine$^{b}$ & $\Delta x_{\rm min}$$^{c}$  &  $t_{\rm max}$$^{d}$  &     \\
	  Model   & $[r_0]$ & Cells & Levels & $[r_0]$ & $ [t_0]$  & $a/\Delta x_{\rm min}$ \\
	  \hline
	  F       & 0.16  & $128^3$ & 5 & 0.0098   & 5000  & 16.3   \\
	  A       & 0.08  & $128^3$ & 6 & 0.0049   & 5000  & 16.3   \\
	  R       & 0.16  & $64^3$  & 5 & 0.0195   & 4000  & 8.2    \\
	  \hline
	\end{tabular}
	     {\raggedright $^{a}$ Number of grid cells in the base, unrefined grid. \\ $^{b}$ Number of SMR refinement levels on top of the base resolution. \\ $^{c}$ Width of each grid cell in highest refinement region. \\$^{d}$ Simulation run time. \par}
\label{tab:convect}
%\end{table}
\vspace{10pt}
%\begin{table}
	\centering
	\caption{Table of collapse simulations.}
	\begin{tabular}{p{9mm}cccc} 
		\hline
		Collapse  & Restart$^{a}$ & Restart~Time$^{b}$     &  Sink Size$^{c}$  &   \\
		Run       & Model   & $t_s \, [t_0]$   & $r_s \, [r_0]$ & $r_s / \Delta x_{\rm min}$\\
		\hline
		1  & F   & 4560  & 0.08 & 8.2 \\
		2  & F   & 4900  & 0.08 & 8.2 \\
		3  & F   & 4270  & 0.08 & 8.2 \\
		4  & F   & 4800  & 0.08 & 8.2 \\
		1s & F   & 4560  & 0.04 & 4.1 \\
		\hline
	\end{tabular}
	{\raggedright $^{a}$ Convection model used to initialize the collapse run. \\ $^{b}$ Convection model snapshot used to initialize the collapse run.  \\ $^{c}$ Radius of low-pressure sink activated at start of collapse run.  \par}

\label{tab:collapse}
\end{table}

Our convection simulations, summarized in Table \ref{tab:convect}, vary the softening length, $a$, and the grid resolution.  Model F is our fiducial model. Model A reduces $a$ by a factor of 2 and adds one level of refinement in order to preserve $a/\Delta x_{\rm min}$. Model R reduces the grid resolution by a factor of 2.  We note that the scale height $H \sim r$ in the simulated envelope, so there are a priori no other length scales that we need to resolve. The numerical challenge is that the simulations have to be run for $\sim 5000 t_0$ and $\sim 3\times 10^6$ timesteps to reach thermal equilibrium. 

The grid configuration for Model F is as follows. The domain covers a spatial volume of $(40\,r_0)^3$,  centered on $x = y = z = 0$, and has a base resolution of $128^3$ cells, which translates to a base cell size of $\Delta x = \Delta y = \Delta z = 0.3125$. We use static mesh refinement (SMR) to increase resolution with decreasing $r$. For Model F, there are 5 levels of refinement above the base resolution and the refinement transitions occur where $x$, $y$, and $z$ have the values $[\pm 0.625, \pm 1.25, \pm 3.75, \pm 7.5, \pm 10.0]$.  At each of these transitions, $\Delta x$, $\Delta y$, and $\Delta z$ each decrease by 2. The highest refinement region, $-0.625 < x,y,z < 0.625$, has a cell size of $\Delta x_{\rm min} = \Delta y_{\rm min} = \Delta z_{\rm min} = 0.3125 / 2^5 = 0.0098$, so that $a/\Delta x_{\rm min} = 16.3$ for the fiducial model.  

Model R has the same domain configuration, but with a reduced base resolution of $64^3$ cells instead of $128^3$.  This translates to a cell size of $\Delta x_{\rm min} = \Delta y_{\rm min} = \Delta z_{\rm min} = 0.625 / 2^5 = 0.0195$ in the highest refinement region ($-0.625 < x,y,z < 0.625$).   The Model A grid structure is the same as Model F, except that we include a sixth level of refinement where $-0.3125 < x,y,z < 0.3125$ so that $\Delta x_{\rm min} = \Delta y_{\rm min} = \Delta z_{\rm min} = 0.0049$.

Convection Model F provides the initial conditions for our collapse simulations, which are summarized in Table~\ref{tab:collapse}.  The collapse simulations vary the sink size, $r_s$, as well as the start time, $t_s$, for the collapse.  The `Restart Time' column indicates which snapshot from Model F was used to start the collapse simulation.  We activate the sink immediately, so $t_s$ is also the time that collapse begins.  The grid resolution of each collapse simulation is inherited from Model F.

\section{Convection Simulations} 
\label{sec:convectresults}
In this section, we present the results of our suite of convection simulations, summarized in Table \ref{tab:convect}.    We discuss our fiducial simulation in detail in Section ~\ref{sec:fid}.  We study the effects of changing the softening length, $a$, and the grid resolution in Section~\ref{sec:softlengthresolution}.   We compare our results to analytical predictions in Section~\ref{sec:localcompare}.

\subsection{Fiducial Model}
\label{sec:fid}
Our fiducial model (convection Model F) has $a = 0.16r_0$ and the heating region extends to $r_{\rm heat, out} = 1.5a = 0.24 r_0$. The cooling radius is at $\rcool = 5r_0$. From $r_{\rm heat,out}$ to $\rcool$, our model covers a factor of $\sim 20$ in radius in the convective zone.  Details about the computational domain were given in Section \ref{sec:parameters}.  We run the model to a maximum time of $5000 \, t_0$ which is $\sim$ 400 dynamical times at $\rcool$ and $\sim 3 \times 10^6$ timesteps.  

\begin{figure}
\begin{center}
	\includegraphics[width=0.9\columnwidth]{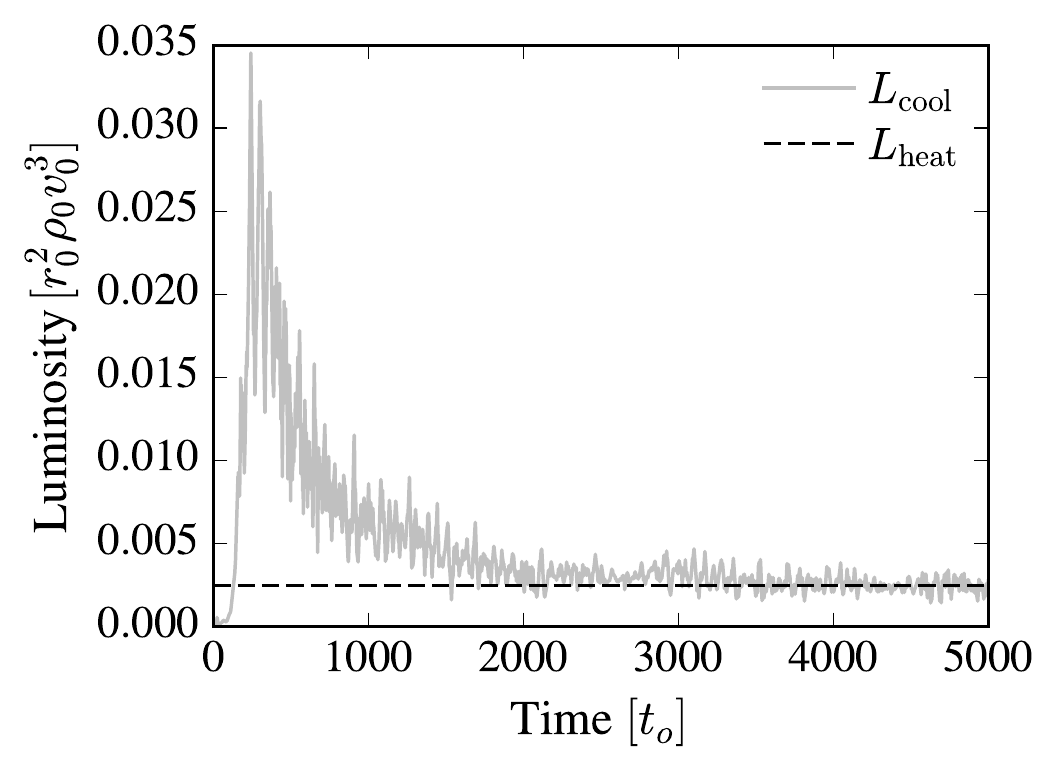}
    \caption{Volume-integrated heating rate (dashed line) and cooling rate (solid line) for Model F. For $t \gtrsim 2000 t_0$, the flow approaches thermal equilibrium with $L_{\rm cool} \approx L_{\rm heat}$.  The flow is qualitatively similar for all $t \gtrsim 2000t_0$.  For quantitative analysis, however, we will restrict our attention to $4000 < t/t_0 < 5000$, when the time average of $L_{\rm cool}$ is very nearly equal to $L_{\rm heat}$ (see Fig.~\ref{fig:Lprof}).}
    \label{fig:edots}
\end{center}
\end{figure}

In Fig.~\ref{fig:edots}, we plot the instantaneous, volume-integrated heating and cooling rates, $L_{\rm heat}$ and $L_{\rm cool} \equiv  \int Q_{\rm cool} dV$, respectively.  $L_{\rm heat}$ is a constant and integrates to eq.~\eqref{eq:Lheat}.  The cooling rate is time-dependent as it depends on the local temperature difference in the cooling region (see eq.~\ref{eq:qcool}). At early times ($t \lesssim 2000t_0$), there is transient convective flow due to the temperature and density drop-off at $\sim \rcool$, resulting in a spike in $L_{\rm cool}$.  After this transient flow subsides, convection is driven by our heating source and the flow is near thermal equilibrium with $L_{\rm cool} \approx L_{\rm heat}$. 

We consider the flow to be in thermal equilibrium if (1) the time- and spherically-averaged convective luminosity, $\langle \overline{L}_{\rm conv} \rangle$, is roughly independent of radius and (2) the time-averaged cooling rate, $\langle L_{\rm cool}\rangle$, is equal to $L_{\rm heat}$.\footnote{Here and in the remainder of this paper, we use $\overline q(r,t)$  or, simply, $\overline q$ to denote the radial profile of the 3D field quantity $q(x,y,z,t)$ and use $\langle \, \rangle$ to denote time-averaged quantities. All radial profiles are computed by finding the volume-weighted mean of the quantity of interest over spherical shells.}  

We compute $\langle \overline{L}_{\rm conv} \rangle$ as follows.  The radial total energy flux is 
\beq
\mathcal{F}_{\rm tot}(x,y,z,t) = \bigg(\frac{1}{2}\rho v^2 + \frac{\gamma}{\gamma -1}P + \rho \phi \bigg)v_r,
\eeq
with spatial mean $\overline{\mathcal{F}}_{\rm tot}(r,t)$. Following \citet{2008ApJ...688..905P}, we decompose the density, temperature, and velocity fields into a mean radial profile and a local deviation from the mean
\begin{align}
 \rho & = \overline{\rho}(r,t) + \delta\rho  \\
 T &=\overline{T}(r,t) + \delta T  \\
 \bmath{v} &= \overline{\bmath{v}}(r,t) + \delta\bmath{v}.
\end{align}
Inserting the decomposed fields into $\overline{\mathcal{F}}_{\rm tot}(r,t)$ yields advective terms proportional to $\overline{\rho}\cdot\overline{v_r}$ and $\overline{\delta \rho \delta v_r}$ (e.g., eq. (41) of \citeauthor{2008ApJ...688..905P}). Subtracting these advective terms from $\overline{\mathcal{F}}_{\rm tot}(r,t)$ gives the radial convective flux profile $\overline{\mathcal{F}}_{\rm conv}(r,t)$.  The instantaneous convective luminosity profile is $\overline{L}_{\rm conv}(r,t) = 4\pi r^2 \overline{\mathcal{F}}_{\rm conv}(r,t)$ whose time average is denoted $\langle \overline{L}_{\rm conv} \rangle$. In general, we find that the advective terms are small so $\langle \overline{L}_{\rm tot} \rangle \approx \langle \overline{L}_{\rm conv} \rangle$.

\begin{figure}
\begin{center}
	\includegraphics[width=0.9\columnwidth]{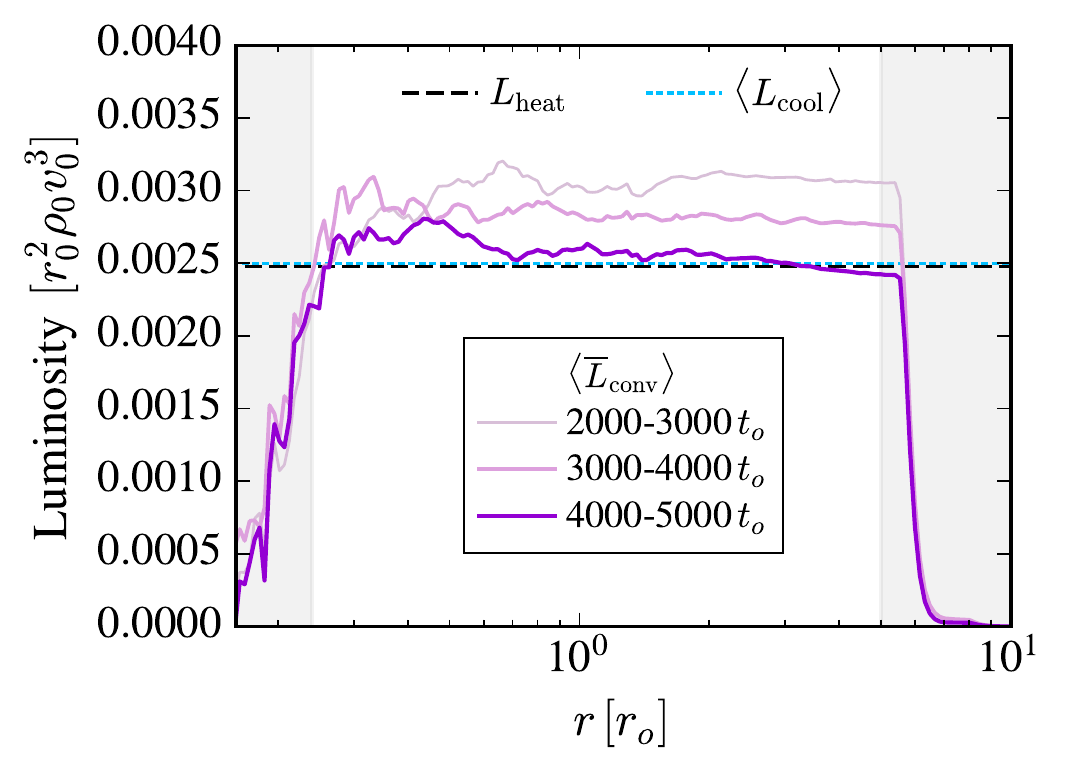}
    \caption{Time- and spherically-averaged convective luminosity (purple lines; times used for each average are given in the legend), heating rate (black, dashed line), and time-averaged cooling rate (blue, dotted line; time average over $4000-5000$~$t_0$) for Model F.  Grey shading is the same as in Fig. \ref{fig:initial_profiles}. The convective luminosity is roughly independent of radius and the flow is dynamically similar for $t > 2000 t_0$.  The flow is in thermal equlibrium with $\langle \overline{L}_{\rm conv} \rangle  \simeq \langle L_{\rm cool}\rangle \simeq L_{\rm heat}$ for $4000 \lesssim t/t_0 \lesssim 5000$.}
    \label{fig:Lprof}
\end{center}
\end{figure}

The purple, solid lines in Fig.~\ref{fig:Lprof} show $\langle \overline{L}_{\rm conv} \rangle$ for the time ranges listed in the lower legend.  The constant heating rate, $L_{\rm heat}$, is indicated with a dashed, black line.  The cooling rate, averaged over $4000 \le t/t_0 \le 5000$, is shown with the dotted, blue line.  After $t \sim 2000 t_0$, $\langle \overline{L}_{\rm conv}\rangle$ is roughly independent of radius and we will see in the next sections that the flow is dynamically similar over these times. Our conditions for thermal equllibrium are met after $t \sim 4000 t_0$, when $\langle \overline{L}_{\rm conv} \rangle \simeq \langle L_{\rm cool}\rangle \simeq L_{\rm heat}$.  The black, dotted lines in Fig.~\ref{fig:initial_profiles} show radial profiles of the simulation upon reaching thermal equilibrium.

\begin{figure*}
	\includegraphics[width=0.9\textwidth]{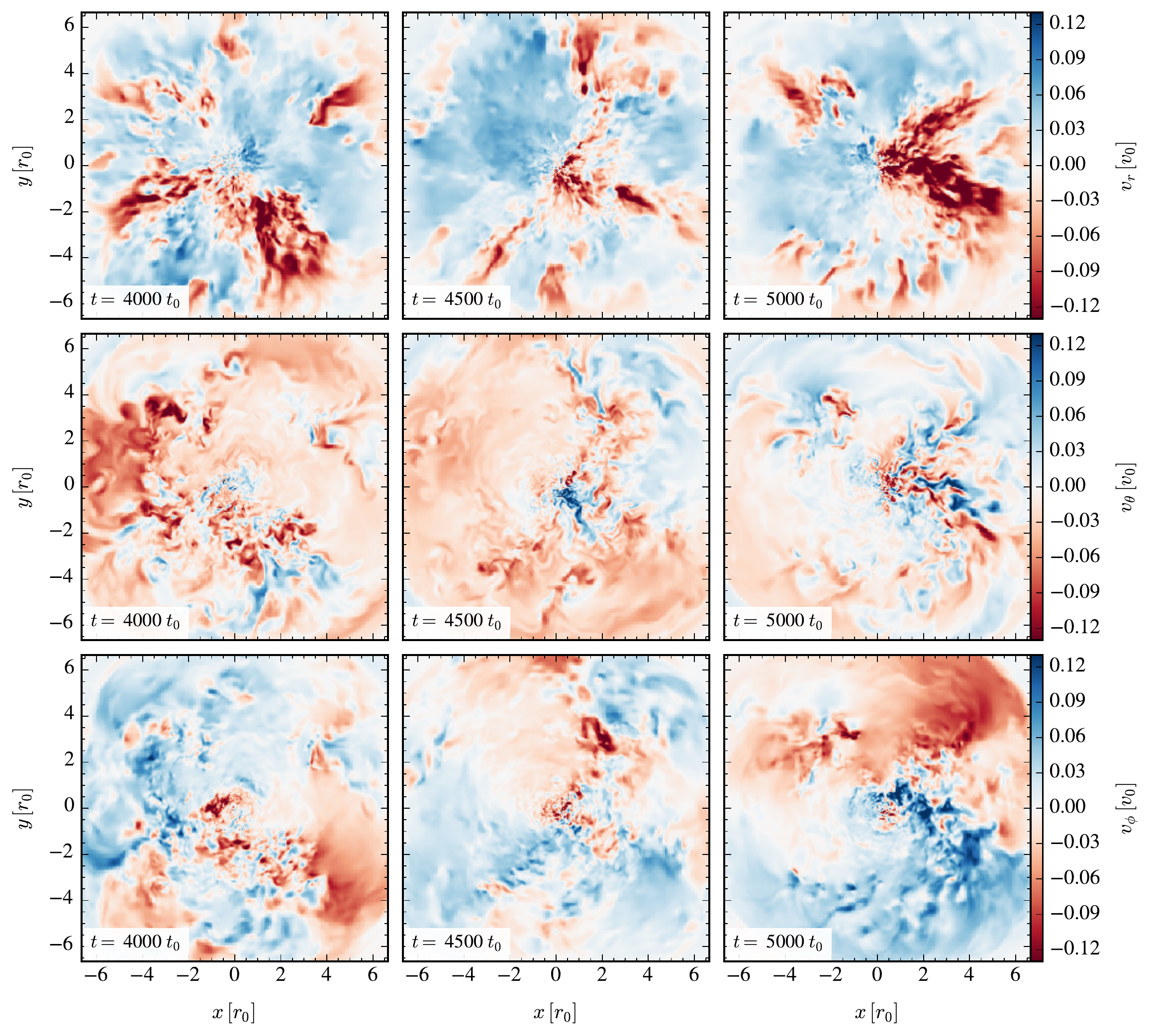}
    \caption{Velocity field slices through the $z=0$ plane of our 3D simulation of convection at three simulation times for Model F. The $z$-axis is arbitrary as our setup has no preferred direction. Slices through other planes that include the origin look similar. Top, middle, and bottom rows show the radial, polar, and azimuthal components, respectively, of the fluid velocity field. With $\theta = 0$ along the $+z$ axis, $v_\theta >0$ indicates material flowing into the page. The cooling radius, $\rcool = 5r_0$, begins the transition to an exponential fall-off in gas density.  In $v_r$, large plumes form and subside throughout the simulation, bringing high entropy material out to $\rcool$. Cooled gas falls back towards the center in streams of varying sizes that collide with material rising from the center. Flows perpendicular to $r$ ($v_\theta$ and $v_\phi$)  have a more obvious radial dependence with smaller-scale structures at small $r$, where the scale height is smallest.}
    \label{fig:convect_flow}
\end{figure*}

Figure \ref{fig:convect_flow} shows snapshots of the total velocity field of the flow at $4000$, $4500$, and $5000$ $t_0$ for fiducial Model F. Each panel is a slice through the $z = 0$ plane. The $z=0$ plane is an arbitrary choice; our setup has no preferred axis and slices through other planes that include the origin are similar. The top row shows radial velocity, $v_r(x,y,z=0)$, the second row shows velocities in the $\theta$ direction, $v_\theta(x,y,z=0)$, and the bottom row shows velocities in the $\phi$ direction, $v_\phi(x,y,z=0)$.  Our angles $\theta$ and $\phi$ are defined in the usual way, relative to the (arbitrary) $z$-axis.   Although our simulation domain extends to $\pm 20 r_0$ in each Cartesian direction, we restrict the slices to $-6.5 < x/r_0 < 6.5$ and $-6.5 < y/r_0 < 6.5$. The density falls off exponentially beyond $r \approx 6 r_0$, so there is little momentum outside of this radius.  The physically relevant part of the domain is  $r_{\rm heat,out} = 0.24 r_0 \lesssim r \lesssim \rcool = 5r_0$.

The panels of Figure \ref{fig:convect_flow} show turbulent flow at all scales. The radial flows (top row) exhibit plumes that individually occupy a large fraction of the star in both solid angle and radius. This morphology is consistent with other 3D simulations of RSG envelopes \citep{2009A&A...506.1351C,2021arXiv211003261G} and with the large convective cells that have been inferred from observations of Betelgeuse \citep{2010A&A...515A..12C,2020ApJ...899...68D}.  The large, rising plumes are often interrupted by narrower streams of cool material falling back towards the origin from $\rcool$. Though not apparent in this figure, the infalling streams colliding at small $r$ often launch waves that move out through the rising plumes. Features of the flow are qualitatively consistent with analytical expectations.  Given the density scale height in the envelope of $H = r/b \approx 0.48 r$, mixing-length-theory predicts large eddies at large $r$. On average, the smallest-scale structures indeed exist at the smallest $r$. 

For a more quantitative analysis of our fiducial model, we compute radial profiles over time to characterize the convective Mach number and angular momentum content of the flow in the following subsections.

\subsubsection{Mean Turbulent Mach Number}
\label{sec:fid_mach}
\begin{figure*}
\begin{center}
	\includegraphics[width=0.9\textwidth]{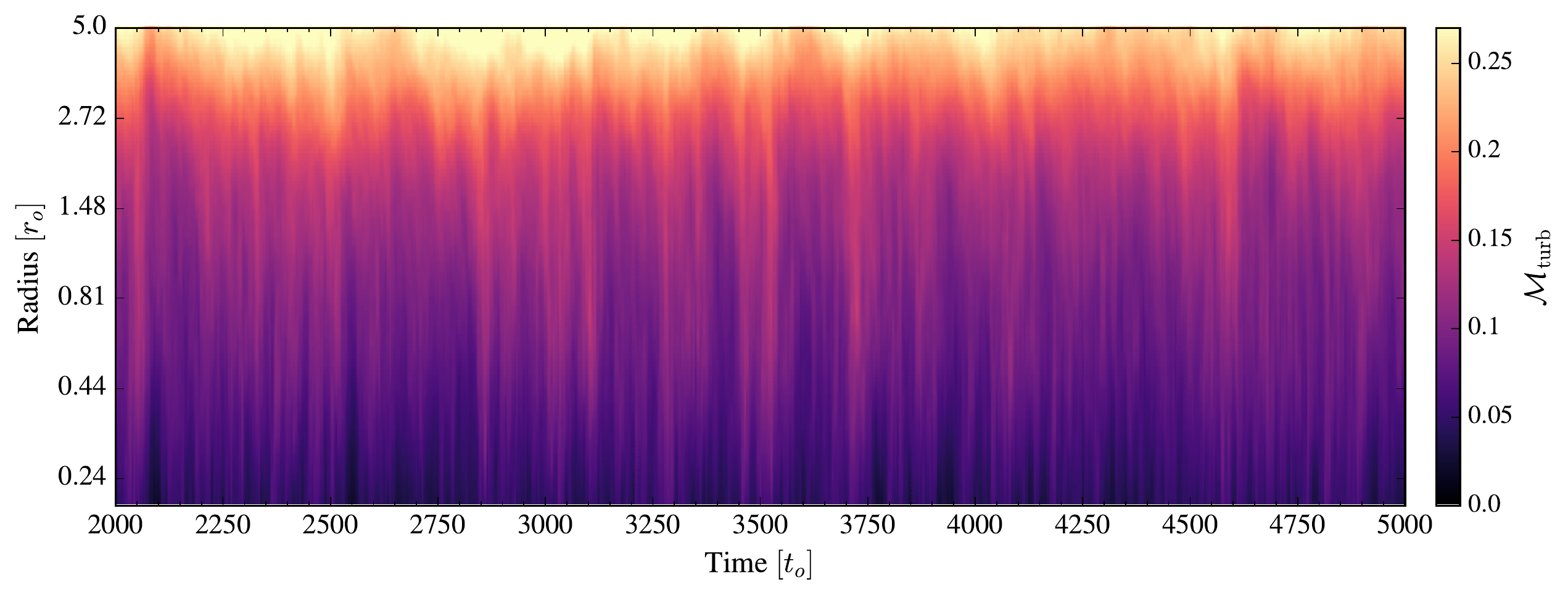}
    \caption{Spherically-averaged turbulent Mach number, eq.~\eqref{eq:mconv}, as a function of radius and time for Model F. The flow is dynamically similar for $t \gtrsim 2000 t_0$ with Mach numbers of $\approx 0.05 - 0.25$. Comparing to Fig. \ref{fig:pts_jrand_mag}, times with larger-than-average $\mconv$ translate to higher-than-average specific angular momentum.}
    \label{fig:mturb_pts}
\end{center}
\end{figure*}

We first consider profiles of the turbulent Mach number, $\mconv$, which is the root-mean-square turbulent velocity profile divided by the mean sound speed profile
\beq
\mconv(r,t) = \frac{v_{\rm turb}(r,t)}{c_s(r,t)}.
\label{eq:mconv}
\eeq
To find $v_{\rm turb}(r,t)$, we first compute the random velocity field 
\beq
\delta \bmath{v}(x,y,z,t) = \bmath{v}(x,y,z,t) - \overline{v_r}(r,t)\hat{r}
\eeq
where $\bmath{v}(x,y,z,t)$ is the total velocity field and  $\overline{v_r}(r,t)\hat{r}$ is the mean radial velocity profile. The turbulent velocity profile, $v_{\rm turb}(r,t) \equiv \overline{\|\delta \bmath{v}\|}(r,t)$, is the average of the magnitude of the random velocity in each shell.  We note that subtracting $\overline{v_r}(r,t)\hat{r}$ is only necessary during the initial transient phase (see. Fig.~\ref{fig:edots}).  After $\sim 1500 t_0$, $\overline{v_r}(r,t)$ is negligible. 

Fig.~\ref{fig:mturb_pts} shows $\mconv(r,t)$ for our fiducial model for $2000 \le t/t_0 \le 5000$, when $\langle \overline{L}_{\rm conv} \rangle$ is independent of radius and when the kinetic energy in the box is constant to within $\approx 10-15\%$. Each column in the figure plots the mean radial profile $\mconv(r)$ at fixed time. We limit the $y$-axis to $r \lesssim 5 r_0$, which is the region of interest in our domain. The $y$-axis is log-scaled to show features at small radii.  The flow achieves turbulent Mach numbers of $\approx 0.05 - 0.25$. The profiles are qualitatively similar over the times shown, though there are periods of higher- and lower-than-average Mach number.  
\begin{figure}
\begin{center}
	\includegraphics[width=0.9\columnwidth]{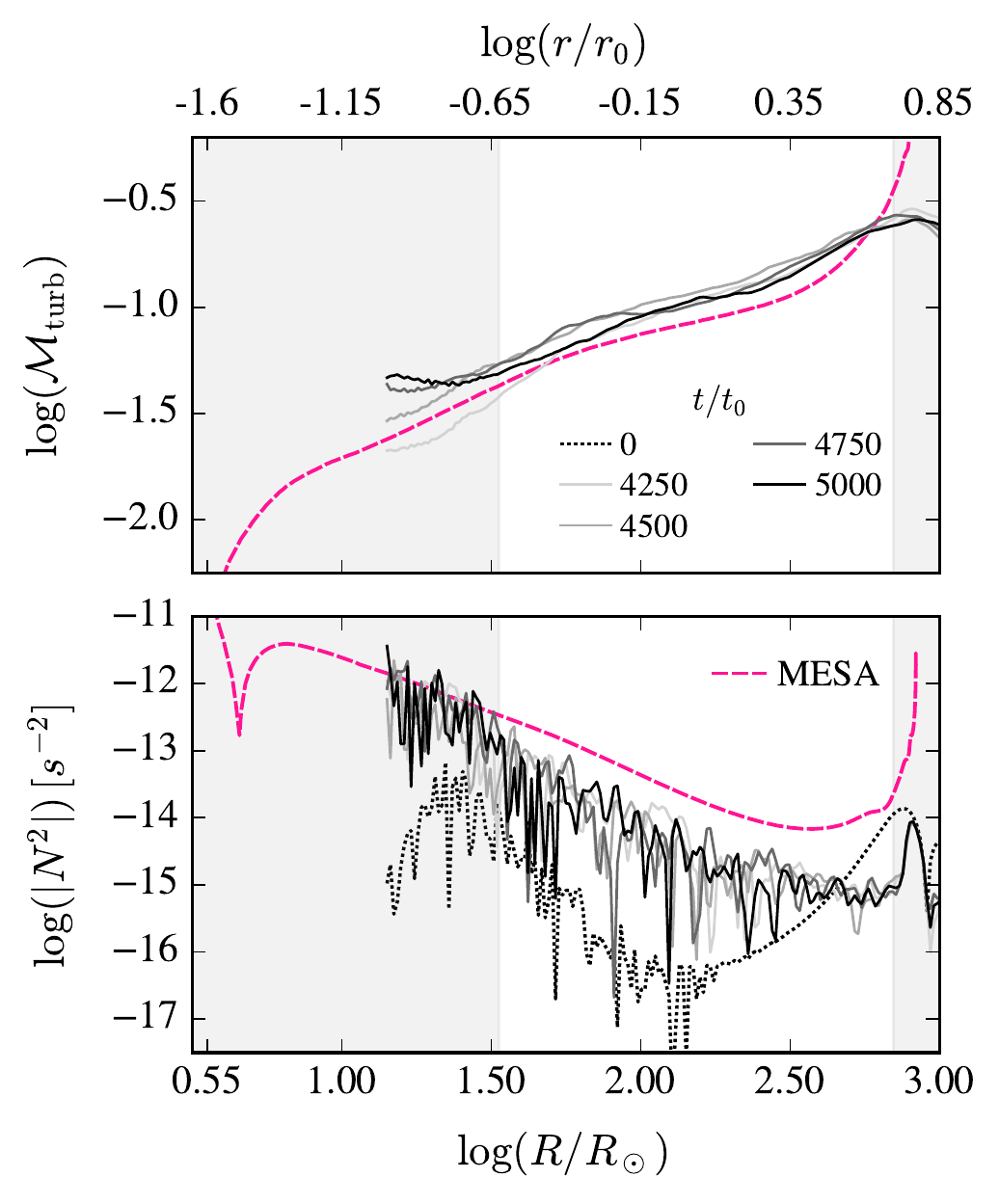}
    \caption{Turbulent Mach number (top panel) and Brunt-V\"ais\"al\"a frequency (bottom panel) profiles for our fiducial \ath model at the simulation times listed in the legend (solid and dotted curves).  The pink, dashed curves show our \mesa RSG model.   The $x$-axes and grey shading are the same as in Fig.~\ref{fig:initial_profiles} and $N^2$ for the \ath model has been scaled to physical units as in that figure.  With our choice of $f_{\rm heat} = 6\times 10^{-4}$, we capture the convective Mach number profile of supergiant envelopes well.}
    \label{fig:brunt_mturb_mesa}
\end{center}
\end{figure}

The top panel of Fig.~\ref{fig:brunt_mturb_mesa} compares $\mconv(r)$ for the simulation times listed in the legend (solid, grey curves) to the convective Mach number\footnote{The convective Mach number in \mesa derives from the radial convective flux from mixing length theory while our definition of $v_{\rm turb}$ uses all three components of $\delta \bmath{v}$.} profile of our \mesa RSG model (pink, dashed curve). The bottom panel of the figure compares the Brunt-V{\"a}is{\"a}l{\"a} frequency instead.  The $r$ values of our simulation have been scaled as in Fig.~\ref{fig:initial_profiles}. Our simulation achieves envelope turbulent Mach numbers that are similar to this \mesa model.  The $\mconv$ profile for the \mesa model shown is representative of RSGs and YSGs and, therefore, our simulation is broadly applicable to a wide range of supergiant progenitors. 

\subsubsection{Total Angular Momentum}
\begin{figure}
\begin{center}
	\includegraphics[width=0.85\columnwidth]{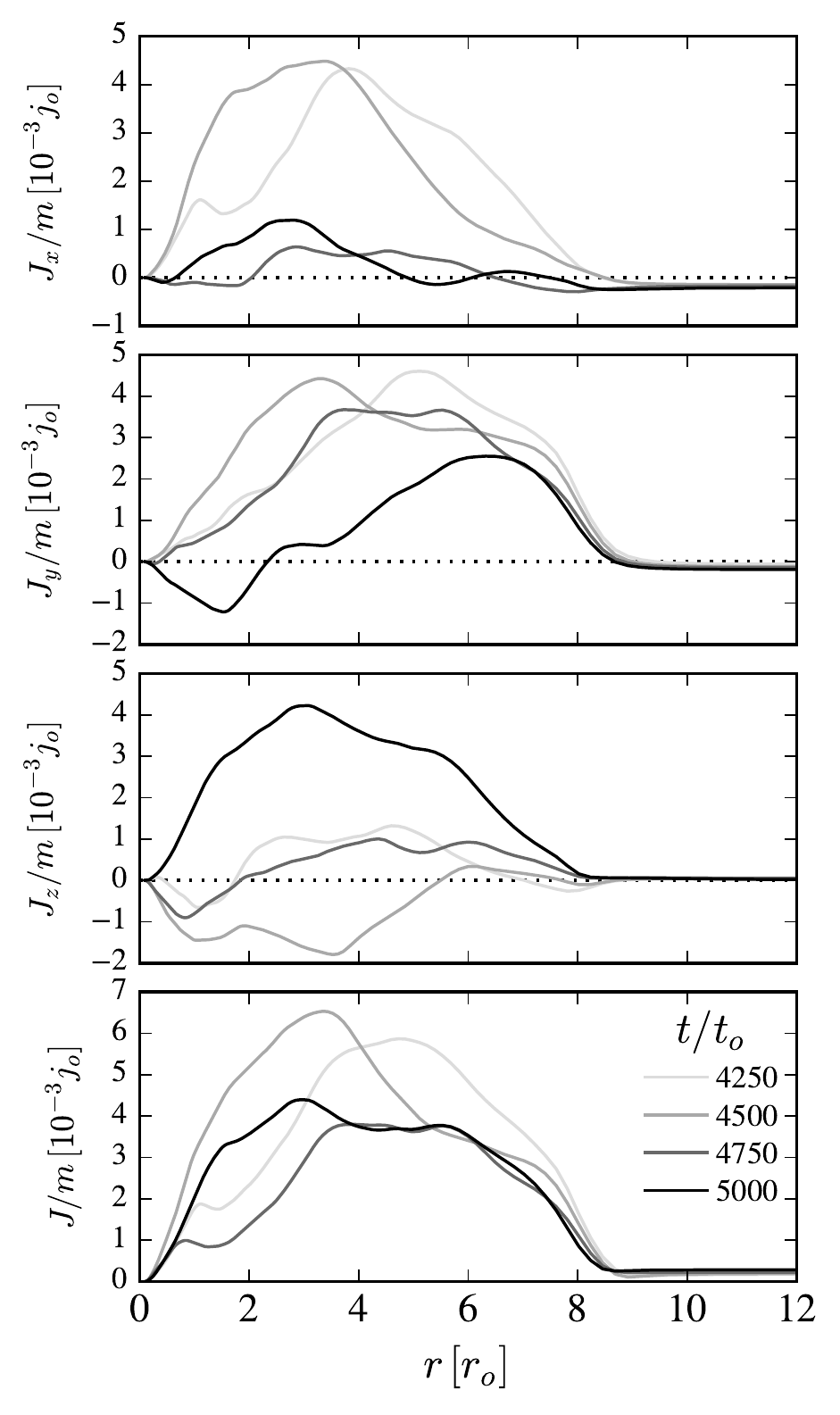}
    \caption{Components of $\bmath{J}(r,t)$, the total angular momentum enclosed at $r$, divided by the total gas mass in the domain.  Although the total angular momentum is very small, there is a non-zero angular momentum enclosed at each $r$ $\lesssim 8 r_0$ due to random convective flows. Accretion of material during collapse leads to the accretion of finite angular momentum as a function of time even though the star has roughly zero net angular momentum.}
    \label{fig:cumulative_J}
\end{center}
\end{figure}

An important objective of our convection simulations is to quantify the (non-zero) mean specific angular momentum at each radius that arises due to the turbulent convective flows. Before turning to that question in Section \ref{sec:convect_specificAM}, it is important to first emphasize that the star is not rotating, so the net angular momentum in the domain must be close to zero.  We explore this in Fig.~\ref{fig:cumulative_J}, where we plot the cumulative Cartesian components of the total angular momentum vector, which are computed as follows.

Let $\ell_i = \rho(\bmath{r}\times\bmath{v})_i$ be the $i$th component of the angular momentum density in each grid cell at time $t$.  We compute the mean radial profile of $\ell_i$ to give $\overline{\ell}_i(r,t)$.  Then we compute the $i$-th component of the total angular momentum vector enclosed at a radius $r$ at fixed time,
\beq
J_i(r,t) = \int_0^r 4\pi r^2 \overline{\ell}_i(r,t) dr.
\label{eq:cumulativeJi}
\eeq
We also compute the total gas mass in the box at each time, $m(t)$.  The top three panels of  Fig.~\ref{fig:cumulative_J} plot $J_i(r,t) / m(t)$ for $i = x,y,$ and $z$. The bottom panel shows the magnitude $J/m = \sqrt{\sum_i J_i^2}/m$.  The curves in each panel correspond to the different simulation times listed in the legend.  We have divided by $m(t)$ in order to put the $y$ axis in units of specific angular momentum ($j_0$), as in later figures.

Fig.~\ref{fig:cumulative_J} shows that although each component of the cumulative Cartesian total angular momentum vector, $\bmath{J}(r,t)$, goes to zero\footnote{There is a small non-zero total specific angular momentum in the box due to gas leaving the domain. See Sec.~\ref{sec:damping}.} when integrated out to $r\approx 8.5r_0$, there is a finite $\bmath{J}(r,t)$ at each $r \lesssim 8.5 r_0$.  Consider, for example, the $t = 5000 t_0$ curve for $J_z(r,t)$ in the third panel.  The curve becomes large and positive out to $r \approx 3 r_0$ as all of the inner shells with positive $J_z$ are added to the sum. The curve then drops toward zero as the material with negative $J_z$ is added to the integral.  Considering the $t = 5000 t_0$ curve in each panel, if only the material out to $r \approx 3 r_0$ were accreted, a net $\boldsymbol{J}/m \approx (0.001\hat{x} + 0.0005\hat{y} + 0.004\hat{z})j_0$ with magnitude $J/m = 0.0045 j_0$ would be available to feed a rotationally-supported structure at small radii.  Although infall of the entire envelope implies a total specific angular momentum budget of $\approx \bmath{0}$, this (perhaps surprisingly) does not mean that the spin of the BH would be zero if the entire star collapsed.  We return to this in Section~\ref{sec:bhspin}. 

For this particular simulation, the magnitudes of each component of $\bmath{J}$ are within a factor of a few of one another for the times shown. This is just a coincidence. The turbulent velocity field is random, as is the direction of $\bmath{J}$.  At other times and in other simulations, one or two of the Cartesian components is larger than the other(s) by an order-of-magnitude.  There is no preferred direction in our setup.

\subsubsection{Mean Specific Angular Momentum}
\label{sec:convect_specificAM}
We now turn to the central question of characterizing the angular momentum profile of the convective envelope. We define the specific angular momentum vector of the gas in each grid cell as $\vjrand$.
The angle between $\vjrand$ and the +$z$-axis is denoted $\theta_j$ and the angle between $\vjrand$ and the $+x$-axis (in the $x$-$y$ plane) is $\phi_j$. Finally, we define $\ujrand$ to be the unit vector in the direction of $\vjrand$.

\begin{figure*}
\begin{center}
	\includegraphics[width=0.9\textwidth]{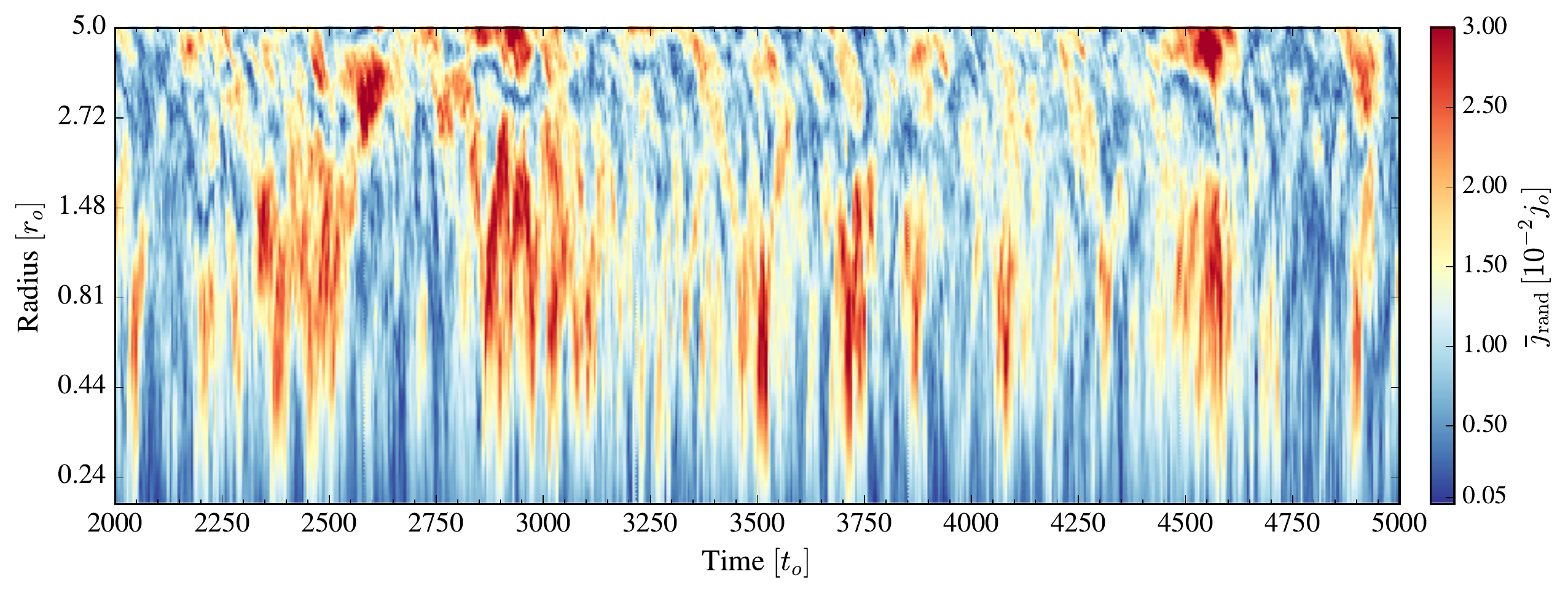}
    \caption{Spherically-averaged specific angular momentum as a function of radius and time for Model F. There is finite mean angular momentum at each radius. When scaled to our \mesa supergiant model, $\avej(r,t) \approx (4 - 28) \times 10^{17}$ cm$^2$ s$^{-1}$ (see eq.~\eqref{eq:j0_star} and accompanying text) implying circularization radii of many hundreds times the ISCO radius for a newly-formed BH (see Fig.~\ref{fig:CONV_pts_rcirc}).}
    \label{fig:pts_jrand_mag}
\end{center}
\end{figure*}

\begin{figure}
\begin{center}
	\includegraphics[width=1\columnwidth]{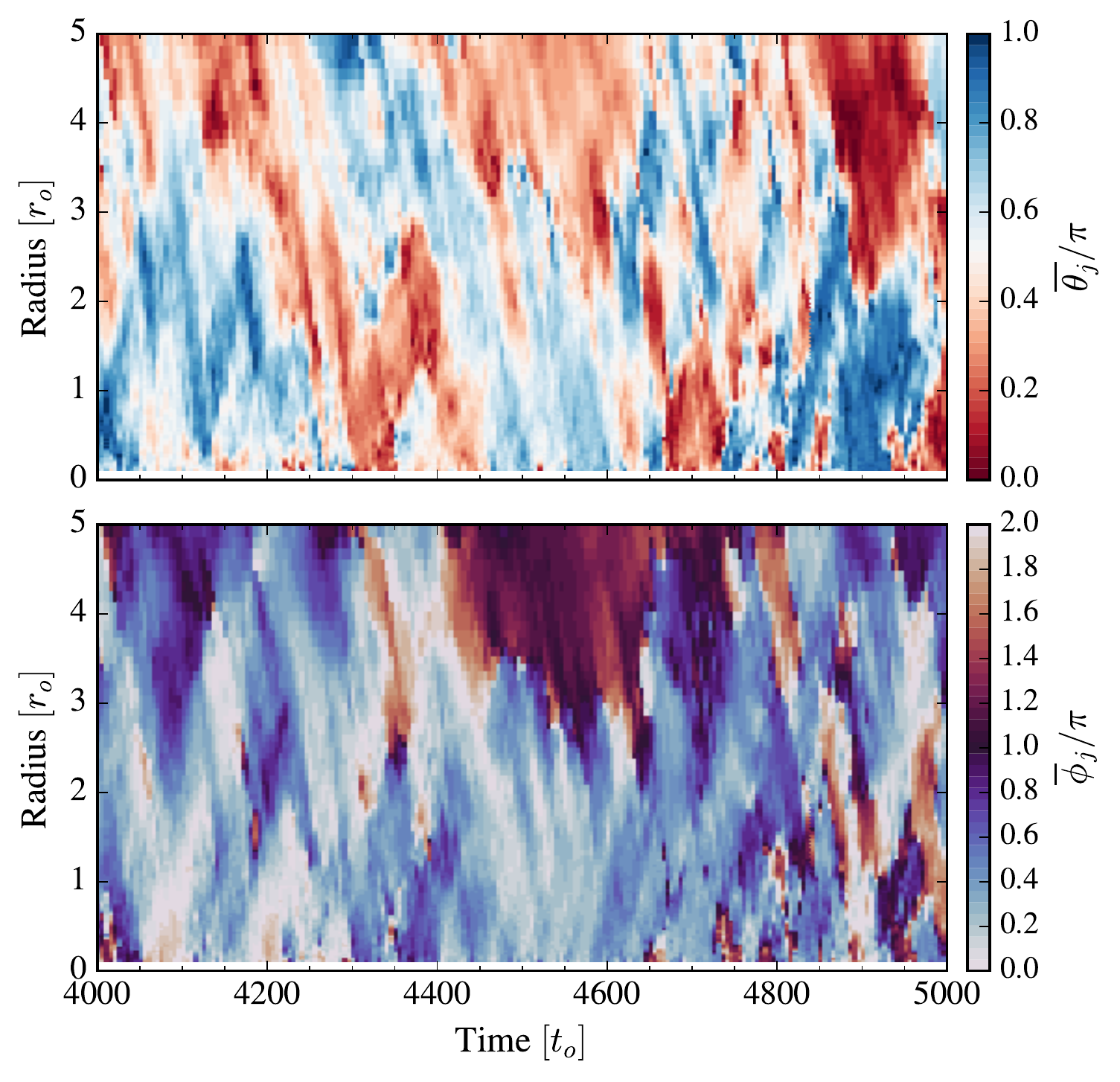}
    \caption{Polar (top panel) and azimuthal (bottom panel) direction angle profiles of the spherically-averaged specific angular momentum vector, $\avejvec$, over time for Model F. In the top panel, white regions show where $\avejvec$ lies near the $x$-$y$ plane while red (blue) colors correspond to a positive (negative) $z$-component of $\avejvec$. Two regions of similar brightness but opposite color have anti-parallel $z$-components of $\avejvec$.  In the lower panel, colors that differ by $\pi$ correspond to anti-parallel projections of $\avejvec$ in the $x$-$y$ plane.
Material from different regions in $r$ and fixed $t$ show features with opposite direction angles, as is required for $\mathbf{J} \approx \mathbf{0}$ when integrated over the box.  The direction of $\avejvec$ at each radius is not fixed in time but flips in sign over long timescales.  The large-scale structures in each panel show connected regions in space and time where the direction of $\avejvec$ is changing slowly.  This coherence of $\avejvec$ over radius and time suggests coherent accretion of $\avejvec$  as material falls in towards the BH.}
    \label{fig:pts_theta_phi}
\end{center}
\end{figure}

Fig.~\ref{fig:pts_jrand_mag} shows $\avej(r,t) \equiv \|\avejvec(r,t)\|$ for our fiducial model. Although the net angular momentum in the envelope is nearly zero, convection gives rise to fluid motions that are perpendicular to $r$, resulting in a finite mean angular momentum at each radius.  The distribution of momentum as a function of $r$ varies in time. Each time (that is, each column in the figure) is a specific realization of the stochastic angular momentum distribution. The time that the star collapses is arbitrary with respect to the flow, and the accretion rate of angular momentum over time will depend on the state of the convective flow over the time that it takes for the envelope to collapse.  We will explore this idea in greater detail in the context of our collapse simulations in Section~\ref{sec:collapseresults}.

The spherically-averaged specific angular momentum profiles of Fig.~\ref{fig:pts_jrand_mag} vary within $0.005  \lesssim \avej/ j_0 \lesssim 0.032$, in our code units with $j_0 =\sqrt{GMr_0}$ (for comparison, the specific angular momentum for rotation at break-up is $0.5 - 2.0$ $j_0$ between $0.24r_0$ and $5r_0$; so the random flows have specific angular momenta that are $0.3 - 6$ \% of breakup). We will apply this result to a specific supergiant model in Fig.~\ref{fig:CONV_pts_rcirc}, but for now we make a rough comparison to eq.~\eqref{eq:jrand_quataert}. Using eq.~\eqref{eq:j0_star} with $M_\bullet=10\msun$ and $r_{\rm ph} = 1000\rsun$, our simulated envelope achieves specific angular momenta of $\overline{\jmath}_{\rm rand} \approx (6 - 40) \times 10^{17}$ cm$^2$ s$^{-1}$ (note that $\overline{\jmath}_{\rm rand}  > \jisco = 1.5 \times 10^{17}$ cm$^2$ s$^{-1}$ for a non-spinning $10\msun$ BH). The values of $\jrand$ that we find are a factor of a few $-$ 10 times the estimate of eq.~\eqref{eq:jrand_quataert} that was derived by \citet{Quataert2019}; we return to this later in Fig.~\ref{fig:CONV_fig_janalytical}.  

For comparison, the $15\msun$ model of \cite{2005ApJ...626..350H}, which has a birth rotational velocity of $200$ km s$^{-1}$, has rotational specific angular momentum of $5 \times 10^{18}$ cm$^2$ s$^{-1}$ at its surface at birth. By the pre-SN phase, the star has rotational specific angular momentum of $\sim 8 \times 10^{17}$ cm$^2$ s$^{-1}$ at its surface, declining to $\sim 5 \times 10^{16}$ cm$^2$ s$^{-1}$ near the base of the hydrogen envelope. The net rotational angular momentum is small compared to the random angular momentum of the convection zone except at the surface of the star where the two are comparable.  

The direction of $\avejvec$ over time is important for determining the dynamics at smaller radii when the envelope eventually accretes onto the newly-formed BH. The upper and lower panels of Fig.~\ref{fig:pts_theta_phi} plot $\overline{\theta}_j(r,t)$ and $\overline{\phi}_j(r,t)$, respectively, in units of $\pi$. In the top panel, white regions show where $\avejvec$ lies near the $x$-$y$ plane. The large regions of dark red show where $\overline{\theta}_j \lesssim \pi/4$. Similarly, the darkest blue regions of space and time are where the $-z$ component of $\avejvec$ is largest. In the lower panel, the darkest colors correspond to $\overline{\phi}_j \approx \pi$ and lightest colors correspond to $\overline{\phi}_j \approx 2\pi$. To interpret the implications of Fig.~\ref{fig:pts_theta_phi}, we note that if $\ujrand$ is rapidly varying on timescales of the dynamical time at the circularization radius, then the gas could not all circularize in the same plane and one might expect a more spherical accretion flow at scales of the circularization radius, $\rcirc$. Additionally, rapidly varying $\ujrand$ in the radial profile could mean greater cancelation of $\vjrand$ as shells (and individual parcels) each reach their respective $\rcirc$ and interact with one another.  Contrast this with the case of a rotating star in which $\ujrand$ is constant and material can build up into a disk as all of the rotationally-supported flow orbits the same axis. 

For material with $\jrand \sim 10^{-2} j_0$ (Fig.~\ref{fig:pts_jrand_mag}), the circularization radius is $\rcirc \sim 10^{-4} r_0$ where the dynamical time is $\sim 10^{-6} t_0$.  Periodograms of $\avej(r,t)$, $\overline{\theta}_j(r,t)$, and $\overline{\phi}_j(r,t)$ show that the most power is on timescales larger than   $80 t_0$ $ \gg 10^{-6} t_0$. This implies that the direction of angular momentum is quite coherent on timescales relevant to the circularization of the disk at small radii.  In Section \ref{sec:predict}, we will explore how much these mean profiles are modified as the envelope collapses.

\subsection{Dependence on $a$ and Resolution}
\label{sec:softlengthresolution}
\begin{figure}
\begin{center}
	\includegraphics[width=0.92\columnwidth]{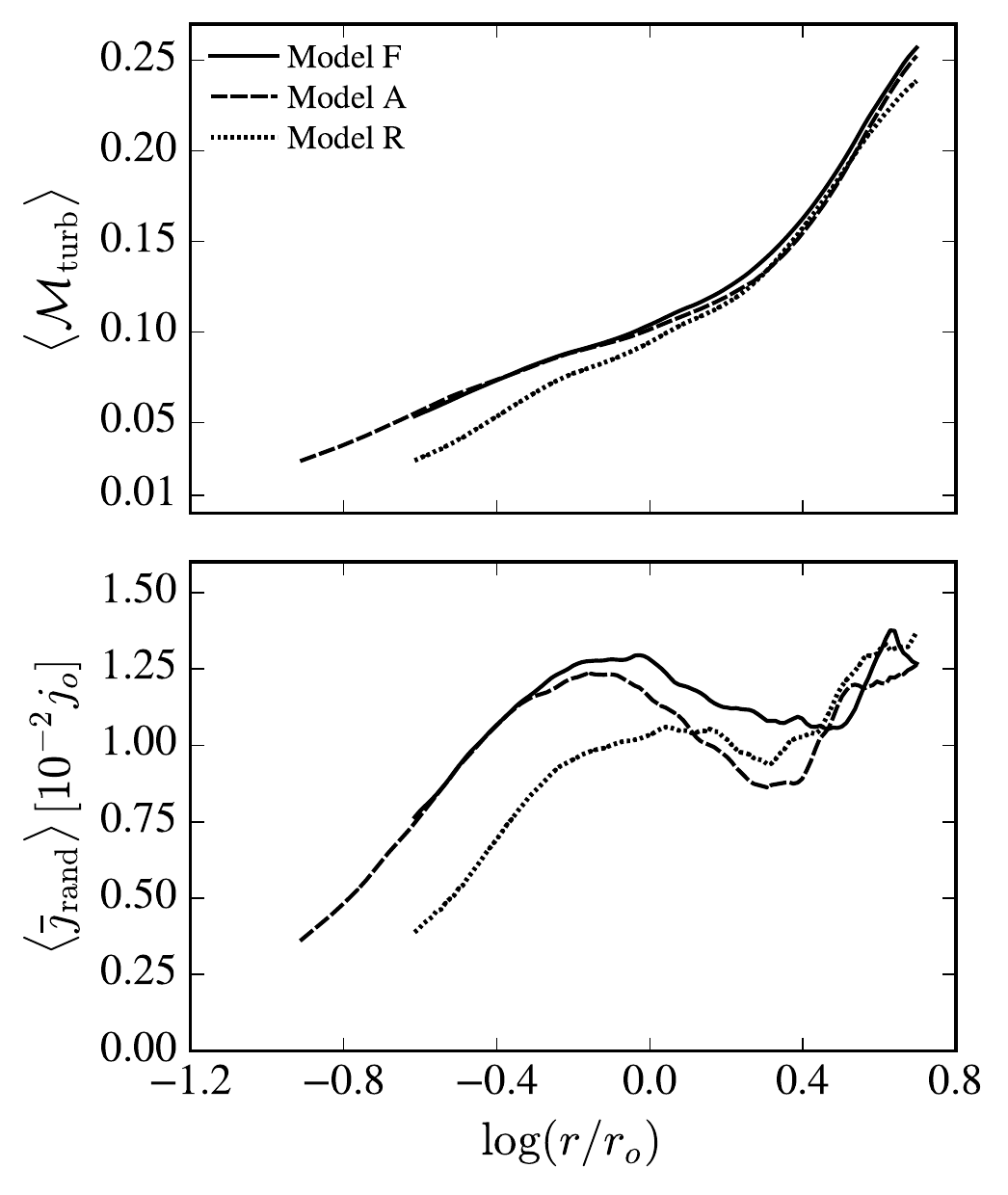}
    \caption{Time- and spherically-averaged convective Mach number profile (top) and specific angular momentum profile (bottom) for our fiducial Model F (solid lines), Model A with smaller Plummer smoothing length $a$ (dashed lines), and Model R with $2\times $ lower base resolution (dotted lines). The time averages are taken over $4000 < t/t_0 < 5000$ (Models F and A) or $2500 < t/t_0 < 3500$ (Model R), when the simulations are in thermal equilibrium. Comparing Model F ($a =0.16r_0$) to Model A ($a = 0.08r_0$):  $\langle\mconv\rangle$ is very similar and where $\langle\mconv\rangle$ differs by $\approx 5\%$, $\langle\avej\rangle$ differs by $\approx 20\%$.  Comparing Model F to Model R (the half-resolution run):  The lower-resolution simulation has lower Mach numbers at small $r$, which results in a $\approx 50\%$ lower specific angular momentum at small radii.}
    \label{fig:CONV_acompare_ave_profiles} 
\end{center}
\end{figure}

To study the influence of the softening length, $a$, on the angular momentum associated with convective flows, we run a simulation with $a =0.08 r_0$ (Model A) for comparison to our fiducial run with $a = 0.16r_0$ (Model F).  The grid is the same for the two models except that we add a sixth SMR level to Model A (see Section~\ref{sec:parameters}). The two models are otherwise identical.   To study the influence of resolution, we run Model R, which is identical to Model F except that it adopts a base resolution of $64^3$ cells instead of $128^3$ cells.  Both simulations have 5 SMR levels above the base resolution with the refinement transitions placed at the same $x$,$y$, and $z$ values.   To compare the different simulations, we take time averages of the instantaneous profiles $\mconv(r,t)$ and $\avej(r,t)$ over the simulation times that our conditions for thermal equilibrium are satisfied.  For Models F and A, these are $4000 < t/t_0 < 5000$.  For Model R, these are $2500 < t/t_0 < 3500$.  The top panel of Fig.~\ref{fig:CONV_acompare_ave_profiles} shows the resultant time-averaged convective Mach number profiles, $\langle\mconv\rangle$, and the bottom panel shows the time-averaged specific angular momentum profiles,  $\langle\avej\rangle$. We restrict each curve to $r_{\rm heat, out} < r < r_{\rm cool}$. 

Comparing the solid curves for Model F to the dashed curves for Model A, both simulations achieve nearly identical $\langle\mconv\rangle$ with a maximum difference of $5\%$ at $r \sim 2 r_0$. This corresponds to roughly a $20\%$ difference in $\langle\avej\rangle$ at that radius. At small $r$, $\langle\avej\rangle$ is identical between the two simulations. 

For Model F and Model R (dotted curves), $\langle\mconv\rangle$ differs from $5\%$ to $45\%$ as $r$ decreases from $\sim 2.5 r_0$. Over this same region, $\langle\avej\rangle$ is lower in the lower-resolution simulation by $\approx 10 - 50\%$.  

Our simulation with smaller $a$ (that is, simulating smaller radii closer to the base of the convective envelope) shows very good convergence in $\langle\avej\rangle$ for $r \lesssim 0.6 r_0$ where our grid is most well-resolved.  Although we are not converged in $\langle\avej\rangle$ at all radii, the more physical simulation with higher resolution increases $\langle\avej\rangle$, reenforcing our result that convective flows with $\mconv$ similar to supergiant envelopes give rise to $\jrand > \jisco$.  

\subsection{Comparison to the Analytical Estimate for $\jrand$}
\label{sec:localcompare}
To compare our measured values of specific angular momentum to the analytical estimate of eq.~\eqref{eq:jrand}, we use the instantaneous profiles $\avej(r,t)$ and $v_{\rm turb}(r,t)$ to compute the time-averaged profiles, $\langle\avej\rangle$ and $v_c \equiv \langle v_{\rm turb}\rangle$, respectively, over $4000 < t/t_0 < 5000$. Then, $v_c$  and $H = r/b$ are used in eq.~\eqref{eq:jrand} to give the analytical estimate for $\jrand$.  
\begin{figure}
\begin{center}
	\includegraphics[width=0.95\columnwidth]{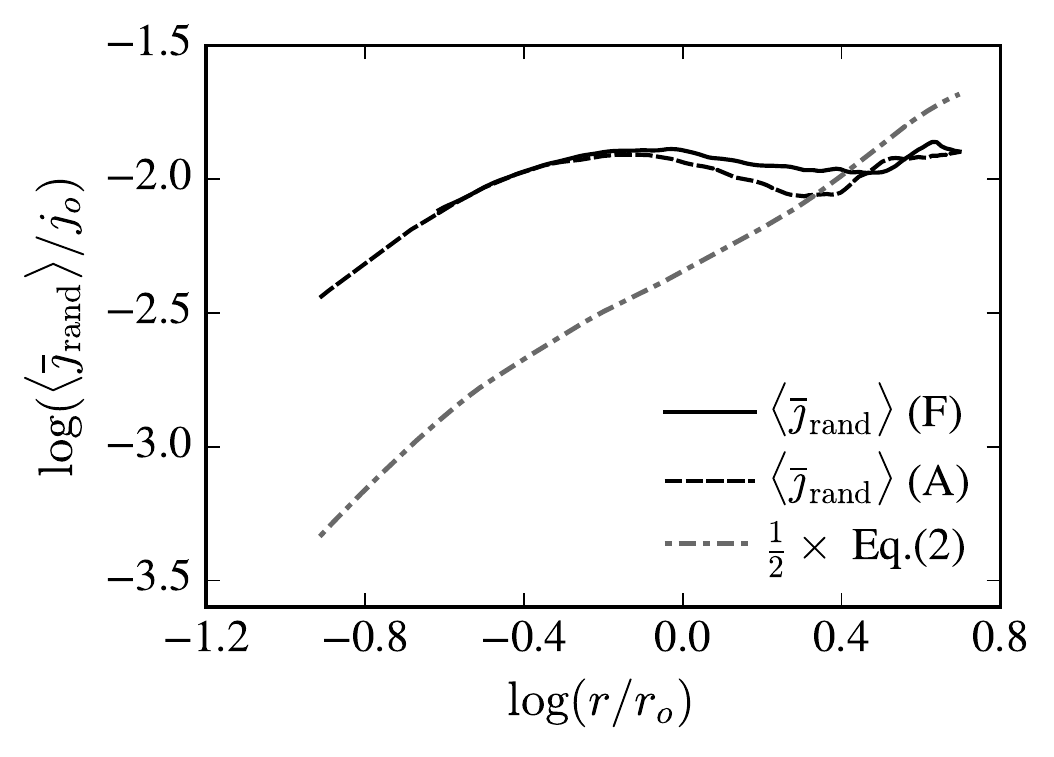}
    \caption{Comparison of the time- and spherically-averaged specific angular momentum profile, $\langle\avej\rangle$, of Model F (solid line) and Model A (dashed line), to the local analytical estimate. The dot-dashed curve uses the time-averaged turbulent velocity, $v_c \equiv \langle v_{\rm turb}\rangle$, to estimate $\jrand$ according to eq.~\eqref{eq:jrand}. We multiply this estimate by the factor of $1/2$ found by \citet{Quataert2019}. Instead of the $\jrand \propto r$ scaling of eq.~\eqref{eq:jrand}, we find that $\langle\avej\rangle$ is flatter in radius, suggesting that the convective flows are global and have roughly constant angular momentum over a range of radii, rather than being set locally at each radius (as is assumed in eq.~\eqref{eq:jrand}).}
    \label{fig:CONV_fig_janalytical} 
\end{center}
\end{figure}

The solid line in Fig.~\ref{fig:CONV_fig_janalytical} plots $\langle\avej\rangle$ versus radius over $r_{\rm heat,out} <  r < \rcool$ for Model F (solid line) and Model A (dashed line). The dash-dotted line is the analytical estimate for $\langle\avej\rangle$ using $v_c$ computed from Model A. In the plot, we include the factor of $1/2$ that was found to describe the local simulations of \citet[][see their equation 4]{Quataert2019}.  Eq. ~\eqref{eq:jrand} is a local estimate for $\jrand$ in a given shell as it assumes the value of $\jrand$ just depends on the local values of $v_c$ and $H$.  The local estimate seems to capture our simulation results at large $r$ well.   However, the $H\propto r$ scaling in the power-law convective envelope gives eq.~\eqref{eq:jrand} a strong radial dependence that is absent from our simulated $\langle\avej\rangle$ over a large range in $r$.  Instead, $\langle\avej\rangle$ is essentially flat in radius above $r \approx 0.6r_0$. At smaller $r$, $\langle\avej\rangle$ falls off with decreasing $r$, though with a shallower slope than the local estimate. By $r = 0.12 r_0$, there is an order-of-magnitude difference between $\langle\avej\rangle$ and the local estimate.   

We interpret our simulation results to indicate that $\jrand$ is being set at large $r$ (where there is good agreement with the local estimate) and $\jrand$ is roughly conserved as material flows to smaller $r$.  We noted in the context of Fig. ~\ref{fig:convect_flow} that the convective cells in our simulation appear coherent over a large fraction of the stellar radius. Our results suggest that these large structures approximately conserve $\jrand$, significantly increasing the angular momentum content of the flows at smaller radii. The recent radiation hydrodynamic simulations of \citet{2021arXiv211003261G} find similar results (see their sec. 4.2 for a detailed comparison).

As shown in Fig.~\ref{fig:initial_profiles}, the base of the convective hydrogen envelope for our \mesa RSG model is located at $r=0.03r_0$ in our code units. If we extrapolate the $\langle\avej\rangle$ curves of Fig.~\ref{fig:CONV_fig_janalytical} to $\log(r/r_0) = -1.52$, we find $\langle\avej\rangle = 10^{-3} j_0$.  Using eq.~\eqref{eq:j0_star} with $r_{\rm ph}= 840\rsun$ and $M_\bullet = 6 \msun$, then $\langle\avej\rangle \approx 8.8 \times 10^{16}$ cm$^2$ s$^{-1}$.  For comparison,  for a non-spinning $6\msun$ BH, $\jisco \approx 9 \times 10^{16}$ cm$^2$ s$^{-1}$.   This extrapolation of our results shows that $\jrand$ is important even near the base of the hydrogen envelope.  Material near the base of the convective hydrogen zone is the most bound and the most likely to survive the weak explosion in a FSN, though we reiterate that most of the convective envelope we simulate will remain bound during a FSN (see panel (d) of Fig.~\ref{fig:initial_profiles}).  

The time-averaged profiles considered here and in the previous section are useful for characterizing the properties of the flow, but the instantaneous specific angular momentum in the envelope is what matters for the collapse of the star.  The collapse begins from a particular state of the flow and the specific angular momentum accreting to small radii as a function of time depends on the state of the envelope at the start of collapse, as we explore with our collapse simulations in the next section.

\section{Collapse Simulations}
\label{sec:collapseresults}
In this section, we present the results of our collapse simulations, which are listed in Table \ref{tab:collapse}.  Each collapse run is initialized from a restart file that was output during our fiducial convection run, Model F.  The time associated with the restart file, $t_s$, is the start time of the collapse run, at which time the sink with radius $r_s$ is activated at the origin.  The physically relevant portion of our domain extends to $r = 5r_0$ so we stop the collapse simulation before shells that began at $r \gtrsim 5r_0$ begin to accrete. The rarefaction wave that is launched when the sink is introduced travels out at the local sound speed.  Once the rarefaction wave reaches a particular $r$, that shell begins to fall in towards the sink. The integrated sound-crossing time from the sink radius to $r = 5r_0$ is $10.8t_0$ (eq.~\ref{eq:integrated_cs} with $b=2.1$ and $\gamma = 1.4762$). The infall time (from rest) from $5r_0$ back to $r_s$ is $14.1t_0$ for a total accretion time of $\approx 25 t_0$ (see eq.~\ref{eq:acctime}). We thus run our collapse simulations until $t- t_s = 25t_0$. To be conservative, given that the material has variations in initial $v_r$, we only use data to a time of $t - t_s = 20t_0$.  

Collapse Runs 1 and 2 are initialized at $t_s = 4560$ and $4900$~$t_0$, respectively, when the convective flow has somewhat higher specific angular momentum with maximum $\avej(r,t_s)$ of $\approx 0.028 - 0.032$~$j_0$  (see Fig.~\ref{fig:pts_jrand_mag}).  Collapse Runs 3 and 4, with $t_s = 4270$ and $4800$~$t_0$, respectively, begin from states of somewhat lower specific angular momentum with maximum $\avej(r,t_s)$ of $\approx  0.015 - 0.02$~$j_0$.  Section \ref{sec:instant} presents the results of collapse Runs 1-4.  Collapse Run 1s is identical to collapse Run 1, except that we adopt a smaller sink size and, in Section \ref{sec:sinksize}, we show that our results are converged with respect to $r_s$.  Section \ref{sec:predict} considers the extent to which material is restructured during the collapse by comparing the measured accretion rates to rates predicted from assuming ballistic infall of the convective material.

\subsection{Flow and Instantaneous Accretion Rates}
\label{sec:instant}
\begin{figure*}
	\includegraphics[width=1\textwidth]{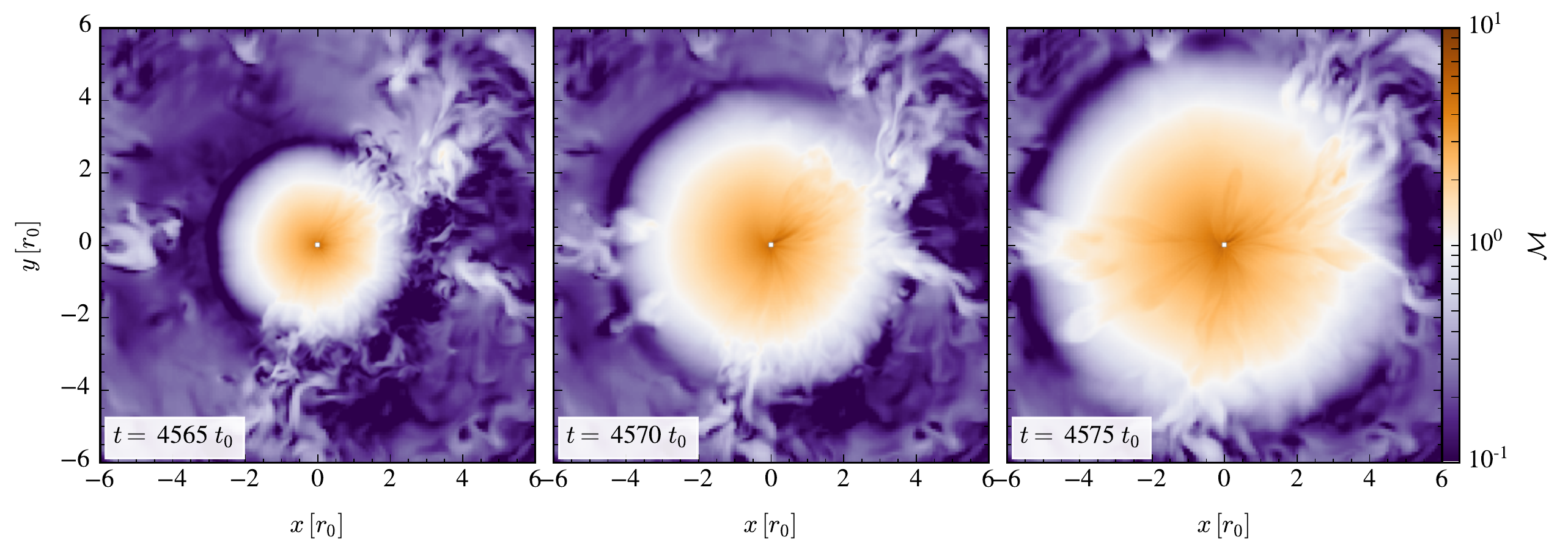}
    \caption{Slices of local Mach number, $\mathcal{M} = v/c_s$, through the $z=0$ plane for a simulation with an absorbing sink at small radii that initiates collapse of the convective envelope (collapse Run 1 in Table \ref{tab:collapse}). The snapshots show $5$, $10$, and $15t_0$ after the sink is activated at $t_s = 4560t_0$.  The transonic rarefaction wave separates the subsonic convective background flow from the supersonic inflow of material roughly in free-fall towards the low-pressure sink. The somewhat non-spherical rarefaction wave and infalling material carry the imprints of the initial state of the convective flows.}
    \label{fig:collapse_flow}
\end{figure*}
Fig.~\ref{fig:collapse_flow} shows snapshots of Run 1 at $5$, $10$, and $15$~$t_0$ after the collapse begins at $t_s = 4560t_0$.  Here we plot slices of the gas Mach number, $\mathcal{M} = v/c_s$, through the $z=0$ plane.  The outgoing rarefaction wave (roughly the outer edge of the white region) moves out through the subsonic convective flow, enclosing a region of supersonic infall as material falls back towards the sink.  The rarefaction wave is nearly but not exactly spherical and instead reflects the differences in the local radial velocity of the turbulent background.  For example, the flow at positive $x$ shows the imprint of two streams of material that were already falling back towards the origin before the introduction of the sink.    As expected for $\gamma < 5/3$ the gas reaching the sink is supersonic.

\begin{figure*}
	\includegraphics[width=0.8\textwidth]{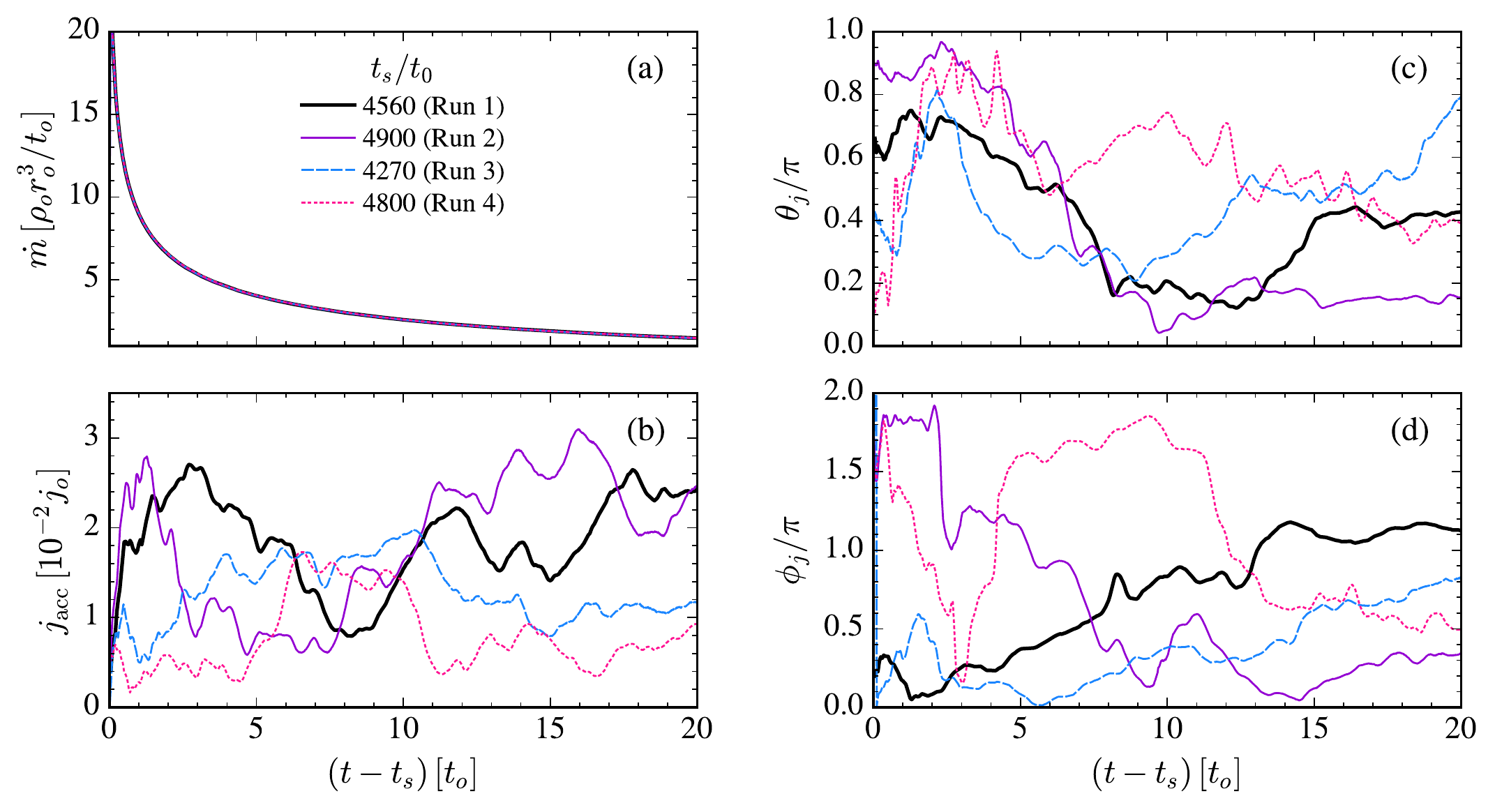}
    \caption{Instantaneous accretion rates of mass and angular momentum for the collapse of Model F starting at four different times.  Panel (a): mass accretion rates.  Panels (b), (c), and (d): magnitude, polar angle, and azimuthal angle, respectively, of $\boldsymbol{j}_{\rm acc} \equiv \boldsymbol{{\dot{J}}}/\dot{m}$, the specific angular momentum vector of the accreted material. The $x$-axis is the elapsed time since the start of the collapse. Panels (a) and (b) can be scaled to physical units via eqs.~\eqref{eq:mdot0_star} and \eqref{eq:j0_star}, respectively. The $\phi_j$ values for Runs 2 and 4 have been given small phase shifts to avoid the discontinuity $\phi_j = 0 = 2\pi$ where the values would otherwise cross the $x$-axis.}
    \label{fig:accrates}
\end{figure*}

Fig.~\ref{fig:accrates} shows the rates of mass and angular momentum accretion for Runs 1 - 4 as a function of elapsed time, $t - t_s$, after introduction of the sink. The rates reported here are computed by measuring the fluxes through the sink surface at $r_{\rm flux} = r_s = 0.08r_0$.   The  mass accretion rate, panel (a), falls off in time because the mean density profile falls off with radius as $\propto r^{-2.1}$. With $\jrand \ll j_0$ and $\mconv <1$ neither centrifugal nor turbulent pressure modify the infall so there is no difference in $\dot m$ between collapse runs started from different convection states, even with a factor of 2 difference in maximum $\avej(r,t_s)$.  

The remaining panels of Fig.~\ref{fig:accrates} characterize the specific angular momentum vector of the accreted material, $\vjacc \equiv \dot{\boldsymbol{J}}/\dot{m}$, over time. Panels (b), (c), and (d) show the magnitude, $\jacc$, the polar angle, $\theta_j$, and the azimuthal angle, $\phi_j$, respectively (the direction angles are defined as in Section~\ref{sec:convect_specificAM}).  Panel (b) shows that, critically, the infall of the envelope has not restructured the material in such a way as to erase or diminish the $\jrand$ resulting from the convective motions in the envelope.  Indeed, $\jacc$  over time for all runs is just as expected from Fig.~\ref{fig:pts_jrand_mag} with $0.005 \lesssim \jacc \lesssim 0.03$ for Runs 1 and 2, and with $0.002 \lesssim \jacc \lesssim 0.02$ for Runs 3 and 4.  

The direction of $\vjacc$ as a function of time is important for determining the ultimate fate of material as it plunges to yet smaller radii. Focusing on the thick, solid line for Run 1, panels (c) and (d) show that the largest variations in $\theta_j$ and $\phi_j$ occur over $\sim 5-10$ $t_0$ while smaller-amplitude changes take place on $\sim t_0$ timescales. Other curves show one or two large-amplitude changes on somewhat shorter timescales.  For example, Run 4 (dotted, pink lines) shows large swings in $\theta_j$ and $\phi_j$ over the first $\sim 3 t_0$ of the collapse before transitioning to a phase where the angles are relatively constant. Recall, however, that the dynamical time at $r_{\rm circ}$ is $10^{-6} t_0 \ll t_0$.   Taken together, these curves show that the magnitude, direction, and variability of $\vjacc$ depend on the state of the envelope at the start of the collapse. Overall, however, the direction of $\vjacc$ tends to be a slowly varying function over the time it takes for a large fraction of the envelope to reach the sink.

\subsection{Comparison to Semi-Analytical Predictions}
\label{sec:predict}
\begin{figure}
\center
	\includegraphics[width=0.8\columnwidth]{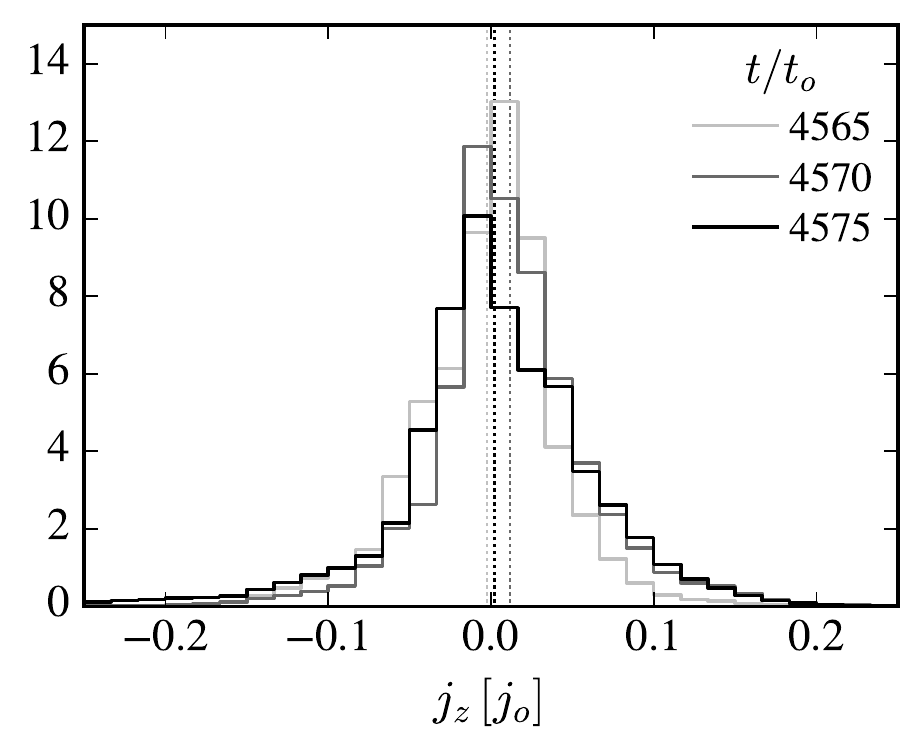}
    \caption{Histogram of $j_z$ for grid cells in a shell just outside the sink ($r_s = 0.08r_0$) during collapse simulation Run 1. Each histogram is for a different simulation time (listed in the legend) and is computed using all grid cells in the shell with $0.1 < r/r_0 < 0.32$. All cells have the same volume and the inflow is supersonic in this region. The times chosen are spaced such that new material has entered the region (and the old material has been accreted by the sink) when each histogram is computed.  While the mean values in each shell (vertical lines) are $-0.0024, 0.0021,$ and  $0.012$ $j_0$, the width of the distribution is a factor of 10 larger. The width of the corresponding distribution in circularization radius is roughly $100$ times the mean circularization radius of the shell.  The parcels of gas in each shell will begin to circularize at a large range of radii, which will result in a complicated flow as centrifugal support becomes important.}
    \label{fig:histogram}
\end{figure}

An important goal of this work is to understand the extent to which the mean angular momentum associated with convective flows is modified during infall.  We explore in this section whether the mean profiles from the convection simulation, that is $\avej$ of Fig.~\ref{fig:pts_jrand_mag} and $\overline{\theta}_j$ and $\overline{\phi}_j$ of Fig.~\ref{fig:pts_theta_phi}, can be used to predict the accretion rates that are realized in the collapse simulation under the assumption of ballistic infall of each shell.  There is significant dispersion in $v_r$, $j_x$, $j_y$ and $j_z$ in each radial shell about the mean profiles that we have shown thus far (see Fig.~\ref{fig:histogram}, which shows histograms of $j_z$ values of material about to enter the sink in our collapse run). One can imagine that if material with the largest $v_r$ consistently has, e.g., $j_z$ that is far from the mean, then material reaching the sink at a given time could be sampling the tail of the $j_z$ distribution of many different shells at the same time, and therefore would not match our ballistic prediction using the mean profiles.

We now briefly explain how we predict accretion rates from the time-dependent mean radial profiles of mass density, $\overline{\rho}(r,t)$, angular momentum density, $\overline{\ell}(r,t) \equiv \overline{\rho}(r,t)\avej(r,t)$, and direction angles $\overline{\theta}_j(r,t)$ and $\overline{\phi}_j(r,t)$ (see Appendix \ref{sec:appxpredict} for a more detailed explanation and a simpler test case in which the profiles are not changing in time).   Upon introduction of the sink, the outgoing rarefaction wave travels at the sound speed and a given shell begins the infall once the rarefaction wave arrives.  The time it takes for the rarefaction wave to reach $r$ is the integrated sound-crossing time, $t_{\rm wave}(r)$ given in eq.~\eqref{eq:integrated_cs}.  The shell falls from rest from $r$ over a time $1.14 t_{\rm ff}(r)$ where $t_{\rm ff}(r)$ is the integrated free-fall time, eq.~\eqref{eq:freefalltime}.  The factor of $1.14$ (supplied by Eric R. Coughlin and based on the self-similar rarefaction wave solutions of \citealt{2019ApJ...874...58C}), comes from the fact that infall begins from a profile in hydrostatic equilibrium. There is thus a pressure gradient behind the rarefaction wave, so the infall is not true zero-pressure `free-fall' as assumed by $t_{\rm ff}(r)$. The total time for the shell initially at $r$ to accrete is the sum $t_{\rm acc}(r) = t_{\rm wave}(r) + {1.14}t_{\rm ff}(r)$.  

In our ballistic prediction, we assume that the mass and angular momentum that arrive at the sink at time $t_{\rm acc}(r_i)$ are just the amount of mass and angular momentum that were contained in the shell at $r_i$ when the rarefaction wave arrived.  These values are computed from the time-dependent mean profiles from the convection simulation.  We then assume the computed value of the quantity of interest in that shell, e.g. the total mass, does not change as that shell falls in over the subsequent time of ${1.14} t_{\rm ff}(r_i)$.   

\begin{figure}
\center
	\includegraphics[width=0.95\columnwidth]{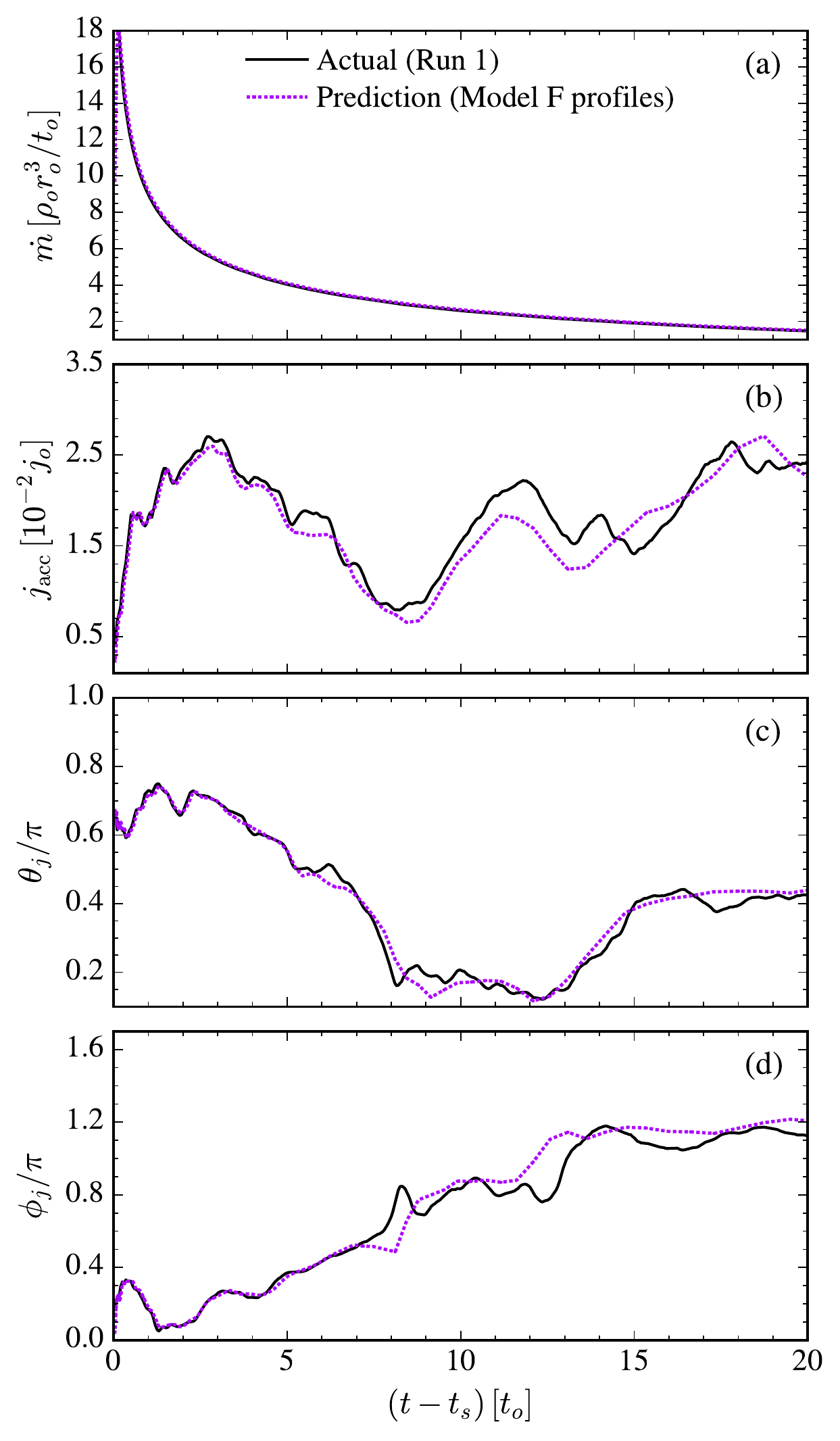}
    \caption{Comparison of the predicted accretion rates of mass and angular momentum with those measured in the collapse simulation. Solid, black curves are the instantaneous accretion results for Run 1 shown in Fig.~\ref{fig:accrates}. The purple, dotted curves use the time-dependent, spherically-averaged profiles from convection Model F to predict the accretion results by accounting for the travel time of the rarefaction wave from the sink to a given shell and assuming each shell then undergoes ballistic infall.  The agreement between the curves in each panel shows that there is little restructuring of $\vjrand$ during this phase of infall in which $r > \rcirc$.}
    \label{fig:collapse_predict}
\end{figure}

The purple, dotted curves in Fig.~\ref{fig:collapse_predict} plot the predicted values of the mass accretion rate, $\dot{m}$, as well as the predicted magnitude and direction angles of the specific angular momentum vector of the accreted material for Model F assuming collapse begins at $t_s = 4560t_0$.  For comparison, the accretion results realized in the equivalent collapse simulation (Run 1) are shown with solid, black lines in each panel.  

Panel (a) shows that $\dot{m}$ is predicted extremely well and serves as independent confirmation of the  \citet{2019ApJ...874...58C} rarefaction wave solutions.  While there are some small differences between the two curves in panels (b)-(d), the magnitude and direction angles of $\vjacc$ are well-predicted using the mean profiles and the assumption of ballistic infall from rest.  This good agreement with the actual accretion rates is non-trivial, especially given the large dispersion of $v_r$ and the components of $\vjrand$ in each shell as well as the fact that all of the material has non-zero $v_r$ rather than starting the infall from rest, as we assumed. 

We interpret the excellent agreement between measured and predicted $\vjacc$  to mean that during the transition from convection to infall and in the subsequent supersonic infall, very little restructuring of the material takes place. This should remain the case until another source of dynamical support becomes important, e.g. centrifugal pressure. This will not occur until much smaller radii than simulated here ($\rcirc \ll r_s$).

Although we do not show the results here, we also computed predictions for comparison to collapse Run 3 (that is, computing the prediction assuming collapse of Model F from $t_s = 4270 t_0$). We found the same level of agreement between predicted and actual accretion rates as in Fig.~\ref{fig:collapse_predict}.

\subsection{Dependence on Sink Size}
\label{sec:sinksize} 
In this section, we compare our measurement of the circularization radius of the accreted material, $r_{\rm circ}(t) = \jacc^2 /GM$, for collapse Runs 1 and 1s which differ only in choice of sink radius ($r_s = 0.08$ and $0.04$ $r_0$, respectively).  For these simulations, we measure mass and angular momentum accretion fluxes at both the sink radius, $r_{\rm flux} = r_s$, and at a larger radius of $r_{\rm flux} = 0.48r_0$. 

\begin{figure}
\begin{center}
	\includegraphics[width=0.9\columnwidth]{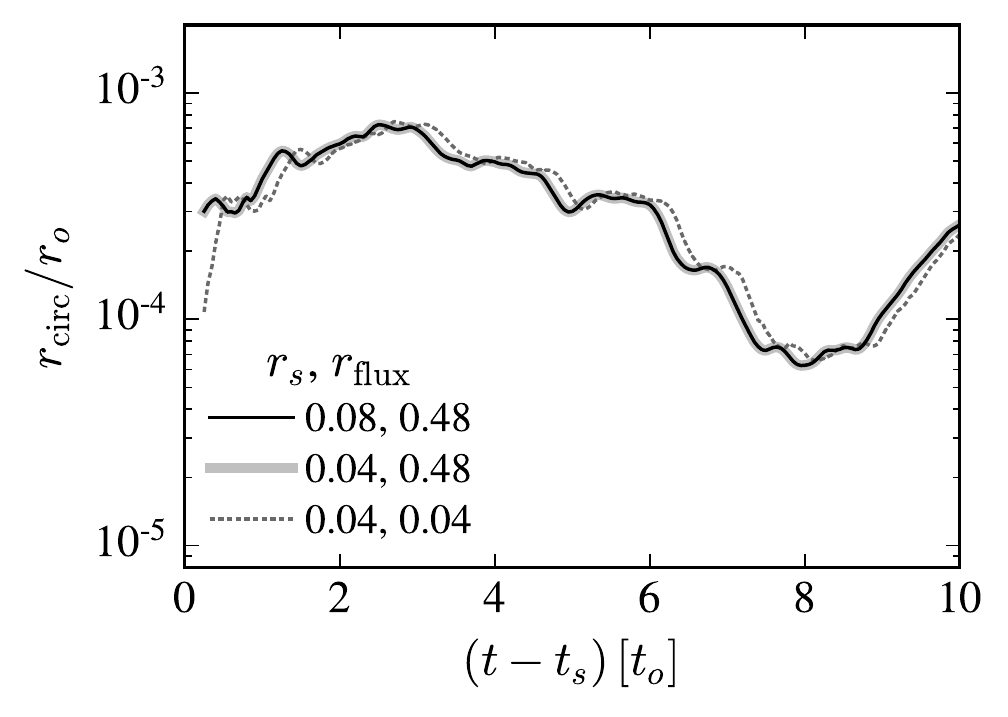}
    \caption{Comparison of the circularization radius of accreted material for simulations with different sink sizes initiated from convection Model F at $t_s = 4560 t_0$. The thick grey and thin black curves,  which lie on top of each other, adopt sink sizes of $r_s = 0.04$ and $0.08$ $r_0$, respectively, and show $\rcirc$ computed using the accretion rates measured at $0.48r_0$ in both simulations.   The dotted and thick grey lines show $\rcirc$ for the simulation with $r_s = 0.04r_0$ computed using the accretion rates measured at $0.04$ and $0.48$ $r_0$, respectively. They are the same except for a small shift in time due to the finite travel time from $0.48$ to $0.04$ $r_0$.}
\label{fig:COLL_sink_size}
\end{center}
\end{figure}

The two solid lines in Fig.~\ref{fig:COLL_sink_size} plot $r_{\rm circ}$ of the accreted material as measured at $r_{\rm flux} = 0.48r_0$. The circularization radii measured at $0.48r_0$ in the two simulations agree to better than $0.15\%$, indicating that the flow has already converged with respect to sink size by $r_s = 0.08r_0$.  

The dotted line in Fig.~\ref{fig:COLL_sink_size} shows the accretion rate for Run 1s measured at $r_{\rm flux} = r_s = 0.04r_0$.  The curve is shifted by the finite travel time from $0.48$ to $0.04$ $r_0$ ($\approx 0.18t_0$) but is otherwise the same as the other two curves. 

\section{Discussion}
\label{sec:discussion}
In this section we consider our results in the context of FSN, scaling our convection results to a typical RSG and discussing the possible character of accretion and outflows during the infall of the convective hydrogen envelope.

\subsection{Application to Supergiants}
\label{sec:application_to_supergiants}
In code units, the circularization radius of material with specific angular momentum $\jrand$ is  
\beq
r_{\rm circ} = ({\jrand}/{j_0})^2 r_0.
\label{eq:rcirc_code}
\eeq  
Making use of $r_0 = r_{\rm ph} /6$ to write the previous expression in physical units, 
\begin{align}
r_{\rm circ} = 1.16\times10^{9} \bigg(\frac{\jrand}{10^{-2} \,j_0}\bigg)^2 \bigg(\frac{r_{\rm ph}}{1000\rsun}\bigg) {\rm cm}.
\label{eq:rcirc_star}
\end{align}
For a non-spinning BH, $\risco = 6 GM_\bullet/c^2$, so the ratio of circularization radius to the radius of the ISCO is 
\beq
\frac{r_{{\rm circ}}}{\risco} =  130 \, \bigg(\frac{\jrand}{10^{-2} \,j_0}\bigg)^2\bigg(\frac{r_{\rm ph}}{1000\rsun}\bigg)\bigg(\frac{10\msun}{M_\bullet}\bigg).
\label{eq:rcirc_rsg}
\eeq
Typical values for the fraction inside the first set of parenthesis are roughly 1-3  (see Fig.~\ref{fig:pts_jrand_mag} and panel (b) of Fig.~\ref{fig:accrates}). 

\begin{figure*}
\begin{center}
	\includegraphics[width=0.8\textwidth]{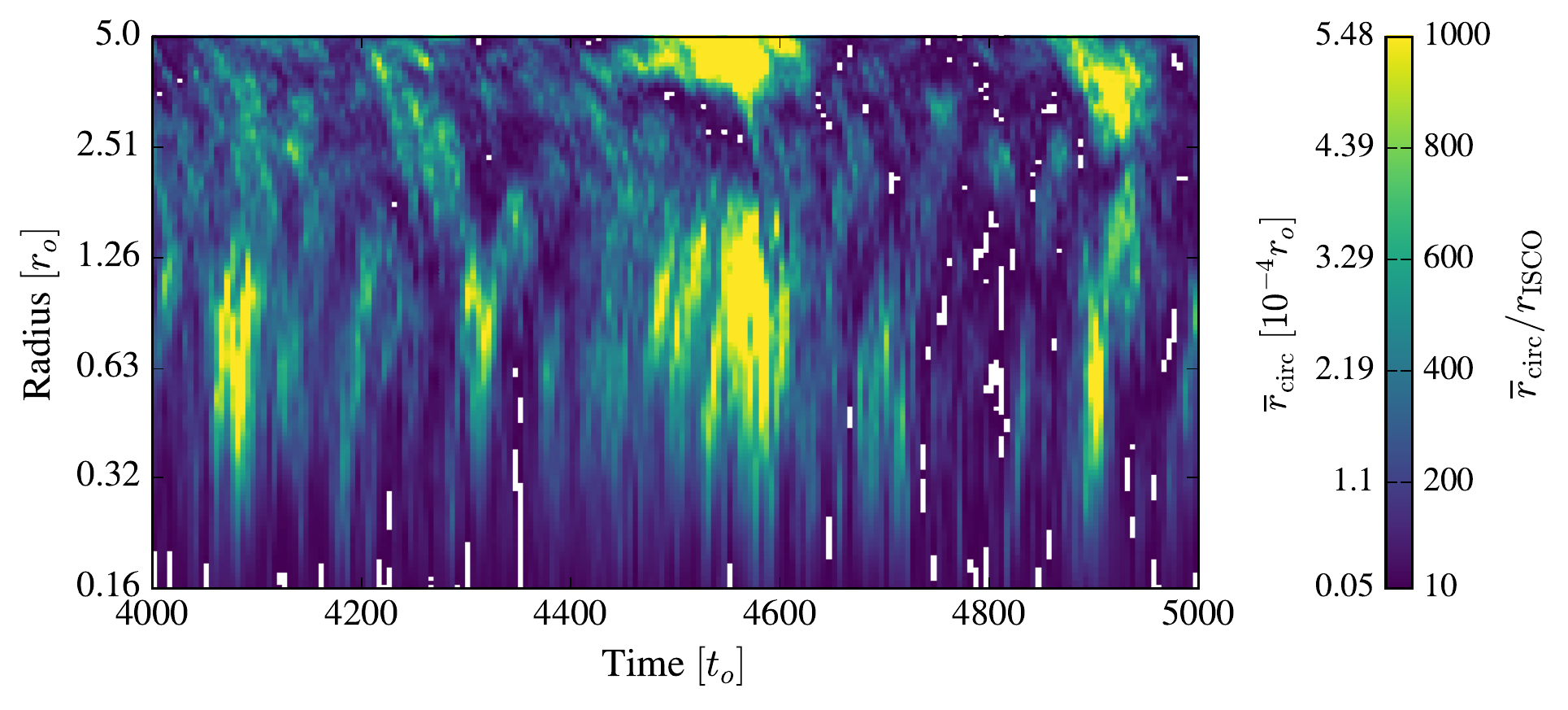}
    \caption{Spherically-averaged circularization radius profile, $\overline{r}_{\rm circ}(r,t)$,  as a function of radius and time for our fiducial convection model.  The left- and right-hand colorbar labels report, respectively, the values in code units (eq.~\ref{eq:rcirc_code}) and as a multiple of the ISCO radius (eq.~\ref{eq:rcirc_rsg}).  We take $r_{\rm ph} = 840 \rsun$ and $M_\bullet = 6\msun$ in eq.~\eqref{eq:rcirc_rsg}, appropriate for the \mesa model of Fig.~\ref{fig:initial_profiles}.  White patches in the plot indicate regions that fall below the colorbar floor of $10$ $\risco$. For nearly all radii and at all times, $\avercirc(r,t)$ of each shell is $\sim 10 - 1500 \times \risco$.}
    \label{fig:CONV_pts_rcirc}
\end{center}
\end{figure*}

Fig.~\ref{fig:CONV_pts_rcirc} shows the mean profiles of circularization radius for the convective material, $\avercirc(r,t)$, which are computed using $\avej(r,t)$ of Fig.~\ref{fig:pts_jrand_mag} in eq.~\eqref{eq:rcirc_code}.  The left side of the colorbar is in code units. The right side converts those code values to $\avercirc/{\risco}$ using eq.~\eqref{eq:rcirc_rsg} and adopting $r_{\rm ph} \approx 840 \rsun$ and $M_\bullet = 6\msun$, as in Fig.~\ref{fig:initial_profiles}. The white regions are where $\avercirc$ is below the colorbar lower-limit of $10\risco$ (if we set this lower bound to $\risco$ instead, only a few white points would remain). The figure shows that turbulent Mach numbers of $\sim 0.1$ (Fig. \ref{fig:mturb_pts}) correspond to circularization radii that are many hundreds to a thousand times larger than $\risco$.  In addition, $\avercirc \gtrsim 10 {\risco}$ at nearly all radii and all times.  So although $\avejvec$ is a function of both radius and time, the convective material begins the collapse with $\avercirc \gtrsim 10 {\risco}$ almost independent of when the collapse of the envelope begins.    

   We further showed in Section \ref{sec:predict} that there is little restructuring of the specific angular momentum as the matter infalls to smaller radii.  Instead, we found that both the magnitude and direction of $\vjacc$ are well-predicted by the mean profiles of the envelope prior to collapse (Fig.~\ref{fig:collapse_predict}).  We therefore conclude that, although the times at which we started the collapse in our collapse runs were arbitrary with respect to the flow, Fig.~\ref{fig:CONV_pts_rcirc} shows that $\rcirc$ of the convective material will be well outside of $\risco$ no matter when the collapse begins. 

\subsection{Implications for Failed Supernovae}
\label{sec:implications}
In Sections~\ref{sec:localcompare} and \ref{sec:application_to_supergiants}, we showed that the entire hydrogen convective zone has $\jrand \gtrsim \jisco$.  Based on our simulation results, we argued that the  local estimate of $\sim\frac{1}{2}$eq.~\eqref{eq:jrand} seems to set $\jrand$ at large radii in the convection zone and that $\jrand$  is roughly conserved down to smaller radii.  The helium layer is also convective, though with smaller convective velocities, $v_c$. If we assume that for the helium convective zone $\jrand$ is related to $v_c$ by $\frac{1}{2}$eq.~\eqref{eq:jrand}, then
 $\avej \le 5 \times 10^{14}$ cm$^2$ s$^{-1}$ or $\avej \le 10^{-2}\jisco$ for our \mesa RSG model (see also fig. 3 of \citealt{Quataert2019}). This value is consistent with simulations of convection in the helium layer performed by \citet{2016ApJ...827...40G} for which the mean profiles of the Cartesian components of $\vjrand$ were $\lesssim 7 \times 10^{14}$ cm$^2$ s$^{-1}$ in magnitude or $\lesssim 0.03\jisco$ for their assumed BH mass. The other convective regions interior to the helium layer have even smaller $\avej$.    We conclude that, for non-rotating stars, convective material interior to the hydrogen convection zone is likely to accrete spherically onto the BH.  We note that this conclusion still holds even if the convective Mach number in the oxygen layer is a factor of $4-5$ larger than captured in the \mesa model, as found in the 3D simulations of \citet{2020ApJ...901...33F}.  
 
One subtlety related to the infall of the helium layer is the large dispersion in $\jrand$ about the mean $\avej$ in each radial shell of the convective zone (see Fig.~\ref{fig:histogram}).  \citet{2016ApJ...827...40G} find that some parcels in their helium layer have $\jrand \sim 10^{16}$ cm$^2$ s$^{-1}$ $\sim \jisco$; they argue that these individual parcels could generate an outflow even though the majority of the infalling material has $\jrand \ll \jisco$.  It is unclear whether localized regions with $\jrand > \jisco$ in the helium layer can reverse the inflow of the bound envelope or are simply advected into the BH along with the bulk of the material. This requires further study.   In what follows, we focus on the infall of the hydrogen envelope onto the newly-formed BH. 

For our \mesa RSG model, panel (d) of Fig.\ref{fig:initial_profiles} shows that the binding energy of the hydrogen envelope is $E_{\rm bind} \approx 2\times10^{48}$ erg.   Some of the hydrogen envelope may be ejected due to weak shocks induced by neutrino-cooling during the proto-NS phase. The arrows in panel (d) of Fig.~\ref{fig:initial_profiles} show the amount of mass that could be unbound based on the range of shock energies found by \citet{2021ApJ...911....6I}.  For the lowest shock energies, only $0.5\msun$ is likely to be unbound (open arrows) while for the highest shock energy, $3.6\msun$ of the envelope could be ejected (filled arrows). In either case, at least $6.9\msun$ of the convective hydrogen zone remains bound to the BH, which represents the bulk of the convection zone in our simulations.   
 
For $\rcirc \approx 500 \risco \approx 0.05\rsun$, as is typical in our simulations,  $E_{\rm bind}$ for the entire hydrogen envelope could be supplied by accretion of an amount of mass 
\beq
\delta m \sim \frac{\rcirc E_{\rm bind}}{GM_\bullet} \sim 3 \times 10^{-3} \msun.
\eeq
(if inflow makes it to $\risco$, then $\delta m \approx 10^{-5}\msun$).   For our \mesa model, the accretion rates of Fig.~\ref{fig:accrates} translate to $\dot{m} \sim 20 \msun$ yr$^{-1}$. At that rate, $\delta m\sim 10^{-3} \msun$ is accreted after roughly half an hour.

At these super-Eddington accretion rates of $\approx$ $10-100$ $\msun$ yr$^{-1}$, the flow is optically thick and unable to cool by radiation. With $\dot{m} \ll 10^{-3}\msun$ s$^{-1}$ there is no neutrino cooling either \citep{2008AIPC.1054...51B}, so accretion of the hydrogen envelope is an optically-thick, advection-dominated accretion flow \citep[optically-thick ADAF;][]{1982ApJ...253..873B,1988ApJ...332..646A,1994ApJ...428L..13N}.  As material falls to small radii without the ability to cool, gravitational potential energy can only be converted into kinetic and thermal energy and the material may just return to large radii in an outflow. In addition to the collimated outflows and disk winds that super-Eddington disks inevitably produce \citep{2019ApJ...880...67J}, accretion of stellar material with fixed rotation axis and roughly uniform circularization radius sets up an accretion shock that sweeps through the infalling material and can unbind the outer parts of the star \citep{2010ApJ...713..800L,2020ApJ...901L..24M}.  

If an accretion shock does not unbind all of the envelope and material can circularize into a disk, then preferential outflow into a region along the angular momentum axis would allow accretion to continue for longer periods than could be possible for more spherical outflow.  The relevant timescale in this case is that over which the direction of $\vjacc$ varies.  For example, if the coherence time for the direction of $\vjacc$ is $\delta t \sim 2 t_0 \approx 25$ days (the approximate time during which $\theta_j$ changes by $\lesssim \pi / 2$ for all curves in Fig.~\ref{fig:accrates}), then $\delta m \sim \dot{m}\delta t \sim 1.4\msun$ would have time to fall in. If a fraction $f$ (possibly $\ll1$) of this material reaches $\risco$ for a $6\msun$ BH, the change in potential energy would be $|\delta U| \sim 4 \times 10^{53}f$ erg. Envelope gas within the funnel region would easily be unbound as this energy drives outflows from small radii.   If the orientation of the disk changes on the same timescale as that of $\vjacc$ ($\gtrsim 2 t_0 \sim 1$ month), then the collimated outflow could sweep out a large fraction of $4\pi$ on this timescale, ejecting most of the material and resulting in an energetic, long-duration transient.  

The timescale and energetics inferred in the previous paragraph are similar to what is needed to power extremely long-duration gamma-ray transients such as Swift 1644+57 \citep{2011GCN.11847....1B} and  Swift J2058.4+0516 \citep{2012ApJ...753...77C} by stellar core-collapse \citep[e.g.,][]{2012MNRAS.419L...1Q, 2012ApJ...752...32W}.   The biggest uncertainty in this application of our results is whether the tenuously-bound hydrogen envelope is unbound before much of it can accrete onto the newly-formed BH.

\subsubsection{BH Spin}
\label{sec:bhspin}
The efficiency of tapping accretion power to unbind the hydrogen envelope determines not only how much material is returned to the star's environment and the nature of the transient produced, but it also determines the final spin of the BH.  We can make a rough estimate of the mass and final BH spin by assuming the BH accretes mass at a rate of $f\dot{m}$, where $\dot{m}$ is the rate that matter falls in from the envelope and $0 < f \le 1$, and by computing the angular momentum accretion rate of the BH according to 
\beq
\dot{\boldsymbol{J}_\bullet} =
\begin{cases}
    f\dot{m} \vjacc       & \text{if } \jacc < \jisco \\
    f\dot{m} \jisco\ujacc & \text{if } \jacc \ge \jisco
\end{cases}
\label{eq:jdotBH} 
\eeq
(here $\ujacc$ is the unit vector in the direction of $\vjacc$ and $\dot{m}(t)$ and $\vjacc(t)$ are computed from the instantaneous profiles from our convection simulations using the methods of Sec~\ref{sec:predict}). By limiting $\jacc$ to $\jisco$, this estimate accounts for the fact that an accretion disk or outflow transports angular momentum to infinity to allow a fraction $f$ of the infalling material to reach $\risco$.

To compute the spin of the BH, we assume that everything interior to the convective hydrogen envelope is accreted and carries no angular momentum, so we initialize the BH mass to $M_\bullet = 6 \msun$ and the BH angular momentum to $\boldsymbol{J}_\bullet = \mathbf{0}$.  Working outward from the base of the hydrogen envelope, we update $M_\bullet$  given $\dot{m}(t)$ and $\boldsymbol{J}_\bullet$ is increased using eq.~\eqref{eq:jdotBH}. Throughout this process, $\jisco$ is evolved with the mass and spin of the BH using  the standard relations (evaluated at the equator) for a Kerr BH.  We do not account for the angle between the spin vector of the BH and the angular momentum vector of the shell except to check whether the dot product is positive or negative.  If the dot product is positive or zero, we assume the material is on a prograde orbit when computing $\jisco$. For a negative dot product, we assume a retrograde orbit.  We also ensure that the spin of the BH never exceeds the maximum value $J_{\rm max} = GM_\bullet^2/c$, though this ends up not being necessary because $J_{\bullet} / J_{\rm max}$ is always $\lesssim 0.8$.  \begin{figure}
\center
%	\includegraphics[width=0.95\columnwidth]{DISC2_spin_parameter}
%\caption{Dimensionless spin parameter of the BH versus the radius up to which the convective envelope is incorporated into the BH.  Here radius refers to the location in the envelope before collapse begins and the spin at each $r$ is the spin that the BH would have if everything exterior to $r$ were unbound. Grey shading indicates where $r> \rcool$. The mass and angular momentum in each accreted shell are calculated using the instantaneous profiles from the convection simulations as described in Sec. \ref{sec:predict}.  The legend gives the convection model and assumed start time for the collapse, $t_s$, that were used to compute each curve. The curves shown here assume all of the mass in each shell is accreted ($f = 1$ in eq.~\ref{eq:jdotBH}). In all cases, the BH spin approaches $\approx 80\%$ of maximum as more of the envelope is accreted before declining (due to a sign change in the accreted $\dot{\boldsymbol{J}}$).  Although the total angular momentum in the envelope is zero, the BH ends up with non-zero spin because only a small fraction of the specific angular momentum in each shell can be accreted by the BH. For partial accretion of the mass in each shell ($f < 1$) the vertical extent of each curve is reduced and the BH achieves smaller maximum spin at the peak of the curve. For $0.03 < f < 1$, the maximum spins are $\sim 0.06 - 0.59$. Weak shocks like those studied by \citet{2021ApJ...911....6I} imply BH spins of $\sim 0.5$, near the peak of the curves.}
	\includegraphics[width=0.95\columnwidth]{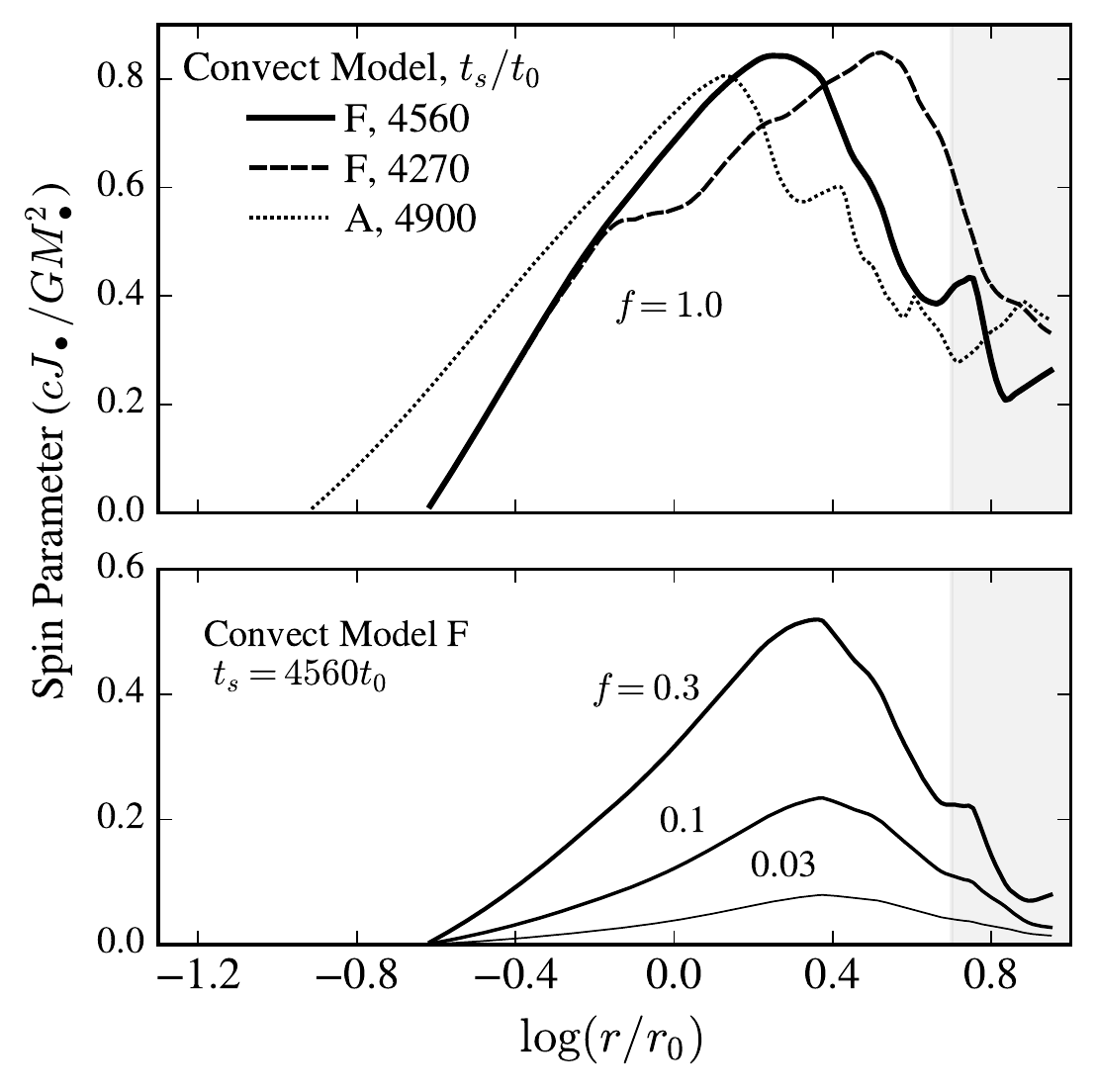}
    \caption{Dimensionless spin parameter of the BH versus the radius up to which the convective envelope is incorporated into the BH.  Here radius refers to the location in the envelope before collapse begins and the spin at each $r$ is the spin that the BH would have if everything exterior to $r$ were unbound. Grey shading indicates where $r> \rcool$. The mass and angular momentum in each accreted shell are calculated using the instantaneous profiles from the convection simulations as described in Sec. \ref{sec:predict}.  The upper panel assumes all of the mass in each shell is accreted ($f = 1$ in eq.~\ref{eq:jdotBH}). The legend gives the convection model and assumed start time for the collapse, $t_s$, that were used to compute each curve. In all cases, the BH spin approaches $\approx 80\%$ of maximum as more of the envelope is accreted before declining (due to a sign change in the accreted $\dot{\boldsymbol{J}}$).  Although the total angular momentum in the envelope is zero, the BH ends up with non-zero spin because only a small fraction of the specific angular momentum in each shell can be accreted by the BH.  For partial accretion of the mass in each shell ($f < 1$, lower panel) the BH achieves smaller maximum spin at the peak of the curve. For $0.03 < f < 1$, the maximum spins are $\sim 0.06 - 0.59$. Weak shocks like those studied by \citet{2021ApJ...911....6I} imply BH spins near the peak of each curve with spins of $\sim 0.5$ when $f \sim 1$ and spins of $\sim 0.1$ when $f \sim 0.1$.}
    \label{fig:spin}  
\end{figure}

In Fig.~\ref{fig:spin}, we plot the dimensionless spin parameter of the BH, $J_\bullet / J_{\rm max}$,  as a function of the radius $r$ out to which the envelope is incorporated into the BH.\footnote{The final density profile in the Athena models differs a bit from the initial profile (see Fig.~\ref{fig:initial_profiles}).   In evaluating the BH spin in Fig.~\ref{fig:spin}, we renormalize the density by a factor of 0.75 to keep the total mass within $r=5r_0$ the same as the initial condition.    The spin calculated without this renormalization is very similar to that shown in Fig.~\ref{fig:spin}, so our conclusions are not sensitive to this choice.} The $r$ coordinate refers to the radius of the material before the collapse. So at each $r$ in the figure, the value of the spin assumes material located at radii $>r$ prior to collapse is unbound and a fraction $f$ of the material that had radius $< r$ before collapse is accreted onto the BH.  The curves in the upper panel correspond to a different convection model and start time for the collapse, as noted in the legend, for the case in which $f = 1$. The spin parameter increases slowly as material with $j\sim \jisco$ is added to the initially non-spinning BH.  For all the models shown, the spin parameter saturates near $\sim 0.8$ before slowly decreasing as material with opposite spin direction is accreted. If material out to $\rcool$ is accreted, the final spin parameter is $0.3 - 0.7$.  There is some angular momentum exterior to $\rcool$ (grey shaded region), so we extend the curves to $r = 9 r_0$, where the cumulative profiles of Fig.~\ref{fig:cumulative_J} approach zero.  If all of this material is incorporated, the final spin parameter is $0.3-0.4$.   The spin parameter does not return to zero (as one might expect from Fig.~\ref{fig:cumulative_J}) because only a small fraction of the total available specific angular momentum in each shell can be incorporated into the BH.  As a result, there is uneven cancelation of the angular momentum vectors across many different shells, unlike the integral over the star at fixed time that is shown in Fig.~\ref{fig:cumulative_J}.

The BH spin versus $r$ curves shown in the upper panel of Fig.~\ref{fig:spin} represent upper limits on the spin for two reasons. First, we assumed that all mass reaching small radii accretes. In reality, some of this mass will go into the outflow and will not accrete onto the BH. Second, once an outflow from small radii occurs, much of the hydrogen envelope will be blown away, thus modifying the $\dot{m}$ that can be fed from large radii by the envelope. \citet{2019arXiv190404835B} considered these two effects in more detail in the context of rapidly rotating collapsars; here we fold both effects into the factor $f$ in eq.~\eqref{eq:jdotBH}.  To explore the influence of a reduced accretion rate on the final BH mass and maximum spin, we repeated each of the calculations of the upper panel of the figure with $f = 0.03, 0.1$, and $0.3$, instead of $f=1$. The lower panel of Fig.~\ref{fig:spin} shows the resulting curves for the particular case of collapse of Convect Model F from $t_s = 4560 t_0$.  Across all models, the lower values of $f$ give maximum spin parameters of $0.06-0.1$, $0.19-0.29$, and $0.45-0.59$, respectively,  and final BH masses (integrating out to $\rcool$) of $\sim 6.6$, $7.3$, and $9.4$ $\msun$, respectively.  For comparison, $f =1$ gives a final BH mass of $\sim 16.8 \msun$ and maximum spin parameters of $0.81 - 0.85$.  If weak shocks, like those studied by \citet{2021ApJ...911....6I}, unbind $\sim$ few $\msun$ of material, the BH spin would be $\sim 0.1 - 0.5$, near the peak of the curves in Fig.~\ref{fig:spin}.  

\subsubsection{Timescale for Angular Momentum Redistribution}
Our calculations neglect self-gravity; thus it is useful to check whether internal gravitational torques can significantly reduce the mean specific angular momentum of a shell before the shell reaches $\rcirc$.  Using snapshots from collapse Run 1, we computed the timescale for angular momentum redistribution for individual shells of material outside of the sink due to torques from the rest of the gas that is in supersonic infall.  At $t = 4578 t_0$, the material is supersonic out to $r \approx 4.5 r_0$, so we compute the vector torque on a shell with $0.1 \le r/r_0 \le 0.2$ due to the gas with $0.08 \le r/r_0 \le 4$.    We do similar calculations for different shell radii, different regions of the background gas, and for different snapshots ($t=4570$ and $4575$ $t_0$). In all cases, the timescale for angular momentum redistribution is a few thousand times the free-fall time of the shell.  So gravitational torques cannot alter the angular momentum vector of the infalling gas before centrifugal pressure becomes important at $\sim \rcirc$.  

In additional to centrifugal pressure, magnetic fields can play an important role as the material continues to infall.  RSGs may have $\sim 1$G magnetic fields at their surface  \citep{2010A&A...516L...2A,2013LNP...857..231P,2017A&A...603A.129T}.  Assuming flux freezing holds, so that $B\propto r^{-2}$, then the magnetic pressure is equal to the ram pressure of the infalling matter (assumed to be in free-fall) at a radius of 
\beq
r_B \sim 0.09 \bigg(\frac{B_0}{1 {\rm G}}\bigg)^{\frac{4}{3}}\bigg(\frac{r_{\rm ph}}{1000\rsun}\bigg)^{\frac{8}{3}}\bigg(\frac{10\msun/{\rm yr}}{\dot{m}}\bigg)^{\frac{2}{3}}\bigg(\frac{10\msun}{M_\bullet}\bigg)^{\frac{1}{3}} \rsun
\eeq
where $B_0$ is the field strength at $r_{\rm ph}$.  For our $18\msun$ RSG and for the range of accretion rates in Fig.~\ref{fig:accrates}, $r_B \sim 0.01 - 0.07$ $\rsun$ which is roughly $500 \risco \sim 0.05\rsun$, comparable to $\rcirc$ realized in our models.  Magnetic fields may thus become dynamically important during the infall.  We note that the field in the interior of the star may be much larger than 1$G$. For example, if $\rho v_{\rm turb}^2 \sim B^2 / 8\pi$, then in the outer $500 \rsun$ of our \mesa RSG model, $B \sim 200 - 800$ $G$, meaning that magnetic fields could be yet more important during collapse.    

\section{Summary and Conclusions}
\label{sec:conclusion}
A fraction of RSGs and YSGs may end their lives in failed supernovae (FSNe), in which core collapse does not lead to a successful, neutrino-powered SN explosion.  Even if weak shocks are launched by radiation of neutrino energy prior to the NS collapsing to a BH, much of the hydrogen envelope remains bound and will fall into the newly-formed BH \citep{2021ApJ...911....6I}; see Fig.~\ref{fig:initial_profiles} panel (d) and associated discussion. Previous work by \citet{2014MNRAS.439.4011G,2016ApJ...827...40G} and \citet{Quataert2019} has shown that, even when the star has zero net angular momentum,  the random velocity field in the convective hydrogen envelope gives rise to finite specific angular momentum (at each radius) that is larger than that associated with the ISCO of the BH. This suggests that accretion disks generically form during the infall of the hydrogen envelope in FSN. 

We perform two sets of 3D hydrodynamical simulations to study the random angular momentum associated with convective flows in the context of FSN.  We first simulate convection in polytropic models that are applicable to the convective envelopes of RSGs and YSGs. Our convection simulations provide the initial conditions for a set of collapse simulations, which follow the infall of the convective material after we introduce a low-pressure sink at the origin that mimics core collapse.  Our convection simulations extend the work of \citet{Quataert2019} by simulating a significant fraction of the hydrogen envelope and in the spherical geometry of a star.  We confirm their finding that the specific angular momentum associated with convective flows is larger than the specific angular momentum of the ISCO of a $10 \msun$ BH, although we find a different scaling with radius than inferred from their results, described in more detail below.  We further show with our collapse simulations that the specific angular momentum of the convective flows is largely conserved during the infall of the material at radii larger than the circularization radius of the gas.  The direction of the specific angular momentum vector is slowly varying over the timescales relevant to the region where the material will circularize.  All of this implies that, during the collapse of a supergiant in a FSN, the random angular momentum of the convective hydrogen envelope is likely to lead to the formation of centrifugally-supported gas at  small radii that could drive outflows and generate an observable transient.  The following summarizes our main results:

\begin{itemize}
	\item Convective flows in supergiant envelopes give rise to finite angular momentum at each radius even when the total angular momentum of the envelope is zero (Fig.~\ref{fig:cumulative_J}).  
	\item For polytropic models with convective Mach numbers of $0.05 < \mconv < 0.25$, consistent with the hydrogen envelopes of RSGs and YSGs (Fig.~\ref{fig:brunt_mturb_mesa}), the convective flows give rise to specific angular momenta of $\jrand \approx (4 - 28) \times 10^{17}$ cm$^2$ s$^{-1}$, where we have used eq.~\eqref{eq:j0_star} to scale our results to a RSG with photosphere radius $840 \rsun$ and an assumed BH mass of $6\msun$. These values correspond to circularization radii relative to the BH ISCO of 10 $\lesssim \rcirc/\risco \lesssim$ 1500 (Fig. \ref{fig:CONV_pts_rcirc}).  
	\item At the largest radii in the convection zone in our simulation, $\jrand$ agrees with the local simulations and analytical scaling (given in our eq.~\ref{eq:jrand}) of \citet{Quataert2019}.  However, we find a different scaling with $r$ than suggested by this local estimate. In our simulations, the convective flows roughly conserve $\jrand$ down to small radii. This results in our simulated $\jrand$ being larger than the estimate of eq.~\eqref{eq:jrand} over a large fraction of the envelope (Fig.~\ref{fig:CONV_fig_janalytical}). This is an important result because the outer radii of the convective zone are more likely to be unbound by weak shocks generated during collapse. Our simulations imply larger $\jrand$ and a larger likelihood of circularization at precisely the radii that are more likely to remain bound and accrete onto the BH.
	\item The specific angular momentum of the convective envelope is not significantly  altered during collapse.  Instead, mean profiles of the specific angular momentum vector $\vjrand$ (in magnitude and direction) from the convection simulations can be used to predict the specific angular momentum vector of the accreted material measured in the collapse simulations ($\vjacc \equiv \dot{\boldsymbol{J}}/\dot{m}$) by assuming ballistic infall and accounting for the finite sound-travel and infall time for each shell (Fig.~\ref{fig:collapse_predict}). 
		\item The specific angular momentum of the accreted material depends on the state of the envelope at the start of collapse.  However, even when collapse begins at one of the lowest angular momentum states (Run 4 with collapse start time $t_s = 4800 t_0$), $\rcirc > \risco$. Indeed, because the mean profiles from the convection simulation can be used to accurately predict the accretion rates of mass and angular momentum flowing to small radii in our collapse calculations, we can conclude from Fig.~\ref{fig:CONV_pts_rcirc} that $\rcirc$ of the accreted hydrogen envelope is always greater than $\approx 10$ $\risco$ for a $6 \msun$ BH.  
	\item The direction of $\vjacc$ is slowly varying (Fig.~\ref{fig:accrates}) on the timescales that we are sensitive to. These timescales, which are set by convective flow times in the envelope, are very long relative to the dynamical time at the circularization radius of the material.  Because the direction of the mean specific angular momenta of the convective envelope are not significantly modified during the infall, it is likely that the direction of $\vjacc$ remains coherent down to $\rcirc$ scales, allowing for the presence of coherent rotation at small radii.  This supports the idea that accretion at small radii will generate outflows that can carry energy to large radii. The resulting accretion dynamics at small radii are likely to be very different from standard accretion simulations because there is a very large dispersion in accreted $\vjacc$ in each spherical shell (Fig.~\ref{fig:histogram}).   As material reaches $\rcirc$, the distribution of $\rcirc$ within each shell will drive interactions between individual parcels, which will generate mixing, torques, and shock-heating as individual parcels are deflected from spherical infall.  The outcome of these flows requires further study.
	\item Accretion of the entire envelope by the BH leads to finite BH spin, even though the total angular momentum of the envelope is zero. The BH spin is $\sim 0.5$ if most of the envelope accretes, but less if outflows at small radii remove significant mass (Fig.~\ref{fig:spin}).  The finite BH spin occurs because $\jacc > \jisco$, the specific angular momentum of the ISCO, in nearly all of the infalling shells. The BH cannot accrete more than $\jisco$ from each shell, so infalling shells with identical but oppositely-oriented specific angular momenta can never cancel.
\end{itemize}

Our results imply that much of the hydrogen envelope that remains bound in FSN is likely to become rotationally-supported during collapse rather than fall directly into the BH.  This should apply both to the initial collapse and to any subsequent fallback.  Sufficient mass is available to accrete onto the newly-formed BH that it could power a month $-$ year long transient with energies up to $\sim 10^{52}$ ergs (Sec.~\ref{sec:implications}).  This could be the origin of unusually long-duration gamma-ray transients such as Swift 1644+57 \citep{2011GCN.11847....1B} and  Swift J2058.4+0516 \citep{2012ApJ...753...77C} and unusually energetic Type II supernovae \citep[e.g., OGLE-2014-SN-073][]{2018MNRAS.475L..11M}. The key uncertainty is that the binding energy of the hydrogen envelope is only $\sim 10^{48}$ ergs and so a small amount of accretion may unbind the envelope before sufficient mass accretes to power an energetic transient.   The FSN candidate discovered by \citet{2015MNRAS.450.3289G} showed a modest brightening up to $\sim 10^6 L_\odot$ for a year before fading to a luminosity significantly below that of the progenitor star \citep{2017MNRAS.468.4968A}.  The initial brightening prior to the star `disappearing' is consistent with a $\sim$ few $\times 10^{48}$ erg shock unbinding the hydrogen envelope \citep[e.g.,][]{2013ApJ...769..109L} but not with a more energetic transient.  Given the highly super-Eddington accretion rates and significant rotational support implied by our results, it is also unclear how rapidly the star is expected to fade away after core collapse, or whether it might remain a near-Eddington source for an extended period of time powered by super-Eddington fallback accretion.   Understanding this better is important for understanding the FSN candidate discovered by \citet{2015MNRAS.450.3289G} and for identifying and interpreting additional candidates in ongoing and future transient surveys.  We expect a large diversity in the observational manifestation of RSG collapse to a BH given the stochastic nature of the angular momentum in the convective envelope.

There are several key aspects of supergiant convection and FSN that are not captured in our study.  The random convective angular momentum in simulations is a more subtle effect than what is usually studied in simulations of convection. Increasing resolution somewhat increases the time-averaged $\jrand$ but it is not known what resolution is required to fully converge.  Additionally, we have neglected the phase of the FSN when a low-energy shock wave sweeps through the convective hydrogen envelope. The outgoing shock could restructure the convective material and will unbind some fraction of the hydrogen envelope.  Simulations that follow the propagation of the weak shock through the convective envelope, the subsequent expansion of the still-bound material to apocenter, and the ultimate infall and accretion onto the BH are needed to self-consistently model FSN.   Finally, future work will extend our simulations by following the inflow to radii where rotational support becomes dynamically important. Such studies are critical for understanding the extent to which the random angular momentum associated with convection leads to flows at small radii that can drive energy and momentum to large radii and may be capable of producing a transient.

\section*{Acknowledgements}
We thank the anonymous referee, Lars Bildsten, Eric R. Coughlin, Paul C. Duffell, Drummond Fielding, Jared A. Goldberg, Yan-Fei Jiang, Wenbin Lu, Morgan MacLeod, Philipp Moesta, Enrico Ramirez-Ruiz, Sean Ressler, and Stephen Ro for useful conversations and suggestions. A.A. gratefully acknowledges support from the University of California, Berkeley Fellowship, the Cranor Fellowship at U.C. Berkeley, and the National Science Foundation Graduate Research Fellowship under Grant No. DGE 1752814. E.Q. was supported in part by a Simons Investigator award from the Simons Foundation. This work benefited from workshops supported by the Gordon and Betty Moore Foundation through Grant GBMF5076. 

The simulations presented in this article were performed on computational resources managed and supported by Princeton Research Computing, a consortium of groups including the Princeton Institute for Computational Science and Engineering (PICSciE) and the Office of Information Technology's High Performance Computing Center and Visualization Laboratory at Princeton University.  Preliminary studies used the Savio computational cluster resource provided by the Berkeley Research Computing program at the University of California, Berkeley (supported by the UC Berkeley Chancellor, Vice Chancellor for Research, and Chief Information Officer).  This project was made possible by the following publicly available software: \texttt{Astropy}\footnote{http://www.astropy.org} \citep{astropy:2013, astropy:2018}, \ath \citep{2020ApJS..249....4S}, \texttt{matplotlib} \citep{Hunter:2007}, \mesa \citep{2011ApJS..192....3P,2013ApJS..208....4P,2015ApJS..220...15P,2018ApJS..234...34P,2019ApJS..243...10P},  \texttt{MESA SDK} \citep{richard_townsend_2020_3706650},  \texttt{NumPy} \citep{harris2020array}, \texttt{yt}\footnote{https://github.com/yt-project/yt} \citep{2011ApJS..192....9T}.

%%%%%%%%%%%%%%%%%%%%%%%%%%%%%%%%%%%%%%%%%%%%%%%%%%
\section*{Data Availability}
Data related to the results of simulations in this article will be shared on reasonable request to the corresponding author via e-mail.



\bibliographystyle{mnras}
\bibliography{references} % if your bibtex file is called example.bib

\begin{thebibliography}{}
\makeatletter
\relax
\def\mn@urlcharsother{\let\do\@makeother \do\$\do\&\do\#\do\^\do\_\do\%\do\~}
\def\mn@doi{\begingroup\mn@urlcharsother \@ifnextchar [ {\mn@doi@}
  {\mn@doi@[]}}
\def\mn@doi@[#1]#2{\def\@tempa{#1}\ifx\@tempa\@empty \href
  {http://dx.doi.org/#2} {doi:#2}\else \href {http://dx.doi.org/#2} {#1}\fi
  \endgroup}
\def\mn@eprint#1#2{\mn@eprint@#1:#2::\@nil}
\def\mn@eprint@arXiv#1{\href {http://arxiv.org/abs/#1} {{\tt arXiv:#1}}}
\def\mn@eprint@dblp#1{\href {http://dblp.uni-trier.de/rec/bibtex/#1.xml}
  {dblp:#1}}
\def\mn@eprint@#1:#2:#3:#4\@nil{\def\@tempa {#1}\def\@tempb {#2}\def\@tempc
  {#3}\ifx \@tempc \@empty \let \@tempc \@tempb \let \@tempb \@tempa \fi \ifx
  \@tempb \@empty \def\@tempb {arXiv}\fi \@ifundefined
  {mn@eprint@\@tempb}{\@tempb:\@tempc}{\expandafter \expandafter \csname
  mn@eprint@\@tempb\endcsname \expandafter{\@tempc}}}

\bibitem[\protect\citeauthoryear{{Abramowicz}, {Czerny}, {Lasota}  \&
  {Szuszkiewicz}}{{Abramowicz} et~al.}{1988}]{1988ApJ...332..646A}
{Abramowicz} M.~A.,  {Czerny} B.,  {Lasota} J.~P.,   {Szuszkiewicz} E.,  1988,
  \mn@doi [\apj] {10.1086/166683}, \href
  {https://ui.adsabs.harvard.edu/abs/1988ApJ...332..646A} {332, 646}

\bibitem[\protect\citeauthoryear{{Adams}, {Kochanek}, {Gerke}, {Stanek}  \&
  {Dai}}{{Adams} et~al.}{2017}]{2017MNRAS.468.4968A}
{Adams} S.~M.,  {Kochanek} C.~S.,  {Gerke} J.~R.,  {Stanek} K.~Z.,   {Dai} X.,
  2017, \mn@doi [\mnras] {10.1093/mnras/stx816}, \href
  {https://ui.adsabs.harvard.edu/abs/2017MNRAS.468.4968A} {468, 4968}

\bibitem[\protect\citeauthoryear{{Astropy Collaboration} et~al.,}{{Astropy
  Collaboration} et~al.}{2013}]{astropy:2013}
{Astropy Collaboration} et~al., 2013, \mn@doi [\aap]
  {10.1051/0004-6361/201322068}, \href
  {http://adsabs.harvard.edu/abs/2013A%26A...558A..33A} {558, A33}

\bibitem[\protect\citeauthoryear{{Astropy Collaboration} et~al.,}{{Astropy
  Collaboration} et~al.}{2018}]{astropy:2018}
{Astropy Collaboration} et~al., 2018, \mn@doi [aj] {10.3847/1538-3881/aabc4f},
  \href {https://ui.adsabs.harvard.edu/abs/2018AJ....156..123A} {156, 123}

\bibitem[\protect\citeauthoryear{{Auri{\`e}re}, {Donati},
  {Konstantinova-Antova}, {Perrin}, {Petit}  \& {Roudier}}{{Auri{\`e}re}
  et~al.}{2010}]{2010A&A...516L...2A}
{Auri{\`e}re} M.,  {Donati} J.~F.,  {Konstantinova-Antova} R.,  {Perrin} G.,
  {Petit} P.,   {Roudier} T.,  2010, \mn@doi [\aap]
  {10.1051/0004-6361/201014925}, \href
  {https://ui.adsabs.harvard.edu/abs/2010A&A...516L...2A} {516, L2}

\bibitem[\protect\citeauthoryear{{Basinger}, {Kochanek}, {Adams}, {Dai}  \&
  {Stanek}}{{Basinger} et~al.}{2020}]{2020arXiv200715658B}
{Basinger} C.~M.,  {Kochanek} C.~S.,  {Adams} S.~M.,  {Dai} X.,   {Stanek}
  K.~Z.,  2020, arXiv e-prints, \href
  {https://ui.adsabs.harvard.edu/abs/2020arXiv200715658B} {p. arXiv:2007.15658}

\bibitem[\protect\citeauthoryear{{Batta} \& {Ramirez-Ruiz}}{{Batta} \&
  {Ramirez-Ruiz}}{2019}]{2019arXiv190404835B}
{Batta} A.,  {Ramirez-Ruiz} E.,  2019, arXiv e-prints, \href
  {https://ui.adsabs.harvard.edu/abs/2019arXiv190404835B} {p. arXiv:1904.04835}

\bibitem[\protect\citeauthoryear{{Begelman} \& {Meier}}{{Begelman} \&
  {Meier}}{1982}]{1982ApJ...253..873B}
{Begelman} M.~C.,  {Meier} D.~L.,  1982, \mn@doi [\apj] {10.1086/159688}, \href
  {https://ui.adsabs.harvard.edu/abs/1982ApJ...253..873B} {253, 873}

\bibitem[\protect\citeauthoryear{{Beloborodov}}{{Beloborodov}}{2008}]{2008AIPC.1054...51B}
{Beloborodov} A.~M.,  2008, in {Axelsson} M.,  ed.,  American Institute of
  Physics Conference Series Vol. 1054, Cool Discs, Hot Flows: The Varying Faces
  of Accreting Compact Objects. pp 51--70 (\mn@eprint {arXiv} {0810.2690}),
  \mn@doi{10.1063/1.3002509}

\bibitem[\protect\citeauthoryear{{Bloom}, {Butler}, {Cenko}  \&
  {Perley}}{{Bloom} et~al.}{2011}]{2011GCN.11847....1B}
{Bloom} J.~S.,  {Butler} N.~R.,  {Cenko} S.~B.,   {Perley} D.~A.,  2011, GRB
  Coordinates Network, \href
  {https://ui.adsabs.harvard.edu/abs/2011GCN.11847....1B} {11847, 1}

\bibitem[\protect\citeauthoryear{{Burrows}, {Radice}, {Vartanyan}, {Nagakura},
  {Skinner}  \& {Dolence}}{{Burrows} et~al.}{2020}]{2020MNRAS.491.2715B}
{Burrows} A.,  {Radice} D.,  {Vartanyan} D.,  {Nagakura} H.,  {Skinner} M.~A.,
   {Dolence} J.~C.,  2020, \mn@doi [\mnras] {10.1093/mnras/stz3223}, \href
  {https://ui.adsabs.harvard.edu/abs/2020MNRAS.491.2715B} {491, 2715}

\bibitem[\protect\citeauthoryear{{Cenko} et~al.,}{{Cenko}
  et~al.}{2012}]{2012ApJ...753...77C}
{Cenko} S.~B.,  et~al., 2012, \mn@doi [\apj] {10.1088/0004-637X/753/1/77},
  \href {https://ui.adsabs.harvard.edu/abs/2012ApJ...753...77C} {753, 77}

\bibitem[\protect\citeauthoryear{{Chiavassa}, {Plez}, {Josselin}  \&
  {Freytag}}{{Chiavassa} et~al.}{2009}]{2009A&A...506.1351C}
{Chiavassa} A.,  {Plez} B.,  {Josselin} E.,   {Freytag} B.,  2009, \mn@doi
  [\aap] {10.1051/0004-6361/200911780}, \href
  {https://ui.adsabs.harvard.edu/abs/2009A&A...506.1351C} {506, 1351}

\bibitem[\protect\citeauthoryear{{Chiavassa}, {Haubois}, {Young}, {Plez},
  {Josselin}, {Perrin}  \& {Freytag}}{{Chiavassa}
  et~al.}{2010}]{2010A&A...515A..12C}
{Chiavassa} A.,  {Haubois} X.,  {Young} J.~S.,  {Plez} B.,  {Josselin} E.,
  {Perrin} G.,   {Freytag} B.,  2010, \mn@doi [\aap]
  {10.1051/0004-6361/200913907}, \href
  {https://ui.adsabs.harvard.edu/abs/2010A&A...515A..12C} {515, A12}

\bibitem[\protect\citeauthoryear{{Couch}, {Chatzopoulos}, {Arnett}  \&
  {Timmes}}{{Couch} et~al.}{2015}]{2015ApJ...808L..21C}
{Couch} S.~M.,  {Chatzopoulos} E.,  {Arnett} W.~D.,   {Timmes} F.~X.,  2015,
  \mn@doi [\apjl] {10.1088/2041-8205/808/1/L21}, \href
  {https://ui.adsabs.harvard.edu/abs/2015ApJ...808L..21C} {808, L21}

\bibitem[\protect\citeauthoryear{{Coughlin}, {Quataert}, {Fern{\'a}ndez}  \&
  {Kasen}}{{Coughlin} et~al.}{2018a}]{2018MNRAS.477.1225C}
{Coughlin} E.~R.,  {Quataert} E.,  {Fern{\'a}ndez} R.,   {Kasen} D.,  2018a,
  \mn@doi [\mnras] {10.1093/mnras/sty667}, \href
  {https://ui.adsabs.harvard.edu/abs/2018MNRAS.477.1225C} {477, 1225}

\bibitem[\protect\citeauthoryear{{Coughlin}, {Quataert}  \& {Ro}}{{Coughlin}
  et~al.}{2018b}]{2018ApJ...863..158C}
{Coughlin} E.~R.,  {Quataert} E.,   {Ro} S.,  2018b, \mn@doi [\apj]
  {10.3847/1538-4357/aad198}, \href
  {https://ui.adsabs.harvard.edu/abs/2018ApJ...863..158C} {863, 158}

\bibitem[\protect\citeauthoryear{{Coughlin}, {Ro}  \& {Quataert}}{{Coughlin}
  et~al.}{2019}]{2019ApJ...874...58C}
{Coughlin} E.~R.,  {Ro} S.,   {Quataert} E.,  2019, \mn@doi [\apj]
  {10.3847/1538-4357/ab09ec}, \href
  {https://ui.adsabs.harvard.edu/abs/2019ApJ...874...58C} {874, 58}

\bibitem[\protect\citeauthoryear{{Dupree} et~al.,}{{Dupree}
  et~al.}{2020}]{2020ApJ...899...68D}
{Dupree} A.~K.,  et~al., 2020, \mn@doi [\apj] {10.3847/1538-4357/aba516}, \href
  {https://ui.adsabs.harvard.edu/abs/2020ApJ...899...68D} {899, 68}

\bibitem[\protect\citeauthoryear{{Ertl}, {Janka}, {Woosley}, {Sukhbold}  \&
  {Ugliano}}{{Ertl} et~al.}{2016}]{2016ApJ...818..124E}
{Ertl} T.,  {Janka} H.~T.,  {Woosley} S.~E.,  {Sukhbold} T.,   {Ugliano} M.,
  2016, \mn@doi [\apj] {10.3847/0004-637X/818/2/124}, \href
  {https://ui.adsabs.harvard.edu/abs/2016ApJ...818..124E} {818, 124}

\bibitem[\protect\citeauthoryear{{Fern{\'a}ndez}, {Quataert}, {Kashiyama}  \&
  {Coughlin}}{{Fern{\'a}ndez} et~al.}{2018}]{2018MNRAS.476.2366F}
{Fern{\'a}ndez} R.,  {Quataert} E.,  {Kashiyama} K.,   {Coughlin} E.~R.,  2018,
  \mn@doi [\mnras] {10.1093/mnras/sty306}, \href
  {https://ui.adsabs.harvard.edu/abs/2018MNRAS.476.2366F} {476, 2366}

\bibitem[\protect\citeauthoryear{{Fields} \& {Couch}}{{Fields} \&
  {Couch}}{2020}]{2020ApJ...901...33F}
{Fields} C.~E.,  {Couch} S.~M.,  2020, \mn@doi [\apj]
  {10.3847/1538-4357/abada7}, \href
  {https://ui.adsabs.harvard.edu/abs/2020ApJ...901...33F} {901, 33}

\bibitem[\protect\citeauthoryear{{Gerke}, {Kochanek}  \& {Stanek}}{{Gerke}
  et~al.}{2015}]{2015MNRAS.450.3289G}
{Gerke} J.~R.,  {Kochanek} C.~S.,   {Stanek} K.~Z.,  2015, \mn@doi [\mnras]
  {10.1093/mnras/stv776}, \href
  {https://ui.adsabs.harvard.edu/abs/2015MNRAS.450.3289G} {450, 3289}

\bibitem[\protect\citeauthoryear{{Gilkis} \& {Soker}}{{Gilkis} \&
  {Soker}}{2014}]{2014MNRAS.439.4011G}
{Gilkis} A.,  {Soker} N.,  2014, \mn@doi [\mnras] {10.1093/mnras/stu257}, \href
  {https://ui.adsabs.harvard.edu/abs/2014MNRAS.439.4011G} {439, 4011}

\bibitem[\protect\citeauthoryear{{Gilkis} \& {Soker}}{{Gilkis} \&
  {Soker}}{2016}]{2016ApJ...827...40G}
{Gilkis} A.,  {Soker} N.,  2016, \mn@doi [\apj] {10.3847/0004-637X/827/1/40},
  \href {https://ui.adsabs.harvard.edu/abs/2016ApJ...827...40G} {827, 40}

\bibitem[\protect\citeauthoryear{{Goldberg}, {Jiang}  \& {Bildsten}}{{Goldberg}
  et~al.}{2021}]{2021arXiv211003261G}
{Goldberg} J.~A.,  {Jiang} Y.-F.,   {Bildsten} L.,  2021, arXiv e-prints, \href
  {https://ui.adsabs.harvard.edu/abs/2021arXiv211003261G} {p. arXiv:2110.03261}

\bibitem[\protect\citeauthoryear{Harris et~al.,}{Harris
  et~al.}{2020}]{harris2020array}
Harris C.~R.,  et~al., 2020, \mn@doi [Nature] {10.1038/s41586-020-2649-2}, 585,
  357

\bibitem[\protect\citeauthoryear{{Heger}, {Woosley}  \& {Spruit}}{{Heger}
  et~al.}{2005}]{2005ApJ...626..350H}
{Heger} A.,  {Woosley} S.~E.,   {Spruit} H.~C.,  2005, \mn@doi [\apj]
  {10.1086/429868}, \href
  {https://ui.adsabs.harvard.edu/abs/2005ApJ...626..350H} {626, 350}

\bibitem[\protect\citeauthoryear{Hunter}{Hunter}{2007}]{Hunter:2007}
Hunter J.~D.,  2007, \mn@doi [Computing in Science \& Engineering]
  {10.1109/MCSE.2007.55}, 9, 90

\bibitem[\protect\citeauthoryear{{Ivanov} \& {Fern{\'a}ndez}}{{Ivanov} \&
  {Fern{\'a}ndez}}{2021}]{2021ApJ...911....6I}
{Ivanov} M.,  {Fern{\'a}ndez} R.,  2021, \mn@doi [\apj]
  {10.3847/1538-4357/abe59e}, \href
  {https://ui.adsabs.harvard.edu/abs/2021ApJ...911....6I} {911, 6}

\bibitem[\protect\citeauthoryear{{Jiang}, {Stone}  \& {Davis}}{{Jiang}
  et~al.}{2019}]{2019ApJ...880...67J}
{Jiang} Y.-F.,  {Stone} J.~M.,   {Davis} S.~W.,  2019, \mn@doi [\apj]
  {10.3847/1538-4357/ab29ff}, \href
  {https://ui.adsabs.harvard.edu/abs/2019ApJ...880...67J} {880, 67}

\bibitem[\protect\citeauthoryear{{Lindner}, {Milosavljevi{\'c}}, {Couch}  \&
  {Kumar}}{{Lindner} et~al.}{2010}]{2010ApJ...713..800L}
{Lindner} C.~C.,  {Milosavljevi{\'c}} M.,  {Couch} S.~M.,   {Kumar} P.,  2010,
  \mn@doi [\apj] {10.1088/0004-637X/713/2/800}, \href
  {https://ui.adsabs.harvard.edu/abs/2010ApJ...713..800L} {713, 800}

\bibitem[\protect\citeauthoryear{{Linial}, {Fuller}  \& {Sari}}{{Linial}
  et~al.}{2021}]{2021MNRAS.501.4266L}
{Linial} I.,  {Fuller} J.,   {Sari} R.,  2021, \mn@doi [\mnras]
  {10.1093/mnras/staa3969}, \href
  {https://ui.adsabs.harvard.edu/abs/2021MNRAS.501.4266L} {501, 4266}

\bibitem[\protect\citeauthoryear{{Lovegrove} \& {Woosley}}{{Lovegrove} \&
  {Woosley}}{2013}]{2013ApJ...769..109L}
{Lovegrove} E.,  {Woosley} S.~E.,  2013, \mn@doi [\apj]
  {10.1088/0004-637X/769/2/109}, \href
  {https://ui.adsabs.harvard.edu/abs/2013ApJ...769..109L} {769, 109}

\bibitem[\protect\citeauthoryear{{Lovegrove}, {Woosley}  \&
  {Zhang}}{{Lovegrove} et~al.}{2017}]{2017ApJ...845..103L}
{Lovegrove} E.,  {Woosley} S.~E.,   {Zhang} W.,  2017, \mn@doi [\apj]
  {10.3847/1538-4357/aa7b7d}, \href
  {https://ui.adsabs.harvard.edu/abs/2017ApJ...845..103L} {845, 103}

\bibitem[\protect\citeauthoryear{{MacFadyen} \& {Woosley}}{{MacFadyen} \&
  {Woosley}}{1999}]{1999ApJ...524..262M}
{MacFadyen} A.~I.,  {Woosley} S.~E.,  1999, \mn@doi [\apj] {10.1086/307790},
  \href {https://ui.adsabs.harvard.edu/abs/1999ApJ...524..262M} {524, 262}

\bibitem[\protect\citeauthoryear{{Moriya}, {Terreran}  \& {Blinnikov}}{{Moriya}
  et~al.}{2018}]{2018MNRAS.475L..11M}
{Moriya} T.~J.,  {Terreran} G.,   {Blinnikov} S.~I.,  2018, \mn@doi [\mnras]
  {10.1093/mnrasl/slx200}, \href
  {https://ui.adsabs.harvard.edu/abs/2018MNRAS.475L..11M} {475, L11}

\bibitem[\protect\citeauthoryear{{Murguia-Berthier}, {Batta}, {Janiuk},
  {Ramirez-Ruiz}, {Mandel}, {Noble}  \& {Everson}}{{Murguia-Berthier}
  et~al.}{2020}]{2020ApJ...901L..24M}
{Murguia-Berthier} A.,  {Batta} A.,  {Janiuk} A.,  {Ramirez-Ruiz} E.,  {Mandel}
  I.,  {Noble} S.~C.,   {Everson} R.~W.,  2020, \mn@doi [\apjl]
  {10.3847/2041-8213/abb818}, \href
  {https://ui.adsabs.harvard.edu/abs/2020ApJ...901L..24M} {901, L24}

\bibitem[\protect\citeauthoryear{{Nadyozhin}}{{Nadyozhin}}{1980}]{1980Ap&SS..69..115N}
{Nadyozhin} D.~K.,  1980, \mn@doi [\apss] {10.1007/BF00638971}, \href
  {https://ui.adsabs.harvard.edu/abs/1980Ap&SS..69..115N} {69, 115}

\bibitem[\protect\citeauthoryear{{Narayan} \& {Yi}}{{Narayan} \&
  {Yi}}{1994}]{1994ApJ...428L..13N}
{Narayan} R.,  {Yi} I.,  1994, \mn@doi [\apjl] {10.1086/187381}, \href
  {https://ui.adsabs.harvard.edu/abs/1994ApJ...428L..13N} {428, L13}

\bibitem[\protect\citeauthoryear{{Neustadt}, {Kochanek}, {Stanek}, {Basinger},
  {Jayasinghe}, {Garling}, {Adams}  \& {Gerke}}{{Neustadt}
  et~al.}{2021}]{2021arXiv210403318N}
{Neustadt} J.~M.~M.,  {Kochanek} C.~S.,  {Stanek} K.~Z.,  {Basinger} C.~M.,
  {Jayasinghe} T.,  {Garling} C.~T.,  {Adams} S.~M.,   {Gerke} J.,  2021, arXiv
  e-prints, \href {https://ui.adsabs.harvard.edu/abs/2021arXiv210403318N} {p.
  arXiv:2104.03318}

\bibitem[\protect\citeauthoryear{{O'Connor} \& {Ott}}{{O'Connor} \&
  {Ott}}{2011}]{2011ApJ...730...70O}
{O'Connor} E.,  {Ott} C.~D.,  2011, \mn@doi [\apj]
  {10.1088/0004-637X/730/2/70}, \href
  {https://ui.adsabs.harvard.edu/abs/2011ApJ...730...70O} {730, 70}

\bibitem[\protect\citeauthoryear{{Papish} \& {Soker}}{{Papish} \&
  {Soker}}{2011}]{2011MNRAS.416.1697P}
{Papish} O.,  {Soker} N.,  2011, \mn@doi [\mnras]
  {10.1111/j.1365-2966.2011.18671.x}, \href
  {https://ui.adsabs.harvard.edu/abs/2011MNRAS.416.1697P} {416, 1697}

\bibitem[\protect\citeauthoryear{{Parrish}, {Stone}  \& {Lemaster}}{{Parrish}
  et~al.}{2008}]{2008ApJ...688..905P}
{Parrish} I.~J.,  {Stone} J.~M.,   {Lemaster} N.,  2008, \mn@doi [\apj]
  {10.1086/592380}, \href
  {https://ui.adsabs.harvard.edu/abs/2008ApJ...688..905P} {688, 905}

\bibitem[\protect\citeauthoryear{{Paxton}, {Bildsten}, {Dotter}, {Herwig},
  {Lesaffre}  \& {Timmes}}{{Paxton} et~al.}{2011}]{2011ApJS..192....3P}
{Paxton} B.,  {Bildsten} L.,  {Dotter} A.,  {Herwig} F.,  {Lesaffre} P.,
  {Timmes} F.,  2011, \mn@doi [\apjs] {10.1088/0067-0049/192/1/3}, \href
  {https://ui.adsabs.harvard.edu/abs/2011ApJS..192....3P} {192, 3}

\bibitem[\protect\citeauthoryear{{Paxton} et~al.,}{{Paxton}
  et~al.}{2013}]{2013ApJS..208....4P}
{Paxton} B.,  et~al., 2013, \mn@doi [\apjs] {10.1088/0067-0049/208/1/4}, \href
  {https://ui.adsabs.harvard.edu/abs/2013ApJS..208....4P} {208, 4}

\bibitem[\protect\citeauthoryear{{Paxton} et~al.,}{{Paxton}
  et~al.}{2015}]{2015ApJS..220...15P}
{Paxton} B.,  et~al., 2015, \mn@doi [\apjs] {10.1088/0067-0049/220/1/15}, \href
  {https://ui.adsabs.harvard.edu/abs/2015ApJS..220...15P} {220, 15}

\bibitem[\protect\citeauthoryear{{Paxton} et~al.,}{{Paxton}
  et~al.}{2018}]{2018ApJS..234...34P}
{Paxton} B.,  et~al., 2018, \mn@doi [\apjs] {10.3847/1538-4365/aaa5a8}, \href
  {https://ui.adsabs.harvard.edu/abs/2018ApJS..234...34P} {234, 34}

\bibitem[\protect\citeauthoryear{{Paxton} et~al.,}{{Paxton}
  et~al.}{2019}]{2019ApJS..243...10P}
{Paxton} B.,  et~al., 2019, \mn@doi [\apjs] {10.3847/1538-4365/ab2241}, \href
  {https://ui.adsabs.harvard.edu/abs/2019ApJS..243...10P} {243, 10}

\bibitem[\protect\citeauthoryear{{Perna}, {Lazzati}  \& {Cantiello}}{{Perna}
  et~al.}{2018}]{2018ApJ...859...48P}
{Perna} R.,  {Lazzati} D.,   {Cantiello} M.,  2018, \mn@doi [\apj]
  {10.3847/1538-4357/aabcc1}, \href
  {https://ui.adsabs.harvard.edu/abs/2018ApJ...859...48P} {859, 48}

\bibitem[\protect\citeauthoryear{{Petit}, {Auri{\`e}re},
  {Konstantinova-Antova}, {Morgenthaler}, {Perrin}, {Roudier}  \&
  {Donati}}{{Petit} et~al.}{2013}]{2013LNP...857..231P}
{Petit} P.,  {Auri{\`e}re} M.,  {Konstantinova-Antova} R.,  {Morgenthaler} A.,
  {Perrin} G.,  {Roudier} T.,   {Donati} J.~F.,  2013, {Magnetic Fields and
  Convection in the Cool Supergiant Betelgeuse}.
p.~231, \mn@doi{10.1007/978-3-642-30648-8_9}

\bibitem[\protect\citeauthoryear{{Piro}}{{Piro}}{2013}]{2013ApJ...768L..14P}
{Piro} A.~L.,  2013, \mn@doi [\apjl] {10.1088/2041-8205/768/1/L14}, \href
  {https://ui.adsabs.harvard.edu/abs/2013ApJ...768L..14P} {768, L14}

\bibitem[\protect\citeauthoryear{{Powell} \& {M{\"u}ller}}{{Powell} \&
  {M{\"u}ller}}{2020}]{2020MNRAS.494.4665P}
{Powell} J.,  {M{\"u}ller} B.,  2020, \mn@doi [\mnras]
  {10.1093/mnras/staa1048}, \href
  {https://ui.adsabs.harvard.edu/abs/2020MNRAS.494.4665P} {494, 4665}

\bibitem[\protect\citeauthoryear{{Quataert} \& {Kasen}}{{Quataert} \&
  {Kasen}}{2012}]{2012MNRAS.419L...1Q}
{Quataert} E.,  {Kasen} D.,  2012, \mn@doi [\mnras]
  {10.1111/j.1745-3933.2011.01151.x}, \href
  {https://ui.adsabs.harvard.edu/abs/2012MNRAS.419L...1Q} {419, L1}

\bibitem[\protect\citeauthoryear{{Quataert}, {Lecoanet}  \&
  {Coughlin}}{{Quataert} et~al.}{2019}]{Quataert2019}
{Quataert} E.,  {Lecoanet} D.,   {Coughlin} E.~R.,  2019, \mn@doi [\mnras]
  {10.1093/mnrasl/slz031}, \href
  {https://ui.adsabs.harvard.edu/abs/2019MNRAS.485L..83Q} {485, L83}

\bibitem[\protect\citeauthoryear{{Ro}, {Coughlin}  \& {Quataert}}{{Ro}
  et~al.}{2019}]{2019ApJ...878..150R}
{Ro} S.,  {Coughlin} E.~R.,   {Quataert} E.,  2019, \mn@doi [\apj]
  {10.3847/1538-4357/ab1ea2}, \href
  {https://ui.adsabs.harvard.edu/abs/2019ApJ...878..150R} {878, 150}

\bibitem[\protect\citeauthoryear{{Soker}}{{Soker}}{2010}]{2010MNRAS.401.2793S}
{Soker} N.,  2010, \mn@doi [\mnras] {10.1111/j.1365-2966.2009.15862.x}, \href
  {https://ui.adsabs.harvard.edu/abs/2010MNRAS.401.2793S} {401, 2793}

\bibitem[\protect\citeauthoryear{{Stone}, {Gardiner}, {Teuben}, {Hawley}  \&
  {Simon}}{{Stone} et~al.}{2008}]{2008ApJS..178..137S}
{Stone} J.~M.,  {Gardiner} T.~A.,  {Teuben} P.,  {Hawley} J.~F.,   {Simon}
  J.~B.,  2008, \mn@doi [\apjs] {10.1086/588755}, \href
  {https://ui.adsabs.harvard.edu/abs/2008ApJS..178..137S} {178, 137}

\bibitem[\protect\citeauthoryear{{Stone}, {Tomida}, {White}  \&
  {Felker}}{{Stone} et~al.}{2020}]{2020ApJS..249....4S}
{Stone} J.~M.,  {Tomida} K.,  {White} C.~J.,   {Felker} K.~G.,  2020, \mn@doi
  [\apjs] {10.3847/1538-4365/ab929b}, \href
  {https://ui.adsabs.harvard.edu/abs/2020ApJS..249....4S} {249, 4}

\bibitem[\protect\citeauthoryear{{Sukhbold} \& {Adams}}{{Sukhbold} \&
  {Adams}}{2020}]{2020MNRAS.492.2578S}
{Sukhbold} T.,  {Adams} S.,  2020, \mn@doi [\mnras] {10.1093/mnras/staa059},
  \href {https://ui.adsabs.harvard.edu/abs/2020MNRAS.492.2578S} {492, 2578}

\bibitem[\protect\citeauthoryear{{Sukhbold}, {Ertl}, {Woosley}, {Brown}  \&
  {Janka}}{{Sukhbold} et~al.}{2016}]{2016ApJ...821...38S}
{Sukhbold} T.,  {Ertl} T.,  {Woosley} S.~E.,  {Brown} J.~M.,   {Janka} H.~T.,
  2016, \mn@doi [\apj] {10.3847/0004-637X/821/1/38}, \href
  {https://ui.adsabs.harvard.edu/abs/2016ApJ...821...38S} {821, 38}

\bibitem[\protect\citeauthoryear{{Sukhbold}, {Woosley}  \& {Heger}}{{Sukhbold}
  et~al.}{2018}]{2018ApJ...860...93S}
{Sukhbold} T.,  {Woosley} S.~E.,   {Heger} A.,  2018, \mn@doi [\apj]
  {10.3847/1538-4357/aac2da}, \href
  {https://ui.adsabs.harvard.edu/abs/2018ApJ...860...93S} {860, 93}

\bibitem[\protect\citeauthoryear{{Tessore}, {L{\`e}bre}, {Morin}, {Mathias},
  {Josselin}  \& {Auri{\`e}re}}{{Tessore} et~al.}{2017}]{2017A&A...603A.129T}
{Tessore} B.,  {L{\`e}bre} A.,  {Morin} J.,  {Mathias} P.,  {Josselin} E.,
  {Auri{\`e}re} M.,  2017, \mn@doi [\aap] {10.1051/0004-6361/201730473}, \href
  {https://ui.adsabs.harvard.edu/abs/2017A&A...603A.129T} {603, A129}

\bibitem[\protect\citeauthoryear{Townsend}{Townsend}{2020}]{richard_townsend_2020_3706650}
Townsend R.,  2020, MESA SDK for Linux, \mn@doi{10.5281/zenodo.3706650}, \url
  {https://doi.org/10.5281/zenodo.3706650}

\bibitem[\protect\citeauthoryear{{Turk}, {Smith}, {Oishi}, {Skory}, {Skillman},
  {Abel}  \& {Norman}}{{Turk} et~al.}{2011}]{2011ApJS..192....9T}
{Turk} M.~J.,  {Smith} B.~D.,  {Oishi} J.~S.,  {Skory} S.,  {Skillman} S.~W.,
  {Abel} T.,   {Norman} M.~L.,  2011, \mn@doi [The Astrophysical Journal
  Supplement Series] {10.1088/0067-0049/192/1/9}, \href
  {https://ui.adsabs.harvard.edu/abs/2011ApJS..192....9T} {192, 9}

\bibitem[\protect\citeauthoryear{{Ugliano}, {Janka}, {Marek}  \&
  {Arcones}}{{Ugliano} et~al.}{2012}]{2012ApJ...757...69U}
{Ugliano} M.,  {Janka} H.-T.,  {Marek} A.,   {Arcones} A.,  2012, \mn@doi
  [\apj] {10.1088/0004-637X/757/1/69}, \href
  {https://ui.adsabs.harvard.edu/abs/2012ApJ...757...69U} {757, 69}

\bibitem[\protect\citeauthoryear{{Woosley}}{{Woosley}}{1993}]{1993ApJ...405..273W}
{Woosley} S.~E.,  1993, \mn@doi [\apj] {10.1086/172359}, \href
  {https://ui.adsabs.harvard.edu/abs/1993ApJ...405..273W} {405, 273}

\bibitem[\protect\citeauthoryear{{Woosley} \& {Heger}}{{Woosley} \&
  {Heger}}{2012}]{2012ApJ...752...32W}
{Woosley} S.~E.,  {Heger} A.,  2012, \mn@doi [\apj]
  {10.1088/0004-637X/752/1/32}, \href
  {https://ui.adsabs.harvard.edu/abs/2012ApJ...752...32W} {752, 32}

\bibitem[\protect\citeauthoryear{{Yadav}, {M{\"u}ller}, {Janka}, {Melson}  \&
  {Heger}}{{Yadav} et~al.}{2020}]{2020ApJ...890...94Y}
{Yadav} N.,  {M{\"u}ller} B.,  {Janka} H.~T.,  {Melson} T.,   {Heger} A.,
  2020, \mn@doi [\apj] {10.3847/1538-4357/ab66bb}, \href
  {https://ui.adsabs.harvard.edu/abs/2020ApJ...890...94Y} {890, 94}

\bibitem[\protect\citeauthoryear{{Zhang}, {Woosley}  \& {Heger}}{{Zhang}
  et~al.}{2008}]{2008ApJ...679..639Z}
{Zhang} W.,  {Woosley} S.~E.,   {Heger} A.,  2008, \mn@doi [\apj]
  {10.1086/526404}, \href
  {https://ui.adsabs.harvard.edu/abs/2008ApJ...679..639Z} {679, 639}

\makeatother
\end{thebibliography}

%%%%%%%%%%%%%%%%%%%%%%%%%%%%%%%%%%%%%%%%%%%%%%%%%%

%%%%%%%%%%%%%%%%% APPENDICES %%%%%%%%%%%%%%%%%%%%%

\appendix

\section{Method for Predicting Accretion Rates}
\label{sec:appxpredict}
In this section we discuss how mass and angular momentum accretion rates can be predicted from the instantaneous snapshot of the convective flow at the start of collapse. Our interest here is in cases where the gas is roughly in hydrostatic equilibrium at the start of collapse (i.e this is not free-fall-from-rest collapse; the gas has pressure support).  Here we use a simple test problem to illustrate the procedure.  

For the test problem, we use a power-law density profile in hydrostatic equilibrium with a point mass potential. Functionally, the potential, the density profile, and the adiabatic sound speed profile are, respectively,
\begin{align}
\Phi(r) &= -GM/r \\
\rho(r) &= \rho_0 (r_0/r)^b \\
c_s(r) &= \sqrt{\frac{\gamma}{(b+1)} \frac{GM}{r}},
\label{eq:appx_cs}
\end{align}
where $r_0$, $\rho_0$, and $b$ are constants and $\gamma$ is the adiabatic index of the gas. Code units are the same as defined in Section~\ref{sec:units}. We use $b = 2.5$ and $\gamma = 4/3$ for our test problem.  At initialization, we give all of the gas in the domain the same angular velocity, $\Omega$, about the z-axis.  This gives a $z$-angular momentum density profile of 
%\beq
%j_z(r) = 
$\frac{2}{3}\Omega r^2 \rho(r)$.
%\eeq
The mass contained in the shell at radius $r$ and thickness $\Delta r$ is
\beq
\Delta m(r) = 4\pi r^2 \rho(r) \Delta r
\eeq
and the z-angular momentum in the shell is 
\beq
\Delta J_z(r) = 4\pi r^2 \bigg[\frac{2}{3}\Omega r^2 \rho(r)\bigg]\Delta r = \frac{8\pi}{3}  \Omega r^4 \rho(r) \Delta r.
\eeq
%\beq
%\Delta J_z(r) = 4\pi r^2 j_z(r) \Delta r.
%\eeq

\begin{figure}
\begin{center}
	\includegraphics[width=0.7\columnwidth]{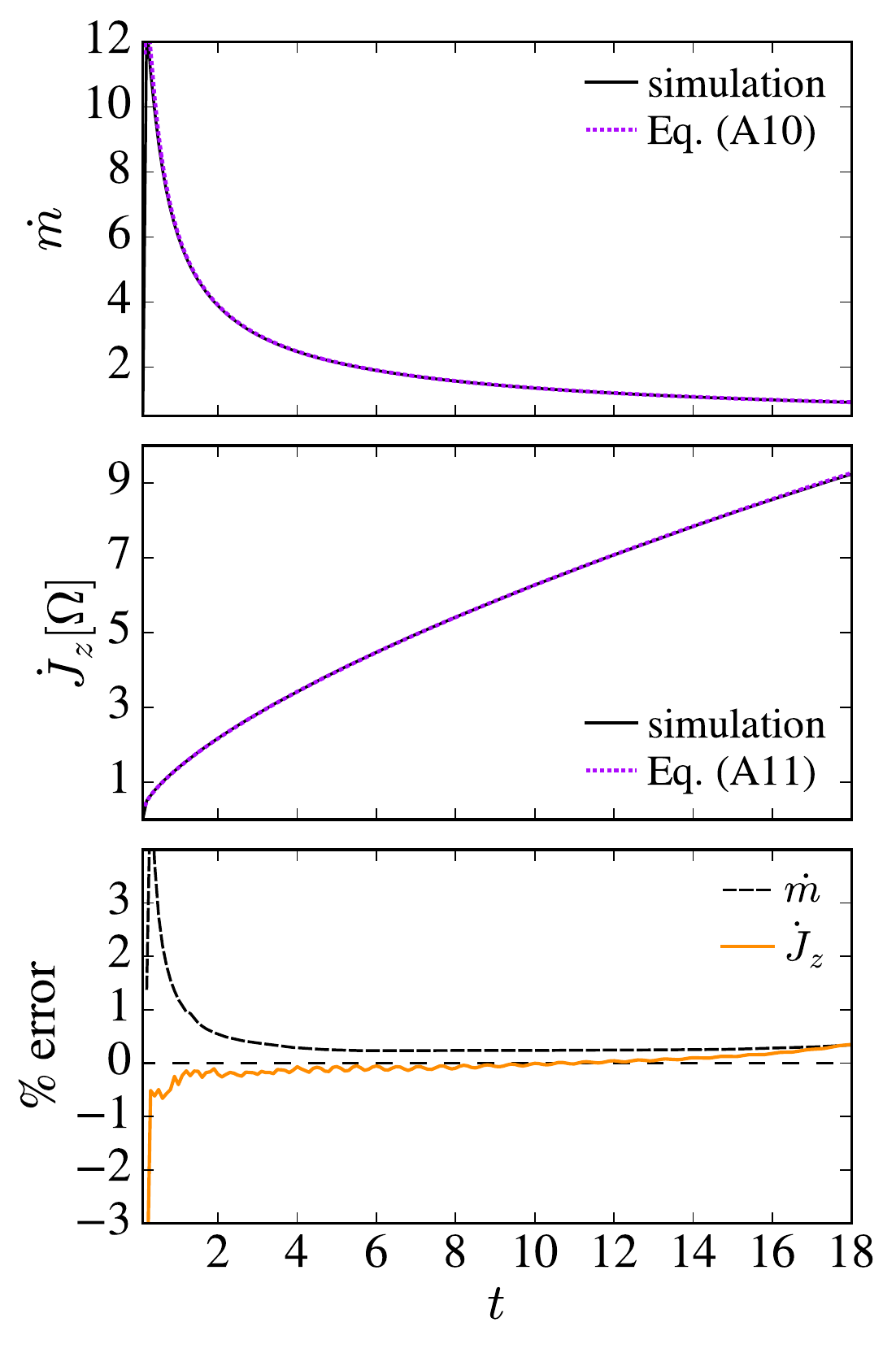}
    \caption{Top and middle panels: Comparison of semi-analytically predicted accretion rates from initial profiles (purple, dotted lines) to the accretion rates measured at the sink radius during the simulation (black, solid lines). Bottom panel: Percent error between semi-analytical and simulated mass (black, dashed curve) and $z$-angular momentum (solid, orange curve) accretion rates. Predicted rates are good to better than a percent after the inner material is accreted. The $< 1\%$ error for the angular momentum prediction shows that angular momentum is conserved to a high degree of accuracy as material falls supersonically through the Cartesian grid.}
    \label{fig:APPX_accretion_prediction}
 \end{center}
 \end{figure}

To predict the accretion rate, we compute the time it takes the shell at $r$ to reach the sink (with radius $r_s$) after the sink is activated. Call this function $t_{\rm acc}(r)$.  We also need to know the time over which that shell accretes, call that $\Delta t_{\rm acc}(r)$.  When we activate the sink, a rarefaction wave begins at $r_s$ and propagates outward in radius.  The time it takes the rarefaction wave to reach $r$ is the integrated sound-crossing time from $r_s$ to $r$,
\beq
t_{\rm wave}(r)  = \int_{r_s}^r \frac{dr'}{c_s(r')} = \frac{2}{3}\bigg(\frac{b+1}{\gamma GM}\bigg)^{1/2}\bigg[r^{3/2} - r_s^{3/2}\bigg].
\label{eq:integrated_cs}
\eeq
If the pressure behind the rarefaction wave were zero, then the shell would fall from rest from its initial position at $r$ to the sink radius $r_s$ over the integrated free-fall time, 
\beq
t_{\rm ff}(r) = \bigg(\frac{r^3}{2GM}\bigg)^{1/2}\Bigg\{\frac{\pi}{2} - \sin^{-1}\bigg(\sqrt{\frac{r_s}{r}}\bigg) + \sqrt{\frac{r_s}{r}\Big(1 - \frac{r_s}{r}\Big)}\Bigg\}.
\label{eq:freefalltime}
\eeq
There is, in fact, a small pressure gradient behind the rarefaction wave \citep{2019ApJ...874...58C}, so the time for the shell to fall back after the arrival of the rarefaction wave is actually $1.14\, t_{\rm ff}(r)$ for our simulated values of $b$ and $\gamma$ (E. R. Coughlin, private communication).  The total time required for the shell at $r$ to reach the sink is thus
\beq
t_{\rm acc}(r) = 1.14 \, t_{\rm ff}(r) + t_{\rm wave}(r).
\label{eq:acctime}
\eeq
and 
\beq
\Delta t_{\rm acc}(r) = t_{\rm acc}(r+\Delta r/2) - t_{\rm acc}(r-\Delta r/2) .
\eeq
We want accretion rates as a function of time, so we invert $t_{\rm acc}(r)$ to obtain $r_{\rm acc}(t)$.   The analytical accretion rate predictions are thus
\begin{align}
\dot{m}(t) &= {\Delta m(r_{\rm acc})}/{\Delta t_{\rm acc}(r_{\rm acc})} \label{eq:mdot_predict}\\
\dot{J}_z(t) &= {\Delta J_z(r_{\rm acc})}/{\Delta t_{\rm acc}(r_{\rm acc})}.
\label{eq:jdot_predict}
\end{align}
Figure~\ref{fig:APPX_accretion_prediction} compares eqs.~\eqref{eq:mdot_predict} and \eqref{eq:jdot_predict} to the accretion rates measured in the simulation.  The predictions reproduce the simulated rates very well. The bottom panel shows the $\%$ error between the predicted and simulated rates. Once the innermost material is accreted, the errors are less than 1\%.
%%%%%%%%%%
%%%%%%%%%

%%%%%%%%%%%%%%%%%%%%%%%%%%%%%%%%%%%%%%%%%%%%%%%%%%
%%%%%%%%%%%%%%%%%%%%%%%%%%%%%%%%%%%%%%%%%%%%%%%%%%
%%%%%%%%%%%%%%%%%%%%%%%%%%%%%%%%%%%%%%%%%%%%%%%%%%
%%%%%%%%%%%%%%%%%%%%%%%%%%%%%%%%%%%%%%%%%%%%%%%%%%

% Don't change these lines
\bsp	% typesetting comment
\label{lastpage}
\end{document}